\documentclass[twocolumn, twocolappendix]{aastex63}

\usepackage{amsmath, amssymb}
\usepackage{bm}
\usepackage{color}
\usepackage{float}
\usepackage{xcolor}
\usepackage[utf8]{inputenc}
\usepackage{mathtools}
\usepackage{enumitem}
\usepackage{epstopdf}
\usepackage{graphicx}
\usepackage{listings}
\usepackage{lineno}

\providecommand{\sorthelp}[1]{}

\definecolor{ol}{rgb}{0.15,0.5,0.15}
\definecolor{mb}{rgb}{0.4,0,0.35}
\definecolor{tb}{rgb}{0,0.3,0.4}
\definecolor{br}{rgb}{0.45,0.15,0.15}
\definecolor{gr}{rgb}{0.5,0.5,0.5}
\definecolor{dg}{rgb}{0.85,0.6,0}

\definecolor{Gray}{gray}{0.9}

\makeatletter
\def\thickhline{%
  \noalign{\ifnum0=`}\fi\hrule \@height \thickarrayrulewidth \futurelet
   \reserved@a\@xthickhline}
\def\@xthickhline{\ifx\reserved@a\thickhline
               \vskip\doublerulesep
               \vskip-\thickarrayrulewidth
             \fi
      \ifnum0=`{\fi}}
\makeatother

\newlength{\thickarrayrulewidth}
\newcommand\cobe{\text{COBE}}
\newcommand\wmap{\text{WMAP}}
\newcommand\planck{\text{Planck}}
\newcommand\litebird{\text{LiteBIRD}}

\def\commander{\texttt{Commander}}
\def\nilc{\texttt{NILC}}
\def\sevem{\texttt{SEVEM}}
\def\smica{\texttt{SMICA}}
\def\xqml{\texttt{xQML}}

\usepackage[encapsulated]{CJK}
\usepackage{ucs}
\newcommand{\cntext}[1]{\begin{CJK}{UTF8}{gbsn}#1\end{CJK}}

\usepackage{devanagari} 

\received{June 1, 2019}
\revised{January 10, 2019}
\accepted{\today}
\submitjournal{\apj}

\shorttitle{CMB Anomalies and Polarization}
\shortauthors{Shi et al.}

\newcommand*\patchAmsMathEnvironmentForLineno[1]{
  \expandafter\let\csname old#1\expandafter\endcsname\csname #1\endcsname
  \expandafter\let\csname oldend#1\expandafter\endcsname\csname end#1\endcsname
  \renewenvironment{#1}
  {\linenomath\csname old#1\endcsname}
  {\csname oldend#1\endcsname\endlinenomath}}
  \newcommand*\patchBothAmsMathEnvironmentsForLineno[1]{
  \patchAmsMathEnvironmentForLineno{#1}
  \patchAmsMathEnvironmentForLineno{#1*}}
  \AtBeginDocument{
  \patchBothAmsMathEnvironmentsForLineno{equation}
  \patchBothAmsMathEnvironmentsForLineno{align}
  \patchBothAmsMathEnvironmentsForLineno{flalign}
  \patchBothAmsMathEnvironmentsForLineno{alignat}
  \patchBothAmsMathEnvironmentsForLineno{gather}
  \patchBothAmsMathEnvironmentsForLineno{multline}
}

\begin{document}

\title{Testing CMB Anomalies in E-mode Polarization with Current and Future Data} 

\correspondingauthor{Rui Shi}
\email{rshi9@jhu.edu}
\author[0000-0001-7458-6946]{Rui Shi (\cntext{时瑞}\!\!)}
\affiliation{The William H. Miller III Department of Physics and Astronomy, Johns Hopkins University, 3701 San Martin Drive, Baltimore, MD 21218, USA}

\author[0000-0003-4496-6520]{Tobias~A. Marriage}
\affiliation{The William H. Miller III Department of Physics and Astronomy, Johns Hopkins University, 3701 San Martin Drive, Baltimore, MD 21218, USA}

\author[0000-0002-8412-630X]{John W. Appel}
\affiliation{The William H. Miller III Department of Physics and Astronomy, Johns Hopkins University, 3701 San Martin Drive, Baltimore, MD 21218, USA}

\author[0000-0001-8839-7206]{Charles~L. Bennett}
\affiliation{The William H. Miller III Department of Physics and Astronomy, Johns Hopkins University, 3701 San Martin Drive, Baltimore, MD 21218, USA}

\author[0000-0003-0016-0533]{David~T. Chuss}
\affiliation{Department of Physics, Villanova University, 800 Lancaster Avenue, Villanova, PA 19085, USA}

\author[0000-0002-7271-0525]{Joseph Cleary}
\affiliation{The William H. Miller III Department of Physics and Astronomy, Johns Hopkins University, 3701 San Martin Drive, Baltimore, MD 21218, USA}

\author[0000-0001-6976-180X]{Joseph Eimer}
\affiliation{The William H. Miller III Department of Physics and Astronomy, Johns Hopkins University, 3701 San Martin Drive, Baltimore, MD 21218, USA}

\author[0000-0002-1708-5464]{Sumit Dahal ({\dn \7{s}Emt dAhAl})}
\affiliation{Goddard Space Flight Center, 8800 Greenbelt Road, Greenbelt, MD 20771, USA}

\author[0000-0003-3853-8757]{Rahul Datta}
\affiliation{The William H. Miller III Department of Physics and Astronomy, Johns Hopkins University, 3701 San Martin Drive, Baltimore, MD 21218, USA}

\author{Francisco Espinoza}
\affiliation{Facultad de Ingenier\'ia, Universidad Cat\'olica de la Sant\'isima Concepci\'on, Alonso de Ribera 2850, Concepci\'on, Chile}

\author[0000-0002-4820-1122]{Yunyang Li (\cntext{李云炀}\!\!)}
\affiliation{The William H. Miller III Department of Physics and Astronomy, Johns Hopkins University, 3701 San Martin Drive, Baltimore, MD 21218, USA}

\author[0000-0002-2245-1027]{Nathan~J. Miller}
\affiliation{The William H. Miller III Department of Physics and Astronomy, Johns Hopkins University, 3701 San Martin Drive, Baltimore, MD 21218, USA}

\author[0000-0002-5247-2523]{Carolina N\'u\~nez}
\affiliation{The William H. Miller III Department of Physics and Astronomy, Johns Hopkins University, 3701 San Martin Drive, Baltimore, MD 21218, USA}

\author[0000-0002-0024-2662]{Ivan~L. Padilla}
\affiliation{The William H. Miller III Department of Physics and Astronomy, Johns Hopkins University, 3701 San Martin Drive, Baltimore, MD 21218, USA}

\author[0000-0002-4436-4215]{Matthew~A. Petroff}
\affiliation{Center for Astrophysics, Harvard \& Smithsonian, 60 Garden Street, Cambridge, MA 02138, USA}

\author[0000-0003-3487-2811]{Deniz A. N. Valle}
\affiliation{The William H. Miller III Department of Physics and Astronomy, Johns Hopkins University, 3701 San Martin Drive, Baltimore, MD 21218, USA}

\author[0000-0002-7567-4451]{Edward~J. Wollack}
\affiliation{Goddard Space Flight Center, 8800 Greenbelt Road, Greenbelt, MD 20771, USA}

\author[0000-0001-5112-2567]{Zhilei Xu (\cntext{徐智磊}\!\!)}
\affiliation{MIT Kavli Institute, Massachusetts Institute of Technology, 77 Massachusetts Avenue, Cambridge, MA 02139, USA}


\begin{abstract}
In this paper, we explore the power of the cosmic microwave background (CMB) polarization (E-mode) data to corroborate four potential anomalies in CMB temperature data: the lack of large angular-scale correlations, the alignment of the quadrupole and octupole (Q-O), the point-parity asymmetry, and the hemispherical power asymmetry.
We use CMB simulations with noise representative of three experiments -- the \planck{} satellite, the Cosmology Large Angular Scale Surveyor (CLASS), and the \litebird{} satellite -- to test how current and future data constrain the anomalies. 
We find the correlation coefficients $\rho$ between temperature and E-mode estimators to be less than $0.1$, except for the point-parity asymmetry ($\rho=0.17$ for cosmic-variance-limited simulations), confirming that E-modes provide a check on the anomalies that is largely independent of temperature data. 
Compared to \planck{} component-separated CMB data (\smica{}), the putative \litebird{} survey would reduce errors on E-mode anomaly estimators by factors of $\sim3$ for hemispherical power asymmetry and point-parity asymmetry, and by $\sim26$ for lack of large-scale correlation. 
The improvement in Q-O alignment is not obvious due to large cosmic variance, but we found the ability to pin down the estimator value will be improved by a factor $\gtrsim100$. 
Improvements with CLASS are intermediate to these.
\end{abstract}

\keywords{\href{http://astrothesaurus.org/uat/322}{Cosmic microwave background radiation (322)}; \href{http://astrothesaurus.org/uat/435}{Early Universe (435)}; \href{http://astrothesaurus.org/uat/1146}{Observational Cosmology (1146)}}

\section{Introduction}\label{sec:intro}
The cosmic microwave background (CMB) anisotropy on large angular scales has been measured with increasing precision over the past few decades by 3 full-sky experiments, namely, the Cosmic Background Explorer (\cobe), the Wilkinson Microwave Anisotropy Probe (\wmap), and the \planck{} satellite.
The CMB data from these experiments largely agree with the six-parameter standard $\Lambda$CDM model.
However, several notable deviations at large angular scales in the temperature data have been identified. 
These include the lack of large-scale correlations \citep{hinshaw1996two, Bennett_2003, copi2009no}, alignment of low multipoles \citep{tegmark2003high, de2004significance, copi2004multipole}, point-parity asymmetry \citep{land2005universe, kim2010anomalous}, hemispherical power asymmetry \citep{eriksen2004asymmetries, monteserin2008low, akrami2014power}, and cold spots \citep{vielva2004detection, cruz2005detection}.
While the lack of large scale correlations with the associated low quadrupole were observed by \cobe{} \citep{hinshaw1996two}, the majority of these temperature deviations were first found in \wmap{}  data \citep{bennett2011seven}, and then were confirmed by \planck{} \citep{akrami2020planck}. 
Although the significance of these large-scale deviations is challenging to pin down \citep[e.g,][]{bennett2011seven}, throughout this paper we will refer to the features by their common name: ``CMB anomalies''. 
For a review of CMB temperature anomalies, see \citet{schwarz2016cmb}. 

For each anomaly, various statistical measures have been formulated to quantify its significance in a standard $\Lambda$CDM universe. 
Most of the statistical tests show mild (2$-$3$\sigma$) deviations compared to the $\Lambda$CDM model, and the a-posteriori nature of the observations requires that care be taken to avoid overstating the significance level.
Consequently, there are three most commonly accepted postulations as to why these anomalies are present in the data.
First, they could have cosmological origins.
For example, there have been several attempts to use ``just-enough'' slow-roll inflation to explain the lack of correlations on large scales \citep{ramirez2012predictions, cicoli2014just}, and introducing modulation fields to the primordial curvature fluctuation spectrum $P(k)$ can explain the hemispherical asymmetry and the quadrupole-octupole alignment \citep{gordon2005spontaneous, dvorkin2008testable}.
The cosmological origin hypothesis is the most exciting explanation because it implies new physics beyond the standard $\Lambda$CDM model.
However, these mechanisms have their corresponding weaknesses: it might not be worth introducing more parameters to explain a single or a few 2-3 sigma level anomalies; the modulation fields are also hard to physically motivate \citep{schwarz2016cmb}.
A second postulate is that these anomalies could result from foreground effects due to the solar system, the Milky Way, and the local supercluster; or systematic effects in the data (see e.g. \citealt{hansen2004asymmetries, abramo2006anomalies, copi2007uncorrelated, groeneboom2008bayesian, hansen2012can, frejsel2015large}).
The primary challenge to these hypotheses is that the same anomalies are observed in both \planck{} and \wmap, which have different sky-scanning strategies, and rely primarily on different frequency bands to estimate the CMB.
The third prevailing view is that the anomalies are merely unlikely fluctuations in our universe \citep{bennett2011seven}.
A better understanding of the anomalies will either support this ``fluke hypothesis'' or deepen our understanding of the universe.
But since CMB temperature measurements have reached the cosmic-variance-limit at large angular scales, information from additional observations is needed.

To extend this investigation, studies have explored  a range of cosmological probes, including 21-cm observations \citep{shiraishi2016violation, li2019testing}, gravitational lensing {\citep{yoho2014probing, mukherjee2016litmus, zibin2017testing}}, the integrated Sachs-Wolfe effect \citep{muir2016reconstructing, copi2016isw}, and the Sunyaev-Zel’dovich effect \citep{cayuso2020towards}. 
In this paper, however, we focus on CMB polarization measurements.
In particular, the dominant E-mode polarization component promises to provide the next most significant signal after the CMB temperature anisotropies.
The \planck{} collaboration recently searched for the existence of such potential anomalies using their polarization data, but due to the low signal-to-noise at large angular scales, different component-separated maps resulted in  different significance levels \citep{akrami2020planck}.
Progress has been made by using single-frequency \planck{} polarization maps \citep{chiocchetta2020lack}. 
Higher sensitivity observations can provide more details about these phenomena.

The situation may improve in the near future with new data from CMB polarization experiments that targeting the large angular scales over greater than 30\% of the sky \cite[e.g.,][]{essinger2014class, lee2020groundbird, perez2016quijote, addamo2021large, gandilo2016primordial, hazumi2020litebird, kogut2011primordial, hanany2019pico}.
Recent studies forecast the constraining power of future polarization experiments on individual anomalies  \citep[e.g.,][]{billi2019polarisation, o2020hemispherical, chiocchetta2020lack}.
It is also possible to separate polarization data into parts that are fully correlated or uncorrelated with temperature, as explored in \citet{frommert2010axis}.
Exploration of multiple anomalies and their correlations has been done in temperature \citep{muir2018covariance}, but not in polarization.

In this paper, we evaluate the significance of anomalies in temperature and polarization taking into consideration the correlation among anomalies in temperature and polarization. 
We first analyze \planck{} data to establish  baseline constraints and validate our pipeline using existing results. 
Next we forecast constraints from the ground-based Cosmology Large Angular Scale Surveyor (CLASS) \citep{harrington2016cosmology,dahal2021four} and from \litebird{} \citep{hazumi2020litebird}. 
Both CLASS and \litebird{} anticipate sample-variance-limited E-mode measurements. 
Therefore, the primary difference between CLASS and \litebird{} in this paper is the partial sky coverage of CLASS ($\sim 70\%$ of the sky fraction of \litebird
).
The four CMB large-scale anomalies we considered in our work are: the lack of large-angular correlations, the alignment of the quadrupole and octupole, the point-parity asymmetry, and the hemispherical power asymmetry.

The structure of this paper is as follows: in Section \ref{sec:methods}, we describe the maps, masks, and simulations we used. Section \ref{sec:anomalies}  introduces the different large-scale anomalies and their statistics. 
We present our results in Section \ref{sec:results} and conclude in Section \ref{sec:conclusion}.

\section{Methods}\label{sec:methods}
Most analyses in this work are based on the temperature map $T(\hat{\bm{n}})$ and maps of the linear polarization Stokes parameters $Q(\hat{\bm{n}})$ and $U(\hat{\bm{n}})$. 
They can be expanded with spherical harmonics $Y_{\ell m}(\hat{\bm{n}})$ and spin-2 spherical harmonics ${}_{\pm 2}Y_{\ell m}(\hat{\bm{n}})$ as
\begin{equation}
\begin{aligned}
T(\hat{\bm{n}}) & = \sum_{\ell,m} a_{\ell m}^{T}Y_{\ell m}(\hat{\bm{n}}),\\
(Q\pm iU) (\hat{\bm{n}}) &= \sum_{\ell,m} a_{\ell m}^{(\pm 2)}{}_{\pm 2}Y_{\ell m}(\hat{\bm{n}}),
\end{aligned}\label{eq:spharm}
\end{equation}
where $a_{\ell m}^{T}$'s and $a_{\ell m}^{(\pm 2)}$'s are the expansion coefficients \citep{zaldarriaga1997all}.

For the polarization analysis of the hemispherical power asymmetry, we constructed the rotationally invariant E-mode map, which can be expanded with spherical harmonics as
\begin{equation}
\begin{aligned}
    E(\hat{\bm{n}}) &= \sum_{\ell, m} a_{\ell m}^EY_{\ell m}(\hat{\bm{n}}).
\end{aligned}\label{eq:Emap}
\end{equation}
The expansion coefficients $a_{\ell m}^E$'s are related to $a_{\ell m}^{(\pm 2)}$'s by
\begin{equation}
\begin{aligned}
    a_{\ell m}^E &= -\frac{1}{2}\left(a_{\ell m}^{(+2)} + a_{\ell m}^{(-2)}\right).
\end{aligned}
\end{equation}
Denoting temperature and E-mode maps as $X(\hat{\bm{n}})$ where $X\in\{T,E\}$, the variance and covariance of their expansion coefficients give the power spectra $C_\ell^{XY}$. 
In the $\Lambda$CDM model, this can be written as
\begin{equation}
    \langle a_{\ell m}^{X}a_{\ell^\prime m^\prime}^{Y*}\rangle = \delta_{\ell\ell^\prime}\delta_{mm^\prime}C_\ell^{XY},\label{eq:defaps}
\end{equation}
where $\langle\cdot\rangle$ refers to taking the ensemble average and $XY\in\{TT, TE, EE\}$.

{The temperature two-point correlation function is defined as $\mathcal C^{TT}(\theta)\equiv\langle T(\hat{\bm n}_1)T(\hat{\bm n}_2)\rangle$ with $\hat{\bm n}_1\cdot\hat{\bm n}_2=\cos\theta$, and is related to the power spectra  (using equations defined above) as
\begin{equation}
    \mathcal{C}^{TT}(\theta)=\sum_{\ell}\frac{2\ell+1}{4\pi}C_\ell^{TT}\mathcal{P}_\ell(\cos\theta),\label{eq:Taps2cf}
\end{equation}
where $\mathcal{P}_\ell$'s are the Legendre Polynomials. 
For the E-mode, as pointed out in \citet{yoho2015microwave}, the physical interpretation of the two-point correlation function defined based on Equation \ref{eq:Emap} is ambiguous because an integral over the full sky is required to extract the $a_{\ell m}^E$ information. 
We followed the path in \citet{yoho2015microwave} and defined the correlation function with local E-modes ($\widehat E(\hat{\bm n})$) that can be calculated from the Stokes Q and U using local spin raising and lowering operators \citep{zaldarriaga1997all, kamionkowski1997statistics}:
\begin{equation}
    \widehat E(\hat{\bm n})=\sum_{\ell, m}\sqrt{\frac{(\ell+2)!}{(\ell-2)!}}a_{\ell m}^EY_{\ell m}(\hat{\bm n}).
\end{equation}
The corresponding E-mode correlation function ($\mathcal C^{EE}(\theta)\equiv\langle \widehat E(\hat{\bm n}_1)\widehat E(\hat{\bm n}_2)\rangle$) is
\begin{equation}
    \mathcal{C}^{EE}(\theta)=\sum_{\ell}\frac{2\ell+1}{4\pi}\frac{(\ell+2)!}{(\ell-2)!}C_\ell^{EE}\mathcal{P}_\ell(\cos\theta).\label{eq:Eaps2cf}
\end{equation}
The estimation of power spectra will be explained in Section \ref{sec:xQML}.}

\subsection{Data and Masks}
\begin{figure*}[t]
    \centering
    \includegraphics[width=\linewidth]{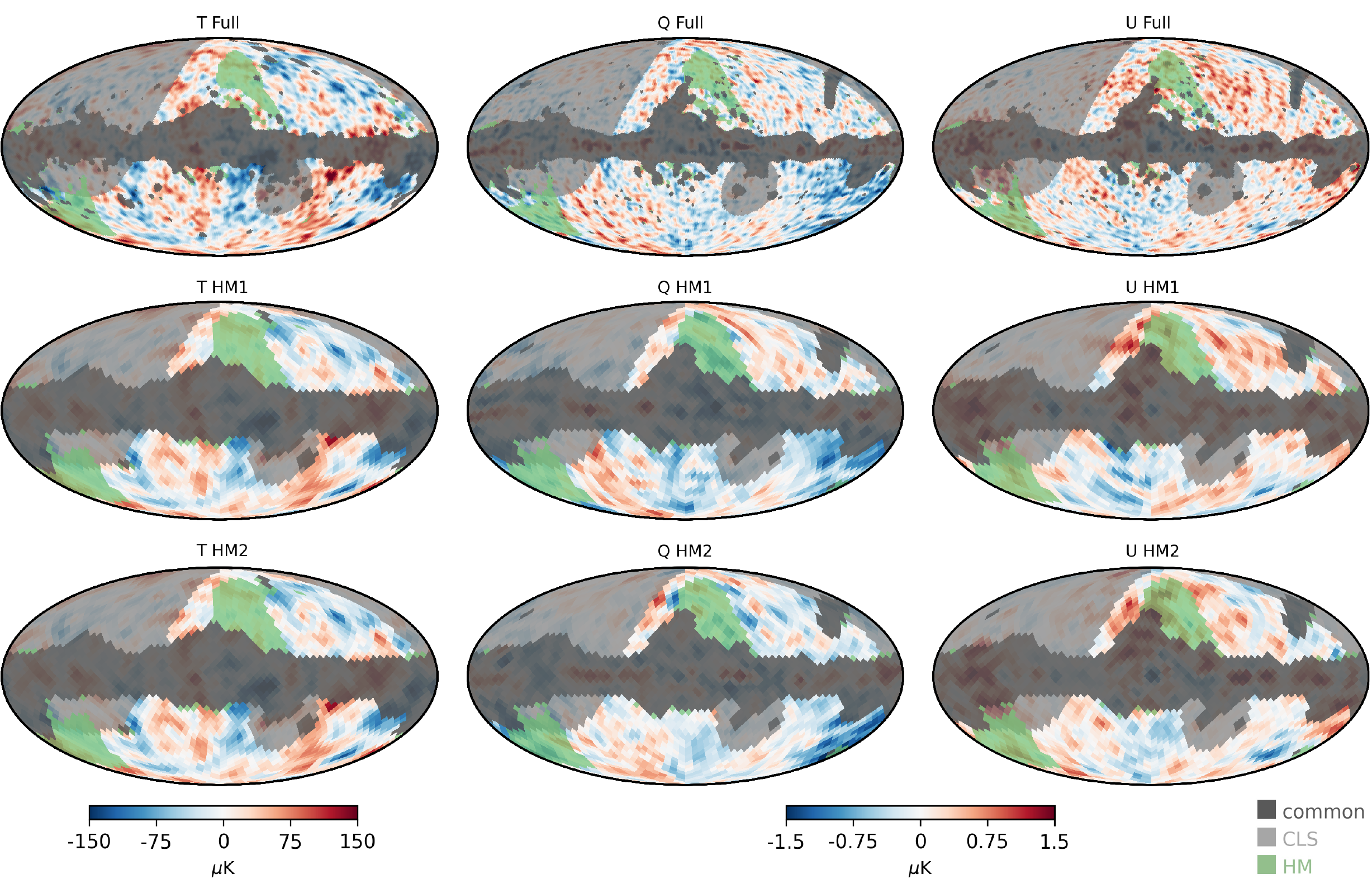}
    \caption{\planck{} \smica{} maps and corresponding masks at \texttt{HEALPix} resolution NSIDE=64 (upper panels, showing the full maps) and 16 (middle and lower panels, middle for half mission (HM) 1 and lower for HM2), with T, Q and U maps from left-to-right. 
    Dark gray regions reflect the \planck{} common masks, light gray regions reflect the CLASS declination limits, and green regions are the parts of the \planck{} HM missing pixel masks that are outside of the \planck{} common masks.
    The HM1\&2 intensity maps agree well with each other, but the polarization maps are different due to noise.}
    \label{fig:maps}
\end{figure*}
We used several data products from the \planck{} 2018 data release.\footnote{Available on \href{https://pla.esac.esa.int/\#home}{Planck Legacy Archive.}} 
We used the \planck{} \smica{} component separation maps, both full mission and half mission (HM) maps\footnote{\texttt{COM\_CMB\_IQU-smica\_2048\_R3.00.fits}, \texttt{ COM\_CMB\_IQU-smica\_2048\_R3.00\_hm1/2.fits}}, in our analyses.
For analyses based on other component separation methods (\commander, \texttt{NILC}, and \texttt{SEVEM}), we refer the reader to \citet{akrami2020planck}.
In this paper, we used \planck{} \smica{}.
Using different component separated maps does not affect our main conclusions.
The \planck{} maps we used in the paper were originally created at \texttt{HEALPix} \citep{gorski2005cosmology} resolution NSIDE~=~2048, at an approximate angular resolution of $5\arcmin$ full width half maximum (FWHM).
Since we are interested in large angular-scale structures, the original versions were downgraded to lower \texttt{HEALPix} resolutions at NSIDE~=~64 and 16, with pixel resolutions being $\sim55\arcmin$ and $\sim220\arcmin$, respectively. 
The NSIDE~=~64 maps were used for map-based analyses of the quadrupole-octupole alignment and the hemispherical power asymmetry. 
For computational efficiency, NSIDE~=~16 maps were used for computing power spectra at low multipoles. 
Power spectra were used in evaluating statistics for anomalies associated with the lack of large scale correlations and the point-parity asymmetry.

The original maps were downweighted in harmonic space according to \citep{muir2018covariance, akrami2020planck}
\begin{equation}
\begin{aligned}
a_{\ell m}^{\mathrm{(out)}} =  \frac{b_\ell^{\mathrm{(out)}}p_\ell^{\mathrm{(out)}}}{b_\ell^{\mathrm{(in)}}p_\ell^{\mathrm{(in)}}}a_{\ell m}^{\mathrm{(in)}},\label{eq:sd}
\end{aligned}
\end{equation}
where $b_{\ell}$'s are Gaussian smoothing functions and $p_\ell$'s are \texttt{HEALPix} window functions. Superscripts $^{\mathrm{(in)}}$ and $^{\mathrm{(out)}}$ stand for input and output maps, respectively. 
For the \planck{} data, the function $b_\ell^{\mathrm{(in)}}$ was the Gaussian approximation to the \planck{} beam transform. 
The Gaussian transform $b_\ell^{\mathrm{(out)}}$ set the output smoothing scale and prevented aliasing of small-scale (high-$\ell$) noise to larger scales when downgrading the resolution of the map.
To downgrade maps to NSIDE~=~64 and NSIDE~=~16, the output Gaussian smoothing scale was chosen to be FWHM~=~160$\arcmin$ and FWHM~=~640$\arcmin$, respectively.
For power spectrum analyses of \planck{} polarization data, instead of using Equation \ref{eq:sd}, we smoothed maps with the filter from the low-$\ell$ \planck{} polarization analysis \cite{aghanim2016planck, aghanim2020planck}:
\begin{gather}
a_{\ell m}^{\mathrm{(out)}} = \frac{p_\ell^{\mathrm{(out)}}}{p_\ell^{\mathrm{(in)}}}f(\ell)a_{\ell m}^{\mathrm{(in)}},\label{eq:newbeam} \\
f(\ell)=\begin{cases}
1, & \ell\leq 16;\\
\frac{1}{2}\left(1+\sin\left(\frac{\pi}{2}\frac{\ell}{16}\right)\right), & 16<\ell\leq 48;\\
0, & \ell>48.
\end{cases} \nonumber
\end{gather}
Finally the downgraded resolution (NSIDE) map is constructed from the set of smoothed $a_{\ell m}^\text{(out)}$.
{The E-mode maps are created from NSIDE=64 maps directly, following Equation \ref{eq:Emap}.}

We used different masks in our work for different purposes.
For the full-sky analysis, we used the 2018 version of the \planck{} common masks\footnote{\texttt{COM\_Mask\_CMB-common-Mask-Int/Pol\_2048\_R3.00.fits}} for temperature and Q/U polarization.
These masks were originally generated at NSIDE=2048.
To obtain binary masks at lower resolutions, we first smoothed and downgraded them according to Equation \ref{eq:sd}, and then set pixel values which were smaller than 0.9 to be 0 (masked), otherwise 1 (unmasked).
For the partial-sky coverage case, we adopted the survey limits for CLASS. 
Specifically, we used the same common masks and survey declination limits of $-70^\circ<\delta<30^\circ$.
When dealing with \planck{} HM data and noise simulations, we additionally combined the \planck{} HM missing pixel masks\footnote{\texttt{COM\_Mask\_CMB-HM-Misspix-Mask-Int/Pol\_2048\_R3.00.fits}} with the \planck{} common masks.
{We first combined the HM missing pixel masks with common masks at NSIDE=2048 then processed the combinations in the same way as for the common masks.}
To reduce impacts from E/B mixing induced by constructing E-mode maps from a set of $a_{\ell m}$'s computed on the masked sky, we created E-mode masks for the hemispherical power asymmetry analyses by setting thresholds on the residual E/B signal. 
Details can be found in Appendix \ref{sec:appendixa}.

The temperature and Q/U polarization maps together with the masks we used in this work are shown in Figure~\ref{fig:maps}. 
E-mode masks are given in Appendix~\ref{sec:appendixa}.
The sky fractions of all the masks used in this work are listed in Table~\ref{tab:fsky}.
\begin{deluxetable}{@{\extracolsep{4pt}}ccccccc}[h]
    \tablenum{1}
    \tablecaption{Sky fractions $[\%]$ for analyses of the Full-sky \planck{} and \litebird{} observations, the Partial-sky observations by CLASS, and the Half Mission (HM) \planck{} data.}\label{tab:fsky}
    \tablewidth{0pt}
    \tablehead{\colhead{} & \multicolumn{2}{c}{Temperature} & \multicolumn{2}{c}{Stokes Q and U} & \multicolumn{2}{c}{E-mode}\\
    \cline{2-3} \cline{4-5} \cline{6-7} \colhead{NSIDE} & \colhead{64} & \colhead{16} & \colhead{64} & \colhead{16} & \colhead{64} & \colhead{16}
    }
    \startdata
    Full & 71 & 64 & 72 & 64 & 62 & --$^a$\\
    Partial & 51$^b$ & 43 & 52$^b$ & 43 & 42$^b$ & --$^a$\\
    HM$^c$ & 65 & 56 & 67 & 56 & 55 & --$^a$
    \enddata
    ~\\[-2pt]
    $^a$We did not use any NSIDE=16 E-mode maps.\\
    $^b$For Q-O alignment, we did not mask out the galaxy for the full and HM case, and only adopted the declination limit for partial sky case. 
    The corresponding sky fraction is 70. More details can be found in Section \ref{sec:Q-O}.\\
    $^c$We combined the corresponding \planck{} HM missing pixel masks with the common masks when processing the masks.
\end{deluxetable}

\subsection{Simulations}
\label{sec:simulations}
We used 4 different levels of noise simulations to estimate the 95\% confidence intervals for the statistical measures of anomalies. 
To indicate the constraining power of \planck{} data, we used 300 realizations of \planck's end-to-end noise simulations\footnote{\texttt{dx12\_v3\_smica\_noise\_mc\_\{realization\}\_raw.fits},\\\texttt{dx12\_v3\_smica\_noise\_hm1/2\_mc\_\{realization\}\_raw.fits}, where \texttt{\{realization\}} is the realization number, between 00000 and 00299.} corresponding to the \smica{} foreground separation method.
These noise simulations were downgraded to NSIDE=64 (16) using the same method as applied to the data.
To estimate the constraining ability of CLASS and \litebird, we used white noise at a level of 15 $\mathrm{\mu K\cdot arcmin}$ {\citep{essinger2014class}}\footnote{{An extra factor of 1.5 is considered for the $10~\mathrm{\mu K\cdot arcmin}$ mentioned in the reference to take into consideration the red noise contributions.}}, and 2 $\mathrm{\mu K\cdot arcmin}$ {\citep{hazumi2020litebird}}, respectively, and generated 300 simulations for both. 
Despite its design to mitigate spurious polarization signals from the atmosphere and the instrument \citep{harrington2020two}, the sensitivity of the CLASS maps at large scales will deviate from the 15 $\mathrm{\mu K\cdot arcmin}$ white noise assumption, the extent of which is still being determined.
Moreover, we present CLASS forecasts for temperature with the same noise prescription despite inaccessibility of the largest scales in temperature from the ground due to the correlated brightness fluctuations from the atmosphere. 
The CLASS temperature results should be taken as a toy case to indicate the impacts of partial sky coverage.
Finally, to compare with the cosmic-variance-limited situation, we added 2 $\mathrm{\mu K\cdot arcmin}$ white noise to temperature maps and 0.02 $\mathrm{\mu K\cdot arcmin}$ white noise to polarization maps.
Noise is needed for taking the \xqml{} power spectra (Section \ref{sec:xQML}), and the impact on anomaly estimators is negligible at these noise levels.
For these cosmic-variance-limited simulations we do not apply any masks, and this specific case is referred to as the ``Ideal" case.
All white noise simulations were generated at NSIDE=64 (16) and smoothed with a 160$\arcmin$ (640$\arcmin$) FWHM Gaussian smoothing function to reproduce the same filtering as in Equation \ref{eq:sd}.

We simulated $10^4$ Gaussian random CMB realizations.
$TQU$ map sets were produced at NSIDE=64 and 16 using the theoretical power spectra based on \planck's best-fit $\Lambda$CDM model\footnote{\texttt{COM\_PowerSpect\_CMB-base-plikHM-TTTEEE-lowl-lowE-\\lensing-minimum-theory\_R3.01.txt}}. The maps were smoothed according to Equation \ref{eq:sd} or \ref{eq:newbeam} so that the properties of the simulated CMB maps are consistent with that of the real data products and/or the noise simulations.
Finally, we randomly combined the 300 noise simulations for \planck{}, CLASS, and \litebird{} with the $10^4$ CMB realizations, forming 3 sets of $10^4$ signal+noise realizations for our analyses. 
The combinations are not completely independent from one another as the number of the noise simulations is much smaller than that of the CMB realizations.
We denote the analyses in which the simulated CMB temperature and polarization signals are allowed to fluctuate freely according to the \planck{}'s best-fit model as the `unconstrained universe' study (Section \ref{ssec:unconstr}). For studies based on constrained simulations, see e.g., \citet{copi2015large,chiocchetta2020lack}.

In addition to the unconstrained universe, we considered two more simulation sets.
The first is the special CMB realization (Special CMB).
The Special CMB was selected from $10^5$ unconstrained CMB simulations such that each E-mode anomaly estimator's probability-to-exceed (PTE) with respect to the Ideal case simulations is either above $\sim95\%$ or below $\sim5\%$.
Tests based on the Special CMB distinguish the contribution of survey noise from that of cosmic variance in the posterior distributions of polarization anomaly estimators.
More details about the Special CMB can be found in Section \ref{ssec:specialCMB} and Appendix \ref{sec:appendixb}.
The second alternative simulation considered is the constrained universe, in which we fix the temperature simulations to be what has been measured from \planck{} \smica{} data and only sample E and B components as Gaussian random fields according to \planck{}'s best-fit $\Lambda$CDM model.
{When generating the constrained simulations, the temperature $a_{\ell m}$'s were derived from the full sky data.
We checked the impact from foreground residuals by comparing the anomaly estimator distributions (Figure \ref{fig:constrainedtest}) constrained by either the full-sky \smica{} temperature (our baseline analysis) or \smica{} temperature with the Galactic regions inpainted as in \citet{planck2016-l04}.
The differences are subtle, which means the existing residual foreground won't significantly impact the conclusions drawn with constrained simulations.}
Looking into distributions from the constrained simulations is useful in making forecasts given the existing temperature measurements.
We found that the predictions from the constrained universe (Appendix \ref{sec:appendixc}) are largely consistent with those from the unconstrained universe.

\subsection{xQML Power Spectra}\label{sec:xQML}
We used \xqml{} to estimate the power spectra.
\xqml{} is a quadratic, maximum-likelihood-based  power spectrum estimator based on cross-correlation between maps \citep{vanneste2018quadratic}. 
A primary advantage of cross-spectra between maps with uncorrelated noise is that such spectra contain no noise bias. 
Compared to pseudo-C$_\ell$ methods \cite[e.g.,][]{Wandelt2001,chon2004fast}, quadratic maximum-likelihood estimators have the advantage of being optimal at large angular scales (low-$\ell$) \citep{Tegmark1997,bond1998estimating,Tegmark2001}.
One disadvantage of \xqml{} spectra is their computational inefficiency limits the $\ell$-range.
All cross spectra in our work were computed based on NSIDE=16 maps, and the maximum $\ell$ used in our analyses is $30$.

\begin{figure}
    \centering
    \includegraphics[width=\linewidth]{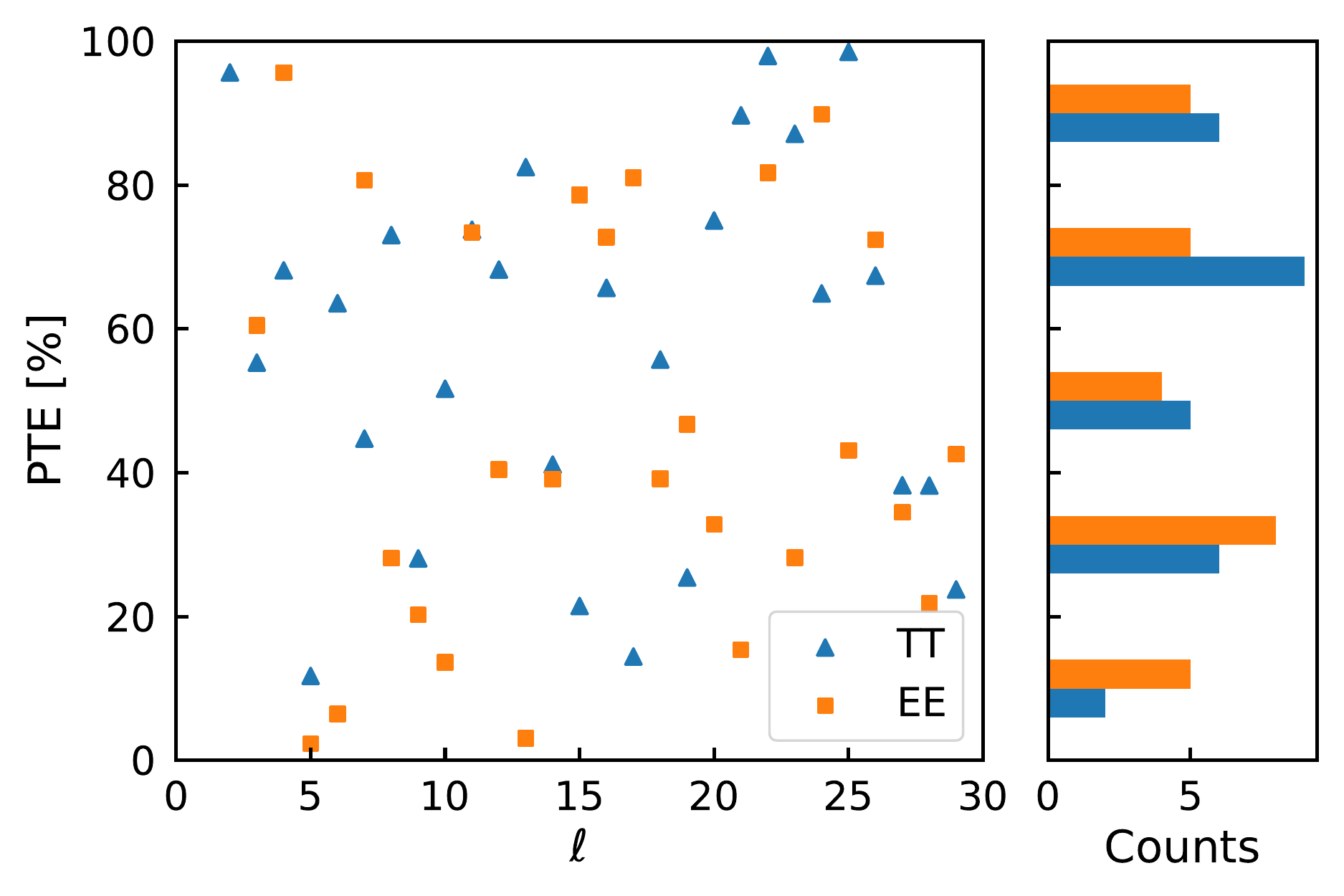}
    \caption{\textit{Left panel}: PTE coefficients (defined as the per-multipole fraction of simulated $C_\ell$'s exceeding that of the data) for TT (blue triangles) and EE (orange squares) HM cross spectra. 
    \textit{Right panel}: distributions of PTE values. 
    The distributions are largely uniform, meaning that \xqml{} power spectra estimation of the simulations are consistent with that of the data.
    }
    \label{fig:pte}
\end{figure}

Quadratic maximum-likelihood estimators require an estimate of the noise covariance matrix (NCVM). 
For \planck, the NCVM for temperature was computed based on 300 SMICA HM noise simulations smoothed and downgraded to NSIDE=16 in the same way as for other \planck{} data.
{The NCVM for estimating EE power spectra was computed based on the 300 SMICA HM noise simulations.}
We found constructing the estimator with only the diagonal component of the NCVMs produced more self-consistent and numerically stable results across the thousands of simulations performed. 
This estimator was then used to compute the cross spectra between \planck{} HM1\&2 maps for both the data and the simulations. 
Although the diagonal-only NCVMs are not optimal, using them does not bias the resulting spectra, which are still more precise than their pseudo-C$_\ell$ counterparts.  
Similarly, we used only the diagonal part of the covariance of the 300 white noise simulations for CLASS and \litebird{}.
We also adopted \planck{} HM masks to eliminate the impact of missing pixels in \planck{} High Frequency Instrument (HFI).
As a check that the simulations were representative of the data, we computed PTE values, defined as the percentage of simulations with power greater than that from the data at each multipole used in this work. 
{Figure \ref{fig:pte} shows that the resulting PTE distribution is largely uniform, meaning that the \xqml{} power spectra estimation of the simulations are consistent with that from the data.}
We did not include the quadrupole for EE analyses because \planck{} MC simulations are not able to characterize the residual systematics at $\ell=2$ \citep{akrami2020planck}.

Throughout the paper we used the constructed estimator to estimate the cross spectra between two maps with the same signal but different noise components.

\section{Anomalies studied}\label{sec:anomalies}
In this section, we give a brief introduction to the four anomalies considered in this work.
Tables \ref{tab:sum-1}, \ref{tab:sum-2}, and \ref{tab:sum-4} summarize masks and noise simulations we used for each anomaly, and are provided in the corresponding subsections.

\subsection{Lack of Large-Angular Correlations}\label{sec:correlation}
\begin{figure}
    \centering
    \includegraphics[width=\linewidth]{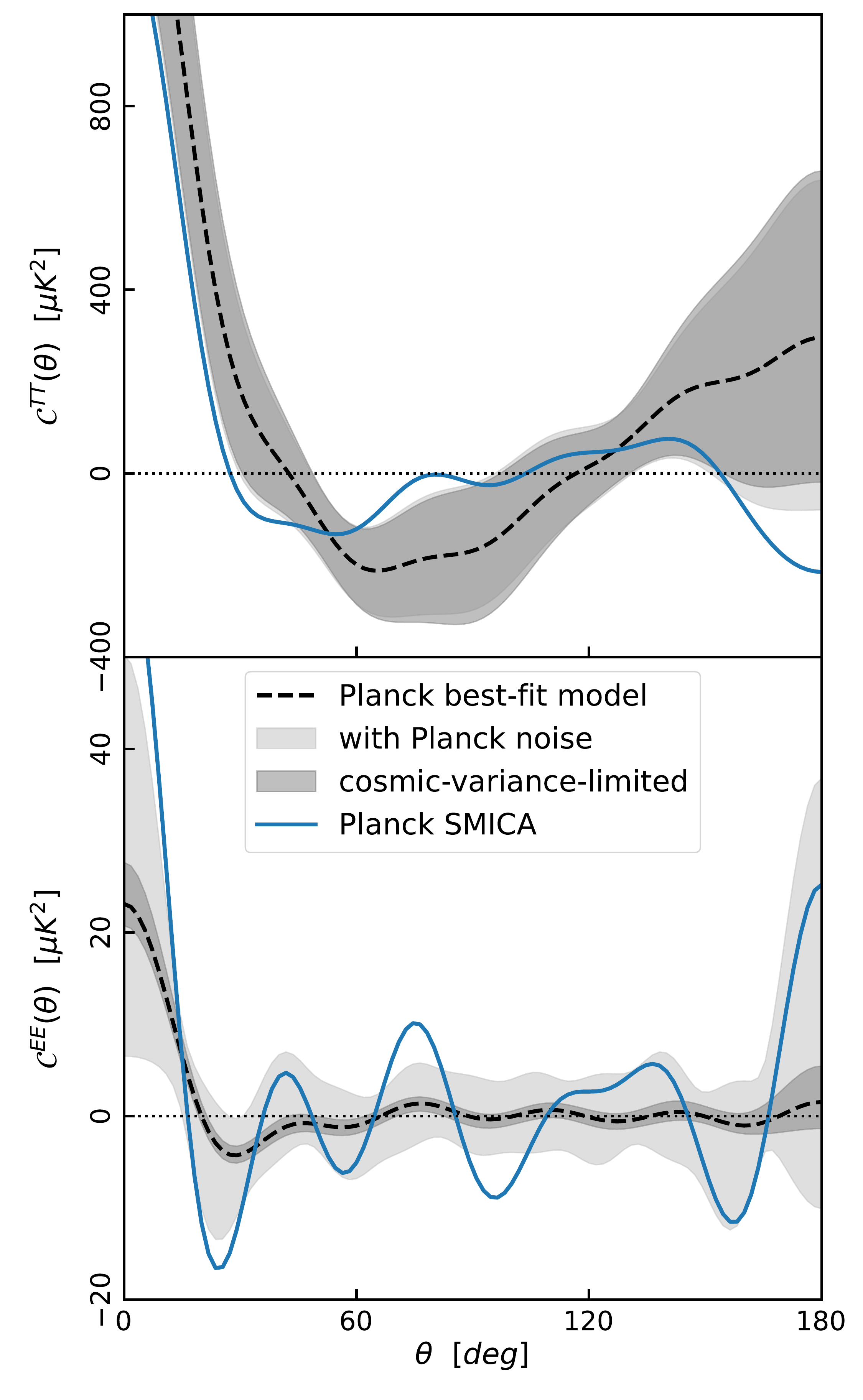}
    \caption{The two-point angular correlation functions computed using equations \ref{eq:Taps2cf} and \ref{eq:Eaps2cf}, with $\ell_\mathrm{max}=30$ for temperature (upper panel) and $\ell_\mathrm{max}=10$ for E-mode polarization (lower panel). 
    Blue solid lines are the correlation functions from \planck{} \smica{}. Black dashed lines are those from \planck{} best-fit model. 
    Light gray regions are the $1\sigma$ range of CMB + \planck{} \smica{} noise simulations, and dark gray regions the $1\sigma$ range of cosmic-variance-limited simulations. 
    A lack of correlation at large angular scales can be seen for the temperature data.
    {The ringing feature for the \planck{} \smica{} E-mode correlation function is mainly due to the systematics and noise in the data.}
    }
    \label{fig:corTT}
\end{figure}

We begin with the lack of temperature correlation on large angular scales: the observed temperature two-point correlation at large angular scales is much closer to zero than predicted from the $\Lambda$CDM model (Figure \ref{fig:corTT}). 
This anomaly was first measured by \cobe{} \citep{hinshaw1996two}, then confirmed by \wmap{} \citep{Bennett_2003} and \planck{} \citep{akrami2020planck}.
Similar tests for, e.g., $TQ$, $QQ$, $UU$ components can be found in, e.g., \citet{copi2013large, yoho2015microwave, chiocchetta2020lack}.

To quantify this anomaly, we chose the statistic in \citet{spergel2003first, akrami2020planck}: 
\begin{equation}
S_{\cos \theta_\mathrm{min}}^{XX}=\int_{-1}^{\cos \theta_\mathrm{min}}[\mathcal{C}^{XX}(\theta)]^2d(\cos\theta),\label{eq:lackcorr}
\end{equation}
where $\mathcal{C}^{XX}(\theta)$ is the two-point correlation function, and $X\in\{T, E\}$.
The integration lower and upper bounds were chosen to be $-1$ ($\theta_\mathrm{max}= 180^\circ$) and 1/2 ($\theta_\mathrm{min}=60^\circ$), respectively, because this combination approximately reflects the greatest possible deviation for the temperature data {\citep{akrami2020planck}}.
The same integration bounds were adopted for EE for consistency.
{Instead of doing numerical integration on the two-point correlation function directly, we construct $S_{1/2}^{XX}$ based on the power spectra by using Equations \ref{eq:Taps2cf} and \ref{eq:Eaps2cf},
which can be expressed as \citep{yoho2015microwave, chiocchetta2020lack}:}
\begin{equation}
\begin{aligned}
    S_{1/2}^{TT}&=\sum_{\ell=2}^{\ell_\mathrm{max}}\sum_{\ell^\prime=2}^{\ell^\prime_\mathrm{max}}\frac{2\ell+1}{4\pi}\frac{2\ell^\prime+1}{4\pi}C_\ell^{TT}I_{\ell\ell^\prime}C_{\ell^\prime}^{TT},\\
    S_{1/2}^{EE}&=\sum_{\ell=3}^{\ell_\mathrm{max}}\sum_{\ell^\prime=3}^{\ell^\prime_\mathrm{max}}\frac{2\ell+1}{4\pi}\frac{(\ell+2)!}{(\ell-2)!}\frac{2\ell^\prime+1}{4\pi}\frac{(\ell^\prime+2)!}{(\ell^\prime-2)!}\\
    &~~~~~~~~~~~~~~~~~\times C_\ell^{EE}I_{\ell\ell^\prime}C_{\ell^\prime}^{EE},
\end{aligned}
\end{equation}
where $I_{\ell\ell^\prime}$ is defined as
\begin{equation}
    I_{\ell\ell^\prime}\equiv\int_{-1}^{1/2}\mathcal{P}_\ell(\cos\theta)\mathcal{P}_{\ell^\prime}(\cos\theta)d\cos\theta.
\end{equation}
The exclusion of $\ell=2$ does not affect the statistics of $S_{1/2}^{EE}$ significantly because the correlation functions do not strongly rely on $C_2^{EE}$ \citep{chiocchetta2020lack}.
We used different values of $\ell_\mathrm{max}$ for temperature and polarization when computing the estimator for the lack of correlation.
For temperature we used $\ell_\mathrm{max}=30$: this provides a good estimation because $\mathcal C^{TT}\sim \ell C_\ell^{TT}\sim 1/\ell$. Therefore, the contribution from $\ell>30$ is negligible for our purpose. 
For the E-mode correlation, we used $\ell_\mathrm{max}=10$. Due to the presence of $(\ell+2)!/(\ell-2)!$ in Equation \ref{eq:Eaps2cf}, the contribution from higher multipoles dominates over that from lower multipoles, and the sum has been found to not converge through $\ell_\mathrm{max}=1500$ \citep{yoho2015microwave}. 
We chose $\ell_\mathrm{max}=10$ based on the cumulative signal-to-noise ratio S/N) as was done in \citep{chiocchetta2020lack}: we found a similar plateau for the cumulative S/N at $\ell>10$ of \planck{} \smica{} data, which indicates noise starts to dominate.
Therefore, we chose $\ell_\mathrm{max}=10$ for the E-mode estimator.
\litebird{} has higher S/N and can go to higher $\ell$, but we chose the same $\ell_\mathrm{max}$ to compare with \planck{} result.
Similar tests based on two-point correlation can be found in, e.g., \citet{copi2007uncorrelated, akrami2020planck}, and these studies provide corresponding checks in the angular space.
It is also possible to understand the lack of correlation via the amplitude of low multipoles as the relationship between them was noticed in \citet{copi2009no, gruppuso2014two, copi2015lack, muir2018covariance} for the temperature.

\begin{deluxetable}{lll}
\tablenum{2}
\tablecaption{Configuration for tests of lack of large-angular correlation and point-parity asymmetry$^a$, NSIDE=16.}\label{tab:sum-1}
\tablewidth{0pt}
\tablehead{\colhead{Case$^b$} & \colhead{Mask$^c$} & \colhead{Noise simulations}}
\startdata
\planck & HM+common & \planck{} \smica{} HM noise \\
CLASS & CLS+common & white noise 15 $\mu K$ \\
\litebird & common & white noise 2 $\mu K$
\enddata
~\\[-2pt]
$^a$We used the same configuration for these two anomalies.\\
$^b$For both intensity and polarization if not mentioned explicitly.\\
$^c$HM refers to \planck{} half mission missing pixel masks; common refers to \planck{} common masks; and CLS refer to declination limit for CLASS. See Figure \ref{fig:maps}.
\end{deluxetable}

We hereafter define the PTE as the percentage of simulations with statistic  values -- $S_{1/2}^{XX}$ in the present case -- greater than the value found in the \planck{} \smica{} data (Section \ref{sec:anomalies} and \ref{ssec:unconstr}) or the Special CMB (Section \ref{ssec:specialCMB}) to reflect statistical significance for the anomalies.
With this prescription, the PTE's for $S_{1/2}^{XX}$ in the \planck{} case (i.e., using \planck{} simulations) are 97.0\% for temperature and 24.0\% for E-modes.
A high PTE value here means the observed power on large angular scales is smaller than expected from the \planck{} simulations.
{The PTE value for the temperature estimator is consistent with the $p$-value obtained with \planck{} public QML spectra in \citet{muir2018covariance} ($p$-value $\sim 6\%$, which corresponds to a PTE value $\sim 94\%$), but lower than reported in \citet{akrami2020planck} ($>99.9\%$).
The inconsistency could be due to the fact that we used a different method to compute the $S_{1/2}$ estimator.
The PTE for the E-mode is different from that in \citet{chiocchetta2020lack} mainly because we used a different dataset.}
See Section \ref{sec:results} for more comments on this.

\subsection{Quadrupole-Octupole Alignment}\label{sec:Q-O}
\begin{figure}
    \centering
    \includegraphics[width=\linewidth]{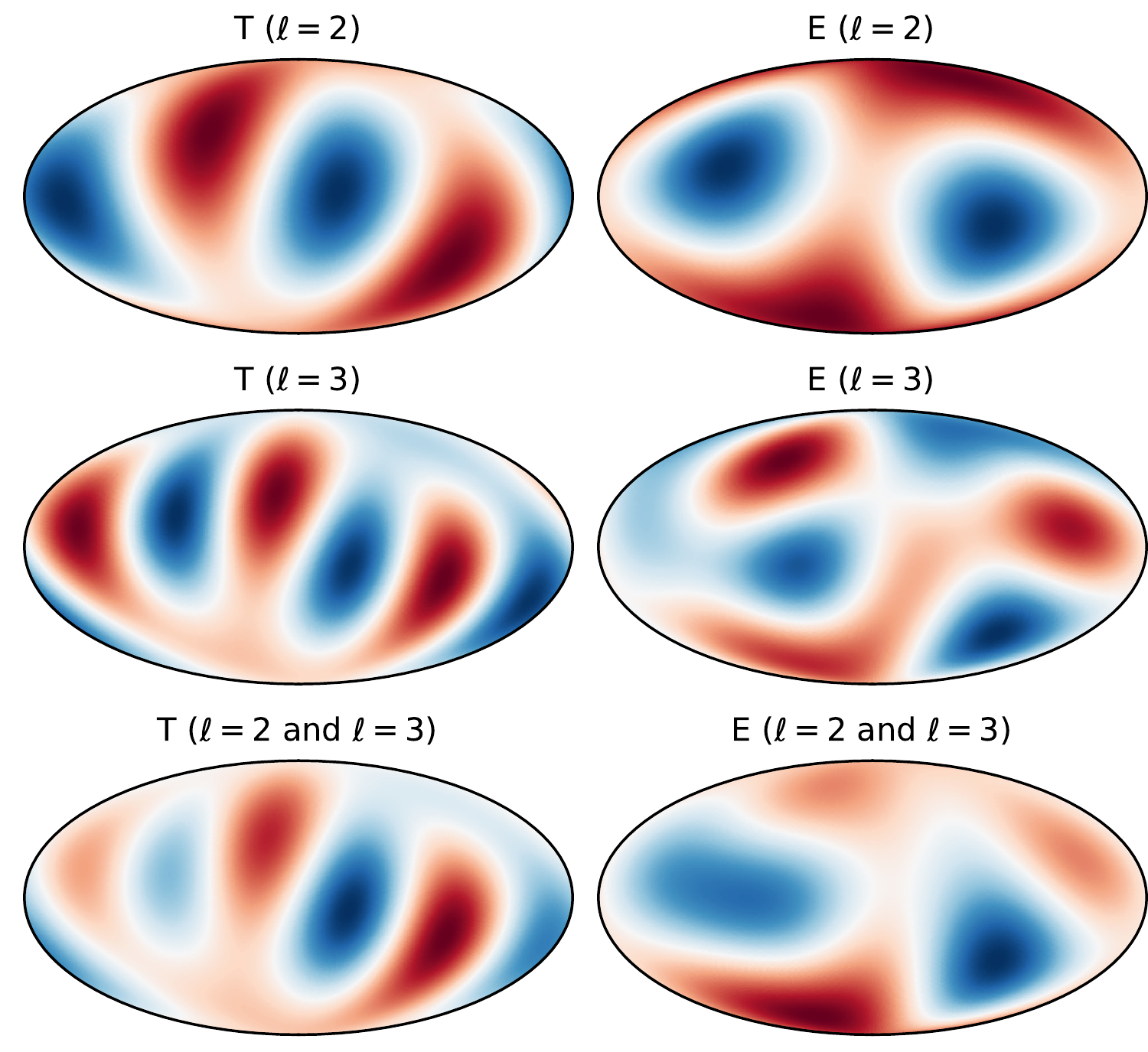}
    \caption{The quadrupole (top), octupole (middle), and their superposition (bottom) of the \smica{} temperature (left) and E-mode (right) maps. 
    Patterns in the left column show clearly that the temperature quadrupole and octupole are both planar and aligned.}
    \label{fig:QO}
\end{figure}
\citet{tegmark2003high} first pointed out that the CMB temperature's quadrupole and octupole are planar and align well with each other using \wmap{} data.
This was later verified in \citet{frommert2010axis, copi2015large, muir2018covariance} and others.
We show the quadrupole, octupole, and their superposition of \smica{} temperature (left) and E-mode (right) maps in Figure \ref{fig:QO}.
It can be seen clearly that the temperature quadrupole and octupole are both planar and aligned.

Probably the most mathematically complete tools to describe directions of multipoles are the Maxwell multipole vectors (MMVs), and there are several studies that apply MMVs to the CMB quadrupole-octupole (Q-O) alignment \citep{copi2004multipole, copi2015large, schwarz2016cmb, muir2018covariance}. 
But here we chose a physically more intuitive method to show the alignment of the quadrupole and octupole following \citet{de2004significance} and \citet{dvorkin2008testable}. 
The relation of this method to the MMVs is discussed in \citet{copi2006large}.

\begin{deluxetable}{lll}
\tablenum{3}
\tablecaption{Configuration for tests of quadrupole-octupole alignment, NSIDE=64.}\label{tab:sum-2}
\tablewidth{0pt}
\tablehead{\colhead{Case} & \colhead{\hspace{15pt}Mask} & \colhead{\hspace{15pt}Noise simulations}}
\startdata
\planck & \hspace{15pt}None & \hspace{15pt}\planck{} \smica{} full noise \\
CLASS & \hspace{15pt}CLS  & \hspace{15pt}white noise 15 $\mu K$ \\
\litebird & \hspace{15pt}None & \hspace{15pt}white noise 2 $\mu K$
\enddata
\end{deluxetable}

Treating the CMB fluctuations as a wave function (using temperature as an example),
\begin{equation}
\frac{\delta T}{T}(\bm{\hat{n}})\equiv\psi(\bm{\hat{n}}),
\end{equation}
the direction for any multipole is defined to be the direction $\hat{\bm{n}}$ of the axis that maximizes the angular momentum dispersion:
\begin{equation}
\langle\psi|(\bm{\hat{n}}\cdot \bm{L})^2|\psi\rangle = \sum_{m=-\ell}^{\ell}m^2|a_{\ell m}(\bm{\hat{n}})|^2,
\end{equation}
where $a_{\ell m}(\hat{\bm n})$ are the spherical harmonic coefficients of the CMB map in a rotated coordinate system with its $z$ axis along the $\hat{\bm n}$ direction \citep{de2004significance}.
By scaling the angular momentum dispersion as
\begin{equation}
L_{\ell}^{2(XX)}(\bm{\hat{n}}) \equiv \frac{\sum_{m}^{~}m^2|a_{\ell m}^X(\bm{\hat{n}})|^2}{\ell^2\sum_{m}^{~}|a_{\ell m}^X(\bm{\hat{n}})|^2},\label{eq:Lell}
\end{equation}
with $X\in\{T,~E\}$, the value of $L_\ell^2$ can be used to indicate how `planar' a multipole is, or mathematically, how large $m=\pm\ell$ are compared with the $|m|<\ell$ components for multipole $\ell$.\footnote{The $m=\pm \ell$ components have only azimuthal variation around $\hat{n}$ whereas $m<\ell$ components have polar variation as well.} 
Finally, the direction $\bm{\hat{n}}_*$ that maximizes
\begin{equation}
L_{23}^{2(XX)}(\bm{\hat{n}}) \equiv \frac{1}{2}\big(L_{2}^{2(XX)}(\bm{\hat{n}})+L_{3}^{2(XX)}(\bm{\hat{n}})\big)\label{eq:L23}
\end{equation}
is treated as the `joint' direction of quadrupole and octupole.
The closer $L_{23}^2(\bm{\hat{n}}_*)$ is to 1, the more planar and aligned the quadrupole and octupole are, and $\bm{\hat{n}}_*$ is normal to that plane.

In practice, we estimated $a_{\ell m}$'s without applying the common masks because otherwise the $a_{\ell m}$'s can be biased \citep{copi2006large}.
{To understand the impacts from foreground residuals in temperature data, we computed the $L_{23}^{2(TT)}$ estimator for all-sky temperature maps from multiple component separation methods: \planck{} PR3 \commander{}, \nilc{}, \sevem{} and \smica{}. 
We also computed $L_{23}^{2(TT)}$ for versions of these maps with regions of the strongest Galactic emission inpainted as in \citet{planck2016-l04}.
The directions and PTEs (with respect to synfast + \planck{} \smica{} noise simulations) of the $L_{23}^{2(TT)}$ for 8 different maps were found to be largely consistent with each other, which suggests that the temperature data are not strongly affected by foreground residuals, unless the foreground is impacting the 8 different maps in a similar way.}
We still adopted the declination limitation for the CLASS case, and the impact of this limitation can be found in Section \ref{sec:results}.
Once the alignment direction for a temperature map is settled, we use that direction for calculating $L_{23}^{2(EE)}$ directly to test whether the same axis and planarity is preferred by the E-modes.
{The alignment estimator relies on the phase information, hence would be more sensitive to issues related to residual foregrounds and systematics.
However, given no existing reliable way to debias the E-mode quadrupole phase, we chose to use the $\ell=2$ phase from \planck{} data directly.
The E-mode estimator value could be biased and is not well-constrained in presence of \planck{} noise, which is reflected in the second panel in Figure \ref{fig:spec_spread}.}

The PTE values for these statistics comparing to \planck{} simulations are 0.3\% for the temperature and 84.1\% for the E-mode.
A low PTE value here means the observed quadrupole and octupole are more planar and aligned in the \planck{} data than in most of the \planck{} simulations. 
{The PTE value for the temperature estimator is consistent with the $p$-value ($\sim0.4\%$) reported in \citet{muir2018covariance}.}

\begin{figure}
    \centering
    \includegraphics[width=\linewidth]{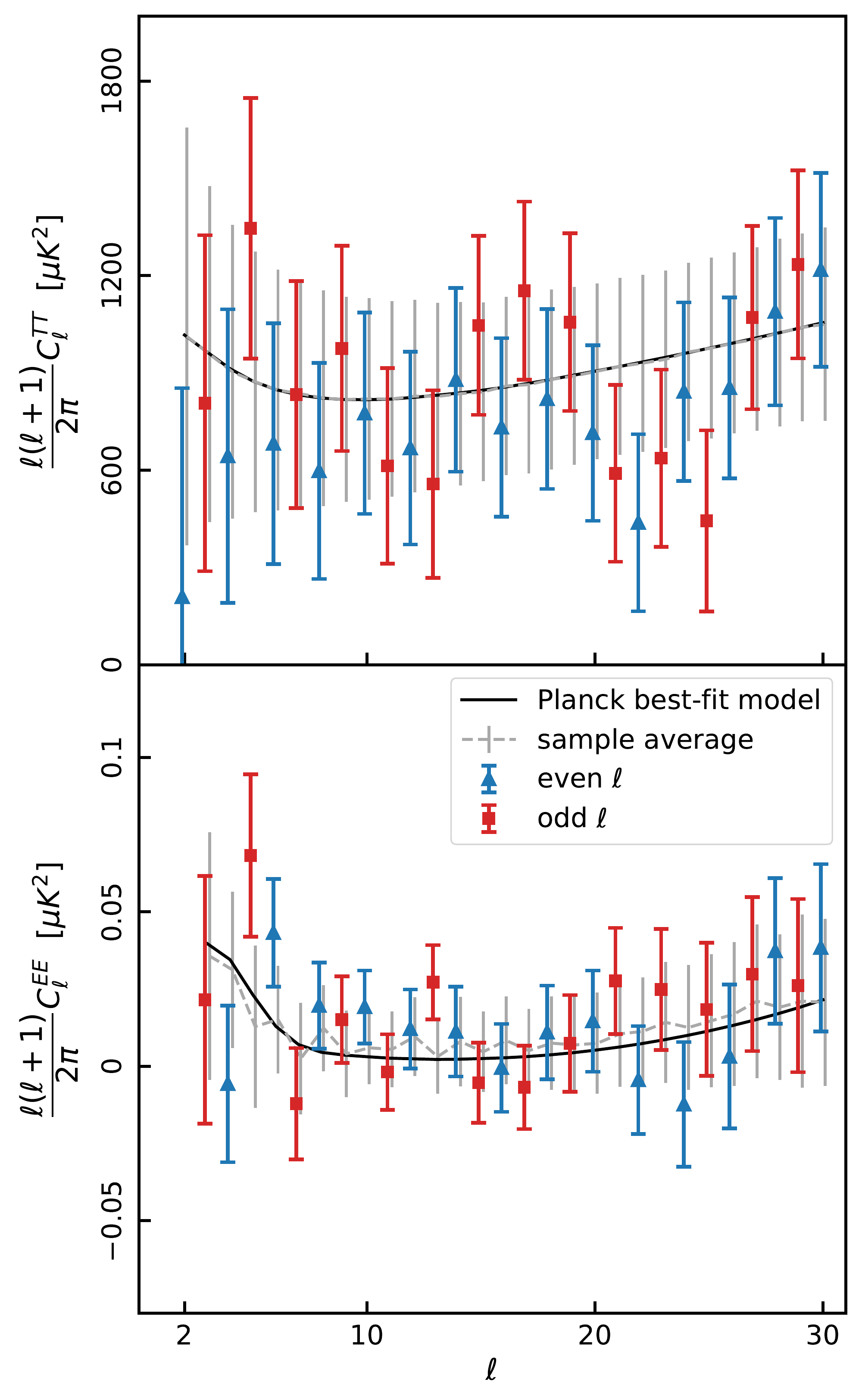}
    \caption{The power spectra for the NSIDE=16 \smica{} maps, showing the first 30 multipoles.
    \textit{Upper panel}: TT power spectra, with even multipoles plotted with blue triangles and odd plotted with red squares. 
    The gray dashed lines are the averages of the CMB + \planck{} \smica{} noise simulations.
    The error bars for both data and simulations are the standard deviations of the simulations.
    For comparison the \planck{} best-fit model is shown with the black curve. \textit{Lower panel}: EE power spectra.
    Both data points and the sample average were horizontally shifted slightly relative to one another to provide a clearer view.
    An offset between odd and even multipoles can be seen in the temperature data.}
    \label{fig:DlTT}
\end{figure}

\subsection{Point-parity asymmetry}\label{sec:pp}
One of the earliest tests of whether temperature anisotropies have a preference for antisymmetric parity on large angular scales was carried out by \citet{land2005universe}, and they did not find a strong preference based on the estimator and data they used.
Equipped with a different estimator, \citet{kim2010anomalous} found the significance level was higher when they applied it to \wmap{} 7-yr temperature anisotropy data.
More work has been done to better understand this anomaly since then  {\cite[e.g.,][]{gruppuso2011new, kim2012symmetry, cheng2016preferred, aluri2017alignments, panda2021parity}}.
The corresponding result in \citet{akrami2020planck} based on the \planck{} 2018 data (significance above 3$\sigma$ for temperature) also indicates that this anomaly deserves investigation.

We quantified this point-parity asymmetry and the corresponding E-mode one using the same estimators as in \citet{akrami2020planck}
\begin{eqnarray}
R^{TT}_{\ell_\mathrm{max}} &=& \frac{C_+^{TT}(\ell_\mathrm{max})}{C_-^{TT}(\ell_\mathrm{max})},\\
D^{EE}_{\ell_\mathrm{max}} &=& C_+^{EE}(\ell_\mathrm{max})-C_-^{EE}(\ell_\mathrm{max}),\label{eq:pp_DEE}
\end{eqnarray}
where $C_{\pm}^{XX}(\ell_\mathrm{max})$ are sums over even ($+$) or odd ($-$) multipoles between $\ell_\mathrm{min}$ and $\ell_\mathrm{max}$, 
\begin{equation}
C_{\pm}^{XX}(\ell_\mathrm{max}) \equiv \frac{1}{\ell^{\pm}_\mathrm{tot}}\sum_{\ell_\mathrm{min}}^{\ell_\mathrm{max}}\frac{\ell(\ell+1)}{2\pi}C_{\ell}^{XX,\pm},
 \end{equation}
and $\ell_\mathrm{tot}^{\pm}$ is the total number of even ($+$) or odd ($-$) multipoles included in the sum.
$C_\ell^{XX,\pm}$ is the T- or E-mode angular power spectrum at even ($+$) or odd ($-$) multipoles.
It follows that $R^{TT}_{\ell_\mathrm{max}}$ and $D^{EE}_{\ell_\mathrm{max}}$ stand respectively for the ratio-of and difference-between the average of band-powers for odd and even multipoles.
The less sensitive $D^{EE}_{\ell_\mathrm{max}}$ was used for E-mode data because the low signal-to-noise ratio in \planck{} \smica{} data resulted in negative estimation of $C_\ell^{EE}$, which causes numerical problems as the denominator of the ratio approaches zero.
Both estimators depend on the cut-off multipoles $\ell_\mathrm{min}$ and $\ell_\mathrm{max}$.
The $\ell_\mathrm{min}$ was chosen to be 2 for TT and 3 for EE, and the $\ell_\mathrm{max}$ was chosen to be 24 for both TT and EE because in \citet{akrami2020planck} the minimum lower-tail PTE for TT was found at $\ell=24$.
This $\ell$-range is motivated by the temperature anomaly and chosen for consistency with past studies. 
However, we note that, unlike with temperature, the E-mode estimator $D^{EE}_{24}$ will be dominated by the reionization bump at the lowest multipoles.
Future studies may well consider modifying the estimator, e.g., starting at higher $\ell_\mathrm{min}$ or using weights, to better capture even-odd variations up to higher E-mode multipoles.
We show power spectra for the first 30 multipoles in Figure \ref{fig:DlTT}, and see Table \ref{tab:sum-1} for a summary of masks and noise simulations used for this anomaly.

The PTE values for $R^{TT}_{24}$ and $D^{EE}_{24}$, comparing to \planck{} simulations, are 97.7\% for temperature and 83.2\% for E-mode.
A high PTE value here means the observed even multipoles are smaller than the odd multipoles in \planck{} data when compared to  the \planck{} simulations.
{Although cut off at a different multipole, the PTE value for the temperature is consistent with the results for $R_{27}^{TT}$ computed with \planck{} public QML spectra reported in \citet{muir2018covariance} ($p$-value $\sim 3\%$, which corresponds to a PTE value $\sim 97\%$, with respect to synfast simulations) and slightly less prominent than reported in \citet{akrami2020planck} (PTE value $\sim 99\%$).}
Additionally, in \citet{muir2018covariance} $R_{27}^{TT}$ was found to be correlated with $S_{1/2}^{TT}$ due to their dependence on the temperature quadrupole, and our results show similar 2-dimensional marginalized distributions with the $R_{24}^{TT}$ estimator (see discussion in Section \ref{sssec:corr}).
{The PTE for the E-mode is less anomalous than reported in \citet{akrami2020planck} (minimum $p$-value $\sim 6\%$ at $\ell_\mathrm{max}=27$ for HM data), but is statistically consistent given the large uncertainty on $D_{24}^{EE}$ introduced by \planck{} noise (Figure \ref{fig:spec_spread}).}

\subsection{Hemispherical Power Asymmetry}
\label{sec:hpa}
Using either \cobe{} or \wmap{} data, \citet{eriksen2004asymmetries} and  \cite{hansen2004asymmetries} first pointed out that the CMB temperature power spectrum is significantly stronger in the southern ecliptic hemisphere than in the northern hemisphere.
Instead of using the power spectrum, one can also quantify the asymmetry in terms of variance in the map \cite[e.g.,][]{monteserin2008low, akrami2014power}.
We adopted the latter approach and refer to this anomaly as ``the hemispherical power asymmetry''.
While the asymmetry is most visually apparent on the largest angular scales, it even persists at scales below 10$^\circ$.

\begin{deluxetable}{lll}
\tablenum{4}
\tablecaption{Hemispherical power asymmetry, NSIDE=64.}\label{tab:sum-4}
\tablewidth{0pt}
\tablehead{\colhead{Case} & \colhead{Mask} & \colhead{Noise simulations}}
\startdata
\planck{} Int. & common & \planck{} \smica{} full noise \\
\planck{} Pol. & E-mode mask$^a$ & \planck{} \smica{} HM noise \\
CLASS & CLS+common & white noise 15 $\mathrm{\mu K}$ \\
\litebird & common & white noise 2 $\mathrm{\mu K}$
\enddata
~\\[-2pt]
$^a$ Construction of the E-mode mask is described in Appendix \ref{sec:appendixa}.
\end{deluxetable}

To quantify this anomaly, we estimated the dipole amplitude and direction in local-variance maps (LVMs) \citep{akrami2014power, akrami2020planck}.
An LVM is a low-resolution map with pixel values capturing the localized variance information of neighbouring pixels in a higher resolution map. 
Therefore, the dipole amplitude of an LVM indicates the preference for the localized variances to be anomalously high in one hemisphere.
We followed the procedure in Section 7.1 of \citet{akrami2020planck} for making LVMs with NSIDE=16 ($\theta_\mathrm{pix}=3.66^\circ$) resolution from temperature and E-mode maps with NSIDE=64 ($\theta_\mathrm{pix}=0.92^\circ$) resolution. 
Each pixel in the LVM map was assigned the variance of temperature or E-mode data within a disk of $\theta=4^\circ$ radius, centered on the LVM pixel. 
(Therefore adjacent LVM pixels, being less than $4^\circ$ in width, are correlated.) 
Masks with resolution NSIDE=64, identified in Table \ref{tab:sum-4}, were used to exclude temperature or E-mode pixels  from the variance calculations. 
Correspondingly, LVM pixels with more than 90\% of the temperature or E-mode data masked in their associated $4^\circ$-radius disks were excluded from further analysis. 

$\mathrm{\Lambda}$CDM and noise simulations, listed in Table \ref{tab:sum-4}, were used to compute the average variance and variance-on-the-variance for each pixel in the LVM. 
We emphasize that these quantities depend both on the $\Lambda$CDM model and the noise properties of each experiment. 
We then fit the monopole and the dipole for each LVM by
\begin{equation}
    d_0,\bm d=\underset{d_0,\bm d}{\mathrm{argmin}}\left\{\sum_{p}\frac{[(v_p-{\bar v_p}) - d_0-\bm d\cdot \hat{\bm r}_p]^2}{\sigma^2_{v_p}}\right\},\label{eq:LVM_dip}
\end{equation}
where $v_p$ is the LVM pixel value at pixel $p$; $\bar v_p$ and $\sigma^2_{v_p}$ are the variance and variance-on-the-variance from each LVM pixel as determined by the simulations; $d_0$ captures the monopole component of the LVM; $\bm d\equiv (d_x, d_y, d_z)$ is the dipole component; and $\hat{\bm r}_p$ is the unit direction vector pointing to the center of pixel $p$. 
The sum is over all LVM pixels that were not excluded due to masking as described in the previous paragraph. 
Subtracting $\bar v_p$ from the LVM removes the variance bias arising from $\Lambda$CDM fluctuations and the experimental noise. 
As with a standard Gaussian-likelihood analysis, $\sigma^2_{v_p}$ down-weights the noisier LVM data. 
This method was applied to both the temperature and E-mode LVMs.
We tested with our simulations that best fit values of $d_x$, $d_y$ and $d_z$ using this likelihood are unbiased (Section \ref{ssec:hpsdir}).

\begin{figure*}
    \centering
    \includegraphics[width=\linewidth]{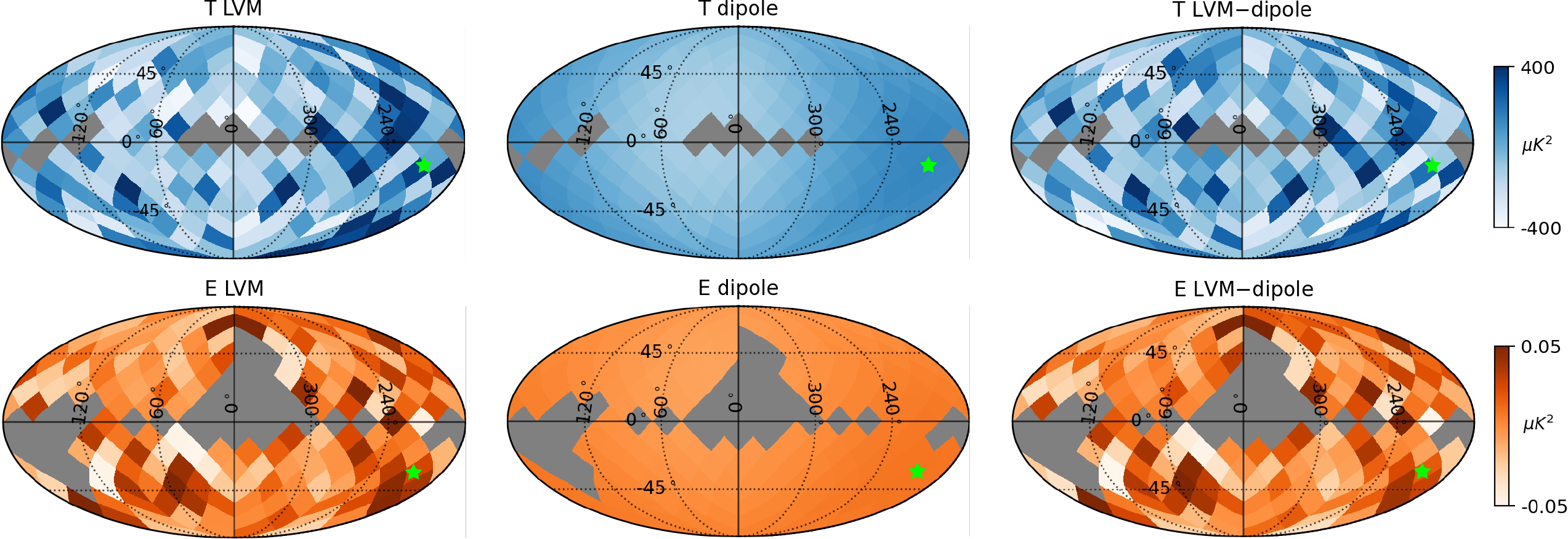}
    \caption{Hemispherical power dipole estimation. The top (blue) and bottom (orange) rows show Local Variance Maps (LVMs) for temperature and E-modes, respectively. 
    \textit{Left panels}: LVMs (downgraded from NSIDE=16 to NSIDE=4) of the \planck{} \smica{} maps, with the average map of LVMs from the corresponding simulations (signal + noise) removed. 
    \textit{Middle panels}: the fitted dipoles at NSIDE=4. 
    \textit{Right panels}: the dipole-subtracted LVMs of \planck{} \smica{} maps. 
    The green stars mark the directions of the fitted dipoles.
    While the proximity of E-mode and temperature dipole directions are compellingly consistent, errors on the E-mode amplitude and direction indicate the agreement may be spurious (Section \ref{ssec:hpsdir})
    }
    \label{fig:LVM}
\end{figure*}

Figure \ref{fig:LVM} shows the LVMs from \planck{} \smica{} data and corresponding dipole information. 
The fitted dipole orientation (Galactic coordinates) and amplitude ($d_4$, with $4$ representing the radius, $4^\circ$, of the disk on which we computed local variances) from \planck{} \smica{} are ${(l,b)=(209^\circ\pm 15^\circ,-14^\circ\pm 13^\circ)}$, ${d_4^{TT}=138.7\pm 35~\mathrm{\mu K}^2}$ for temperature and ${(l,b)=(206^\circ ~^{+38^\circ}_{-32^\circ},-32^\circ~ ^{+21^\circ}_{-22^\circ})}$, ${d_4^{EE}=0.008\pm0.004~\mathrm{\mu K}^2}$ for E-mode.
{The errors are the $1\sigma$ ranges of the dipole orientations obtained from the LVM simulations with dipole estimated from \planck{} \smica{} inserted.
The dipole directions we obtained are consistent with the results in \citet{akrami2020planck} for the \smica{} component-separated maps, where $(209^\circ, -15^\circ)$ for temperature and $(219^\circ, -16^\circ)$ for E-mode was reported.}
The PTE values for the dipole amplitudes are 0.1\% for temperature and 23.7\% for E-mode.
A low PTE value here means the variance dipole from \planck{} is stronger than the expectation from \planck{} simulations.
{The temperature PTE we obtained is consistent with the finding in \citet{akrami2020planck} (none of the 1000 simulations had a dipole amplitude larger than obtained from data), but the E-mode one is higher than the reported $5.5\%$.
We discuss more about our dipole estimator in Section \ref{ssec:hpsdir}.}

\section{Forecast results}\label{sec:results}
\subsection{Unconstrained universe}\label{ssec:unconstr}
We first checked the distributions of estimators including the effects of the cosmic variance, i.e., the distributions from combinations of $10^4$ CMB simulations and 300 noise simulations for each experiment.
We also studied the correlations between estimators caused by the combination of cosmic variance and noise.

\begin{deluxetable*}{@{\extracolsep{4pt}}l|ccccccccc}[!]
\tablenum{5}
\tablecaption{Bias and error based on the main simulation set.}\label{tab:95conf}
\tablewidth{0pt}
\tablehead{\multicolumn{1}{c|}{Anomaly} & \colhead{Estimator} & \colhead{\smica{} (PTE)$^a$} & \multicolumn{2}{c}{\planck} & \multicolumn{2}{c}{CLASS} & \multicolumn{2}{c}{\litebird} & \colhead{Ideal}\\
\cline{4-5} \cline{6-7} \cline{8-9} \multicolumn{1}{c|}{} & \colhead{} & \colhead{} & \colhead{bias$^b$} & \colhead{error$^b$} & \colhead{bias$^b$} & \colhead{error$^b$} & \colhead{bias$^b$} & \colhead{error$^b$} & \colhead{}}
\startdata
Lack of large-angular                       & $\log_{10}S_{1/2}^{TT}$ & $3.63$ $(97.0\%)$   & $0.00$  & $1.0$ & $0.02$  & $1.0$ & $-0.00$  & $1.0$ & $4.54_{{-0.99}}^{{+0.93}}$ \\
correlation (Sect. \ref{sec:correlation})   & $\log_{10}S_{1/2}^{EE}$ & $1.69$ ${(24.0\%)}$   & $1.27$  & $1.4$ & $0.10$  & $1.1$ & $0.00$  & $1.0$ & $0.19_{-0.50}^{+0.48}$ \\
\hline
Q-O alignment$^c$                               & $L_{23}^{2(TT)}$        & $0.97$ $(0.3\%)$    & $-0.00$ & $1.0$ & $0.12$  & $1.0$ & $-0.00$ & $1.0$ & $0.74_{-0.14}^{+0.18}$ \\
(Sect. \ref{sec:Q-O})                       & $L_{23}^{2(EE)}$        & $0.34$ $(84.1\%)$   & $-0.04$ & $1.0$ & $-0.04$ & $1.0$ & $-0.00$ & $1.0$ & $0.52_{-0.28}^{+0.29}$ \\
\hline
Point-parity                       & $R^{TT}_{24}$           & $0.74$ $(97.7\%)$   & $0.00$  & $1.1$ & $-0.01$ & $1.4$ & $-0.00$ & $1.1$ & $1.01_{-0.25}^{+0.32}$ \\
asymmetry (Sect. \ref{sec:pp})                        & $D^{EE}_{24}$           & $-0.005$ ${(83.2\%)}$ & $0.27$   & ${3.0}$ & $-0.00$  & $1.1$ & $0.00$  & $1.0$ & $-0.001_{-0.006}^{+0.005}$ \\
\hline
Hemispherical power                         & $d^{TT}_4~{[\mu K^2]}$    & $139$ $(0.1\%)$     & $0.03$  & $1.1$ & $0.15$  & $1.4$ & $0.03$  & $1.1$ & $51_{-36}^{+51}$ \\
asymmetry (Sect. \ref{sec:hpa})             & $d^{EE}_4~[\mu K^2]$    & $0.008$ $(23.7\%)$  & $1.37$  & $3.6$ & $0.42$  & $1.9$ & $0.14$  & $1.2$ & $0.002_{-0.001}^{+0.002}$ \\
\enddata
\tablenotetext{}{$^a$We quote the PTEs of the \smica{} measurements (column 3) in the brackets. PTEs were obtained by comparing to the \planck{} simulations.}
\tablenotetext{}{$^b$The bias relative to the Ideal value and the error are unitless (normalized by the Ideal 95\% error).}
\end{deluxetable*}

\subsubsection{Constraints on the anomaly estimators}\label{sssec:constr}
Table \ref{tab:95conf} shows the 95\% confidence intervals derived from simulations for the 8 anomaly statistics $\log_{10}S_{1/2}^{XX}$, $L_{23}^{2(XX)}$, $R_{24}^{TT}$, $D_{24}^{EE}$, $d_4^{XX}$ ($XX\in\{TT, EE\}$) defined in the previous section. 
The second column in Table \ref{tab:95conf} shows the estimators with corresponding units, and the third column gives the values of estimators from \planck{} 2018 \smica{} data, with PTEs computed based on simulations with \planck{} noise displayed in the brackets.
In the last column we show the medians and 95\% confidence intervals derived from the Ideal case (cosmic-variance-limited noise level with no mask applied).
The three double columns in between show the unitless bias and error of the distributions with \planck, CLASS and \litebird{} noise relative to those shown in the last column. 
The error is the 95\% confidence interval width of the \planck{}/CLASS/\litebird{} simulation normalized by the same width of the Ideal simulation. The bias is normalized in the same way and is computed as the difference between medians of the \planck{}/CLASS/\litebird{} and the Ideal case.

For temperature, all the biases are below $0.15$, meaning that the results from the three experiments are consistent with each other and the Ideal case.
The errors are increased for the CLASS case for the odd-parity and hemispherical power asymmetry anomaly due to the limited sky coverage.
Otherwise, no significant change from the current \planck{} constraints was found for any temperature anomaly estimator mainly because, on these scales, the uncertainty caused by the instrumental noise is smaller than that due to the cosmic variance.

For E-mode polarization, the errors on the estimator for lack of correlation 
($\log S_{1/2}^{EE}$) are similar between the Ideal, CLASS, and \litebird{} experiments.
Furthermore, the bias is modest for CLASS and negligible for \litebird.
The significant bias for \planck{} (1.27) implies that the current \planck{} value is more than 10 times greater than would be expected from the Ideal E-mode measurement. 
This is because the $S_{1/2}^{EE}$ estimated from \planck{} data is dominated by instrument noise, at least for the \smica{} reduction.
Instrument noise also explains the higher $S^{EE}_{1/2}$ uncertainty from \planck. Similar analysis has been done in \citet{chiocchetta2020lack} with the \planck{} HFI $100\times 143$ power spectra, which has less noise contribution than the \smica{}-based spectra used here.
Table \ref{tab:95conf} shows no improvement for CLASS and \litebird{} compared to \planck{} for the Q-O alignment estimator ($L_{23}^{2(EE)}$) due to the contribution from cosmic variance, which (being isotropic) trivially saturates the error budget. 
Given a fixed Q-O alignment in the simulations, \litebird{} can determine the corresponding Q-O alignment estimator to much higher precision than \planck{}. 
This will be discussed more in Section \ref{ssec:specialCMB}.
The distribution of $D_{24}^{EE}$ from \planck{} \smica{} simulations is not significantly biased (0.27) from the Ideal one, but the error will be improved from 2.8 to 1.1 and 1.0 for CLASS and \litebird, respectively.
For the hemispherical power asymmetry, both the bias and error shrink significantly from the \planck{} case to CLASS and \litebird. 
In particular, the high bias (1.37) and large error (3.6) of \planck{} \smica{} indicate that the current \smica{}-based constraints on the E-mode hemispherical power asymmetry are measurement-noise dominated.

Based on the observations listed above, we conclude that the current \smica{}-based constraints are noise dominated for the commonly used E-mode anomaly estimators of lack of correlation, odd-parity asymmetry and hemispherical power asymmetry. The situation will improve with results from \litebird-like experiments.

\subsubsection{Correlation}\label{sssec:corr}
\begin{figure*}
    \centering
    \includegraphics[width=\linewidth]{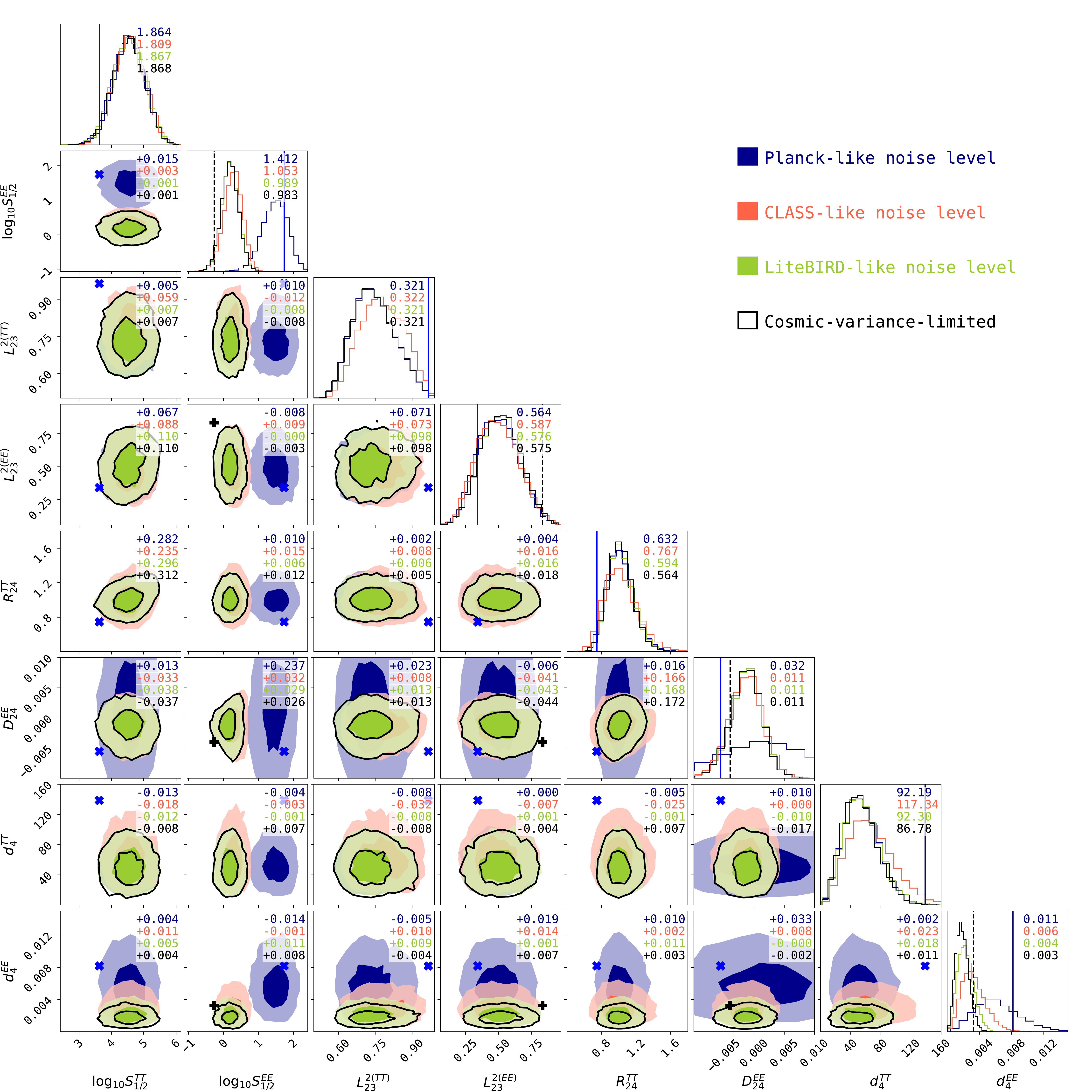}
    \caption{Confidence-curve matrix of temperature and  polarization anomaly estimators, with contours and histograms from \planck{} (blue), CLASS (pink), \litebird{} (green), and the cosmic-variance-limited Ideal case (black); see Table \ref{tab:95conf} for a summary of anomaly names and estimators.
    The numbers on the top-right of off-diagonal panels are the Pearson correlation coefficients, and contours show $1\sigma$ and $2\sigma$ significance levels.
    The numbers on the top-right of diagonal panels are the 95\% interval widths.
    The blue crosses and vertical lines represent \planck{} \smica{} measurements, and black crosses and dashed vertical lines represent E-mode measurements from the Special CMB (see Section \ref{ssec:specialCMB} for more about the Special CMB).}
    \label{fig:corner}
\end{figure*}
Figure \ref{fig:corner} shows the marginalized distributions of the 8-dimensional space spanned by the estimators.
We use blue to denote contours and histograms from \planck{}, pink from CLASS, green from \litebird{}, and black from Ideal simulations.
The diagonal panels display the posterior distribution for each estimator from the four types of simulations. 
In these panels, the blue vertical lines mark the estimator values from \planck{} \smica{} data, the black dashed vertical lines show the E-mode estimators for the Special CMB (Section \ref{ssec:specialCMB}), and the numbers on the top-right are the widths of the 95\% intervals for the corresponding estimators.
The off-diagonal panels show the 2-dimensional marginalized distributions, in which we used black unfilled contours to show distributions from the Ideal case and filled contours for those from the three experiments. 
The blue crosses mark measurements from \smica{} data and black crosses mark the E-mode measurements from the Special CMB. 
The numbers on the top-right are the Pearson correlation coefficients.

Consistent with the conclusions in Section \ref{sssec:constr}, the contours and distributions of \planck{} simulations are significantly biased relative to the Ideal case for $\log_{10}S_{1/2}^{EE}$ and $d_4^{EE}$, and the contours for \litebird{} are similar to the Ideal contours.
The most significant correlation ($\sim 0.3$) is between $\log_{10}S_{1/2}^{TT}$ and $R_{24}^{TT}$. 
(The CLASS simulations show a smaller correlation as the uncertainty in power spectra estimation was larger due to the limited sky coverage.) 
This correlation is consistent with the result in \citet{muir2018covariance} and is mostly due to the correlation between $S_{1/2}^{TT}$ and the quadrupole $C_2^{TT}$, as a larger $S_{1/2}^{TT}$ implies a stronger $C_2^{TT}$ hence a larger $R_{24}^{TT}$. 
A similar correlation is not found between $\log_{10}S_{1/2}^{EE}$ and $D_{24}^{EE}$ because $D_{24}^{EE}$ is largely determined by the octupole $C_3^{EE}$ but $S_{1/2}^{EE}$ is not. 
The second most significant correlation ($\sim 0.17$) is between $R_{24}^{TT}$ and $D_{24}^{EE}$ (except for the \planck{} simulations in which it is diminished by noise). 
This is mostly due to the correlation between $C_\ell^{TT}$and $C_\ell^{EE}$ at low multipoles.
The correlation coefficients in almost all the other off-diagonal panels are at or below the $\sim 0.1$ level for all four cases including the Ideal one\footnote{The 0.2 correlation between $D_{24}^{EE}$ and $\log_{10}S_{1/2}^{EE}$ for \planck{} simulations can be blamed to the low signal-to-noise ratio.}, which means the $\Lambda$CDM model does not predict strong correlations for those pairs of estimators.
Therefore, if an E-mode estimator reinforces the anomaly identified by its temperature counterpart, then this result would not be due to correlation between estimators but would present an independent challenge to the statistical-fluke hypothesis.

\subsection{Tests based on the Special CMB}\label{ssec:specialCMB}
In Section 4.1, we computed the distribution of the anomaly estimators resulting from $\Lambda$CDM cosmic variance and noise from \planck{}, CLASS and \litebird. 
As reflected in Table \ref{tab:95conf} and Figure \ref{fig:corner}, if moving from \planck{} to \litebird{}, the constraining power would increase for three of the E-mode anomaly estimators but no obvious change was found for the Q-O alignment estimator distribution. 
This is because the fluctuations caused by cosmic variance are maximal without additional noise.
In this section, we fix the CMB realization and focus on the bias and spread of the E-mode anomaly estimator distributions caused by noise only.
Therefore, we generated the Special CMB such that the E-mode signal had all four anomalies, and we tested how different experiments will constrain the anomaly estimators for this specific realization.

The way we produced the Special CMB was by first generating $10^5$ Gaussian CMB simulations based on \planck{} 2018 best-fit model and computing their anomaly estimator values in the cosmic-variance-limited case. 
Then the Special CMB was selected by setting thresholds on the PTEs of the E-mode anomaly estimators. The second column in Table \ref{tab:spec_constr} lists the PTEs of the CMB realization we picked. 
This realization has a more than 95\% PTE (or less than a 5\% PTE) for the lack of correlation and Q-O alignment, and an almost 5\% PTE for the hemispherical power asymmetry. 
The odd-parity estimator is the least aberrant because we are seeking one realization that has all of these E-mode anomalies, which means that we are using the conditional probability instead of the marginal one. 
The black triangle-shape contours in the $D_{24}^{EE}-\log_{10}S_{1/2}^{EE}$ panel of Figure \ref{fig:corner} implies that a smaller $\log_{10}S_{1/2}^{EE}$ corresponds to a smaller spread of $D_{24}^{EE}$. 
Consequently, when a realization lacks correlation, the probability for its $D_{24}^{EE}$ to reach the $2\sigma$ level of the marginal distribution is extremely small. 
For the realization that we selected, the PTE of $D_{24}^{EE}$ among simulations that have PTEs for $\log_{10}S_{1/2}^{EE}$ above 95\% is 98.5\%, which is anomalous enough for our purpose. 
The visualization of the 4 E-mode anomalies for the Special CMB in analogy to Figures \ref{fig:corTT}--\ref{fig:LVM} can be found in Appendix \ref{sec:appendixb}.

\begin{figure}
    \centering
    \includegraphics[width=\linewidth]{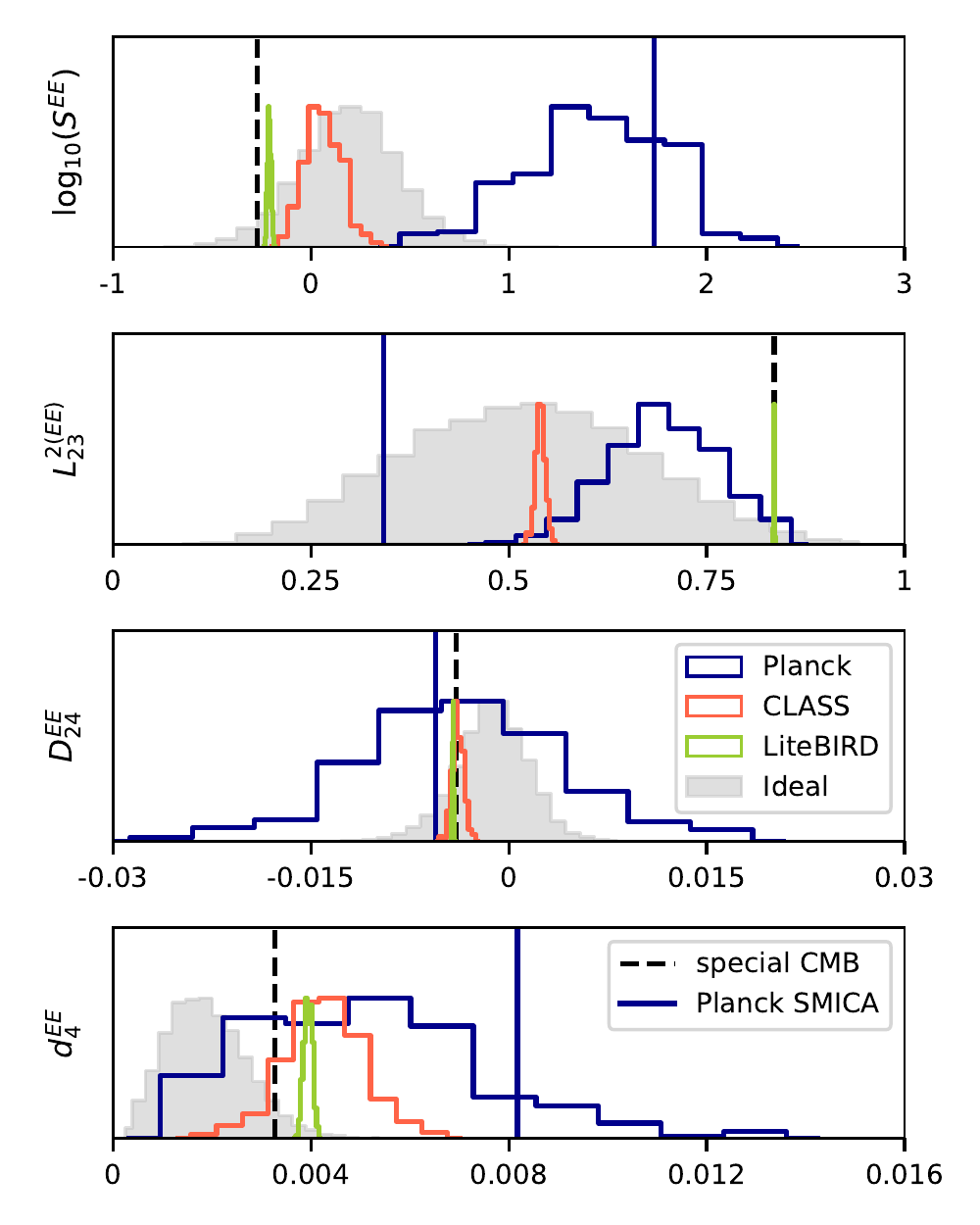}
    \caption{The posterior distributions of different anomaly estimators, from top to bottom: lack of correlation ($\log_{10}S_{1/2}^{EE}$), Q-O alignment ($L_{23}^{2(EE)}$), point-parity asymmetry ($D_{24}^{EE}$) and hemispherical power asymmetry ($d_4^{EE}$). 
    The gray filled histograms represent the spread caused by cosmic variance; other unfilled histograms represent the spread caused by \planck{} (blue), CLASS (pink) and \litebird{} (green) noise, respectively.
    The black dashed vertical lines mark the estimator values of the Special CMB, and blue solid are those from the \planck{} \smica{} measurements.}
    \label{fig:spec_spread}
\end{figure}

Our main analysis results based on the Special CMB are shown in Figure \ref{fig:spec_spread} and Table \ref{tab:spec_constr}.
In Figure \ref{fig:spec_spread}, filled histograms are the posterior distributions from the Ideal simulations from Section \ref{ssec:unconstr} (i.e., its spread reflects cosmic variance). 
We use blue, pink and green unfilled histograms to denote distributions for the Special CMB with \planck{}, CLASS and \litebird{} noise simulations, respectively. 
The black dashed vertical lines mark the estimator value of the Special CMB we picked and blue vertical lines are those from \planck{} \smica{}, staying consistent with Figure \ref{fig:corner}.
Unsurprisingly, we found that the spread of the unfilled histograms reduces as one goes from using \planck{} noise to \litebird{} for all E-mode anomaly estimators. 
The biases are different in different cases: in the first panel for $\log_{10}(S^{EE})$, we noticed that all three unfilled histograms are positively biased, which makes sense because noise contributes in $S_{1/2}^{TT,EE}$ positively by definition as displayed in Equation \ref{eq:lackcorr}, and the bias becomes smaller as the noise level becomes smaller. 
In the second panel, corresponding to the Q-O alignment, the bias for the \planck{} histogram is mainly due to the noise while, for CLASS, bias results from the incomplete sky coverage. 
This emphasizes the importance of having both low noise level and full sky coverage for constraining $L_{23}^{2(EE)}$. 
Nearly no bias was found for all three cases for the odd-parity estimator in the third panel. 
In the last panel for the variance dipole amplitude, the behavior of the \planck{} and CLASS histograms are as expected: the positive bias goes down as noise level decreases. 
The bias of the \litebird{} histogram is mainly related to the shape of the mask: we found that using \litebird{} noise simulations but with full sky coverage only resulted in an $-0.005$ bias on the dipole amplitude.
The unitless bias and error listed in Table \ref{tab:spec_constr} are consistent with our observations of Figure \ref{fig:spec_spread}. 
Bias and error are still normalized by the error on the Ideal simulations as in Section \ref{sssec:constr}, but the bias is now computed relative to the Special CMB estimator values.
We conclude that, compared to the bias and error of \planck{} noise, \litebird{} is promising in providing much stronger (more than an order of magnitude) constraints on the 4 E-mode anomaly estimators, while observation from CLASS is promising in constraining $D_{24}^{EE}$. While the  sky coverage of CLASS limits its constraining power relative to the all-sky \litebird{} measurement, the large-scale E-mode data from CLASS will  be valuable as a high-S/N check of \litebird{}.

\begin{deluxetable*}{@{\extracolsep{4pt}}ccccccccc}[!]
\tablenum{6}
\tablecaption{Bias and error for the Special CMB with noise and masks representing the three experiments}\label{tab:spec_constr}
\tablewidth{0pt}
\tablehead{
\colhead{Estimator} & \colhead{Special CMB (PTE)} & \multicolumn{2}{c}{\planck} & \multicolumn{2}{c}{CLASS} & \multicolumn{2}{c}{\litebird} & \colhead{Ideal}\\
\cline{3-4} \cline{5-6} \cline{7-8} \colhead{} & \colhead{} & \colhead{bias$^a$} & \colhead{error$^a$} & \colhead{bias$^a$} & \colhead{error$^a$} & \colhead{bias$^a$} & \colhead{error$^a$} & \colhead{}}
\startdata
$\log_{10}S_{1/2}^{EE}$ & $-0.27~(96.3\%)$  & $ 1.73$ & $1.30$ & $ 0.32$ & $0.36$ & $ 0.06$ & $0.03$ & $0.19_{-0.50}^{+0.48}$ \\
$L_{23}^{2(EE)}$        & ~\;$0.84~(1.4\%)$ & $-0.25$ & $0.46$ & $-0.53$ & $0.04$ & $-0.00$ & $0.003$ & $0.52_{-0.28}^{+0.29}$ \\
$D^{EE}_{24}$           & $-0.004~(84.6\%)$ & $ 0.03$ & $2.90$ & $ 0.01$ & $0.15$ & $-0.02$ & $0.02$ & $-0.001_{-0.006}^{+0.005}$ \\
$d^{EE}_4~[\mu K^2]$    & $0.003~(5.6\%)$   & $ 0.55$ & $2.81$ & $ 0.33$ & $1.04$ & $ 0.2$ & $0.10$ & $0.002_{-0.001}^{+0.002}$ \\
\enddata
\tablenotetext{}{$^a$The unitless bias and error with definitions given in Section \ref{sssec:constr}. 
In this case the bias was computed with respect to the estimator values of the Special CMB instead of the median of the Ideal distributions.}
\end{deluxetable*}

\subsection{Constraints on the hemispherical power asymmetry orientation}\label{ssec:hpsdir}
\begin{figure*}
    \centering
    \includegraphics[width=\linewidth]{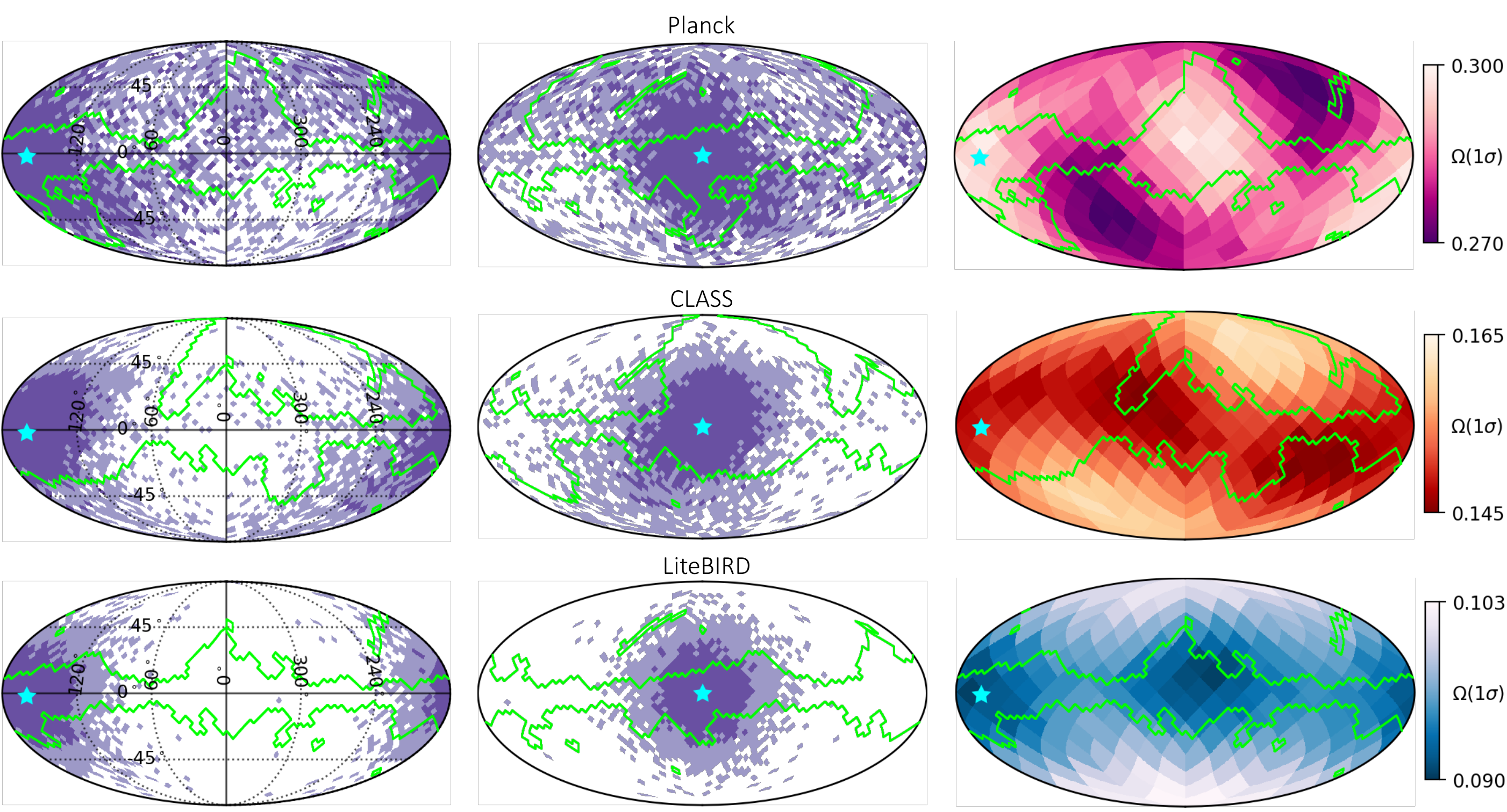}
    \caption{E-mode LVM dipole orientation distributions generated from simulations of different surveys with the dipole from the Special CMB inserted. \textit{Left panels}: Maps show $1,2\sigma$ dipole-orientation uncertainty regions (from dark to light purple) from $10^4$ simulations  for \planck{} \smica{} (top), CLASS (middle) and \litebird{} E-mode (bottom) in Galactic coordinates. 
    The cyan star stands for the direction of the inserted dipole, and the green lines are the borders of the masks used when fitting dipoles from LVMs. 
    \textit{Center panels}: Maps contain the same results as in the left column but are rotated such that the inserted dipole is located at the center. 
    \textit{Right panels}: Maps show the solid angles, $\Omega(1\sigma)$, of the $1\sigma$ dipole-orientation confidence region for additional simulations where the dipole from the Special CMB was recentered on each NSIDE=4 pixel. 
    This result suggests that orientation localization is better when the maximum gradient of the dipole (versus its pole) lies outside the Galactic mask. 
    We used different colormaps to highlight the different ranges of the $\Omega(1\sigma)$.}
    \label{fig:orientation}
\end{figure*}

We did two extra analyses to understand the recovery of the LVM dipole orientation from our method under the impacts of noise and the E-mode mask.
To investigate the impacts of the noise, we added the dipole from the LVM of the Special CMB to the LVM of Gaussian CMB+Noise simulations for all three experiments.
We then fitted for the dipole of these new LVMs using Equation \ref{eq:LVM_dip}, with $\bar v_p$ and $\sigma^2_{v_p}$ computed for each experiment from the simulations without the Special CMB dipole.
The left and middle panels in Figure \ref{fig:orientation} show $1,2\sigma$ confidence regions of dipole orientations on NSIDE=16 maps for each experiment. 
In the left panels we used the Galactic coordinate system. 
To better show the shape of the contours, in the middle panels we rotated the coordinate system so that the inserted dipole is located at the origin.
To obtain the $1,2\sigma$ contours, we first binned the recovered dipole orientations from the $10^4$ simulations into the pixels of an NSIDE=16 map. 
In the resulting spherical histogram of dipole directions, the $1,2\sigma$ regions correspond to the highest-valued pixels whose sum  is equal to (or slightly less than) 68\%, 95\% of the total number of simulations ($10^4$) contributing to the histogram. 

The recovered dipole components $d_x$, $d_y$, and $d_z$ were unbiased.
However, the projection of the elliptical distribution of the recovered dipoles onto the celestial sphere can skew the angular distribution, especially in the case of high noise. 
Nevertheless, for the \litebird{} experiment, we found the distribution of dipole orientations from these simulations centers well on the input dipole direction (cyan stars). 
For CLASS and \planck{}, the apparent bias of the $2\sigma$ region is caused by radially projecting the 3-dimensional ellipsoidal normal distribution corresponding to symmetric distributions for $d_x$, $d_y$, and $d_z$.
The solid angle subtended by the $1\sigma$ ($2\sigma$) region in \planck{} \smica{} is 12180 deg$^2$ (28415~deg$^2$), whereas it is 6191 deg$^2$ (18720~deg$^2$) for CLASS (a $\sim2\times$ reduction) and 3773 deg$^2$ (11870~deg$^2$) for \litebird{} (a $\sim3\times$ reduction).

We then investigated how the dipole fit depends on the shape of the mask by inserting dipoles aligned with the centers of all the NSIDE=4 map pixels.
We used $10^4$ simulations in each direction, and the amplitude was fixed to be that from the Special CMB. 
The maps in the right panels of Figure \ref{fig:orientation} show the value of $\Omega(1\sigma)$, the fraction of the whole sky occupied by the $1\sigma$ orientation confidence region, for each direction.
$\Omega(1\sigma)$ varies across the map mainly due to the presence of the mask, and we noticed that for \litebird{} experiment, the pattern aligns well with the shape of the mask: pixels within the mask seem to have smaller $\Omega(1\sigma)$ values. 
This implies the fitting is more effective if the gradient of the dipole is unmasked.
The alignment is less obvious for the CLASS experiment but follows the same general trend with smaller confidence regions within the Galactic mask or outside of the survey footprint.
The \planck{} experiment result has a quite different pattern due to the lower signal-to-noise ratio. 
Instead of lying in the mask, the best constraints align with the high-noise regions of the \planck{} data.
{We found the mean recovered dipole amplitudes ($d_x$, $d_y$, $d_z$) matched the simulation input to better than 0.0003 times the standard error of the simulation ensemble for all the 192 input directions (NSIDE=4 map pixel centers).
This means our estimation for $d_x$, $d_y$ and $d_z$ are unbiased.}
The medians of $\Omega(1\sigma)$ are 0.285, 0.151, 0.096 for \planck{}, CLASS and \litebird{} respectively, also suggesting a $\sim 2\times$ ($\sim 3\times$) reduction for CLASS (\litebird) comparing to \planck{} \smica{}.
Finally, we emphasize that the uncertainty on the orientation depends on the amplitude of the input dipole, and the one we used from the special realization may not reflect the true case.
Therefore, these tests are merely demonstrations based on a dipole with amplitude larger than that found in $\sim95\%$ of ideal $\Lambda$CDM realizations.

\section{Conclusions}\label{sec:conclusion}
In this paper, we explored how CMB polarization data will improve our understanding of CMB anomalies.
The four anomalies we studied are: the lack of correlation on large angular scales, the alignment of quadrupole and octupole, the point-parity asymmetry, and the hemispherical power asymmetry.
The definitions for estimators of each of the anomalies are in Section \ref{sec:anomalies}.
We forecast constraints from future experiments on the temperature and polarization estimators and explored the correlation between the estimators of different anomalies.
Our main analyses were based on the combination of $10^4$ Gaussian CMB simulations and 3 different types of 300 noise simulations based on \planck{} \smica, CLASS and \litebird{}. 
We also include tests based on a special CMB realization that was selected from Gaussian CMB simulations, of which all 4 E-mode anomaly estimators are at $\sim 2\sigma$ significance.

We found that \planck{} \smica{} does not significantly constrain the four anomalies, but future E-mode measurements look promising.
Our unitless bias and error in Table \ref{tab:95conf} show the constraining power on the lack of correlation $S_{1/2}^{EE}$ ($\log_{10}S_{1/2}^{EE}$) will be improved by factors of $\sim20$ ($\sim1.3$) and $\sim26$ ($\sim1.4$) moving from \planck{} to CLASS and \litebird{}, respectively. 
The improvements on $S_{1/2}^{EE}$ are high compared to the result in \citet{chiocchetta2020lack} as they used a reduction of \planck{} data that has lower noise. 
Future E-mode studies with \planck{} data could yield similar gains for the other anomalies.
While no analogous improvement in constraining power was found for the Q-O alignment based on the main simulation set, the constrained CMB simulation test (with quadrupole and octupole moments fixed to those of the Special CMB) showed significant improvement due to reduced instrumental noise ($\sim 100$ for \litebird) as reflected in Table \ref{tab:spec_constr}. 
The same test shows that the limited sky fraction makes it difficult for CLASS to constrain the Q-O alignment estimator.
A factor-of-3 improvement was found for the point-parity asymmetry for both CLASS and \litebird, while a factor of 2 (3) was found for the amplitude of the hemispherical power asymmetry for CLASS (\litebird). 
The localization of the variance dipole extracted from the Special CMB improves from $\sim30\%$ of the sky ($1\sigma$) for \planck{} \smica{} to $\sim15\%$ for CLASS and $\sim9\%$ for \litebird{}. 
The localization depends on both the dipole amplitude and  orientation with respect to masks and high-noise regions.

The correlation between different anomalies is negligible for most of the estimators except for two pairs: one is $R_{24}^{TT}$ and $\log_{10}S_{1/2}^{TT}$, which has a Pearson correlation coefficient $r\sim 0.3$ in the cosmic-variance-limited result; the other is $R_{24}^{TT}$ and $D_{24}^{EE}$, with $r\sim 0.17$.
We note that the low correlations between this particular set of estimators does not imply that there are no correlations between  potential underlying physical models as each anomaly could have several different estimators.
The general lack of correlation found between temperature and polarization estimators implies that if estimators from future polarization experiments reproduce anomalies in temperature, then the statistical-fluke hypothesis will be challenged.

\section*{Acknowledgments}
We thank G Addison, Y Akrami, AJ Banday, P Bielewicz, C Chiocchetta, C Dvorkin, L Ji, Z Wang, and M Kamionkowski for helpful discussions.
We acknowledge the National Science Foundation Division of Astronomical Sciences for their support under grant numbers 1636634, 1654494, 2034400, and 2109311.
Sumit Dahal is supported by an appointment to the NASA Postdoctoral Program at the NASA Goddard Space Flight Center, administered by Oak Ridge Associated Universities under contract with NASA.
Zhilei Xu is supported by the Gordon and Betty Moore Foundation through grant GBMF5215 to the Massachusetts Institute of Technology.
This study used observational data from \planck{} (\href{http://www.esa.int/Planck}{http://www.esa.int/Planck}), an ESA science mission with instruments and contributions directly funded by ESA Member States, NASA, and Canada.
\software{\texttt{HEALPix} \citep{gorski2005cosmology}; \texttt{matplotlib} \citep{hunter2007matplotlib, caswell2019matplotlib}; \texttt{numpy} \citep{van2011numpy};}

\appendix
\counterwithin{figure}{section}
\section{E-mode mask construction}\label{sec:appendixa}
\begin{figure*}
    \centering
    \includegraphics[width=0.32\linewidth]{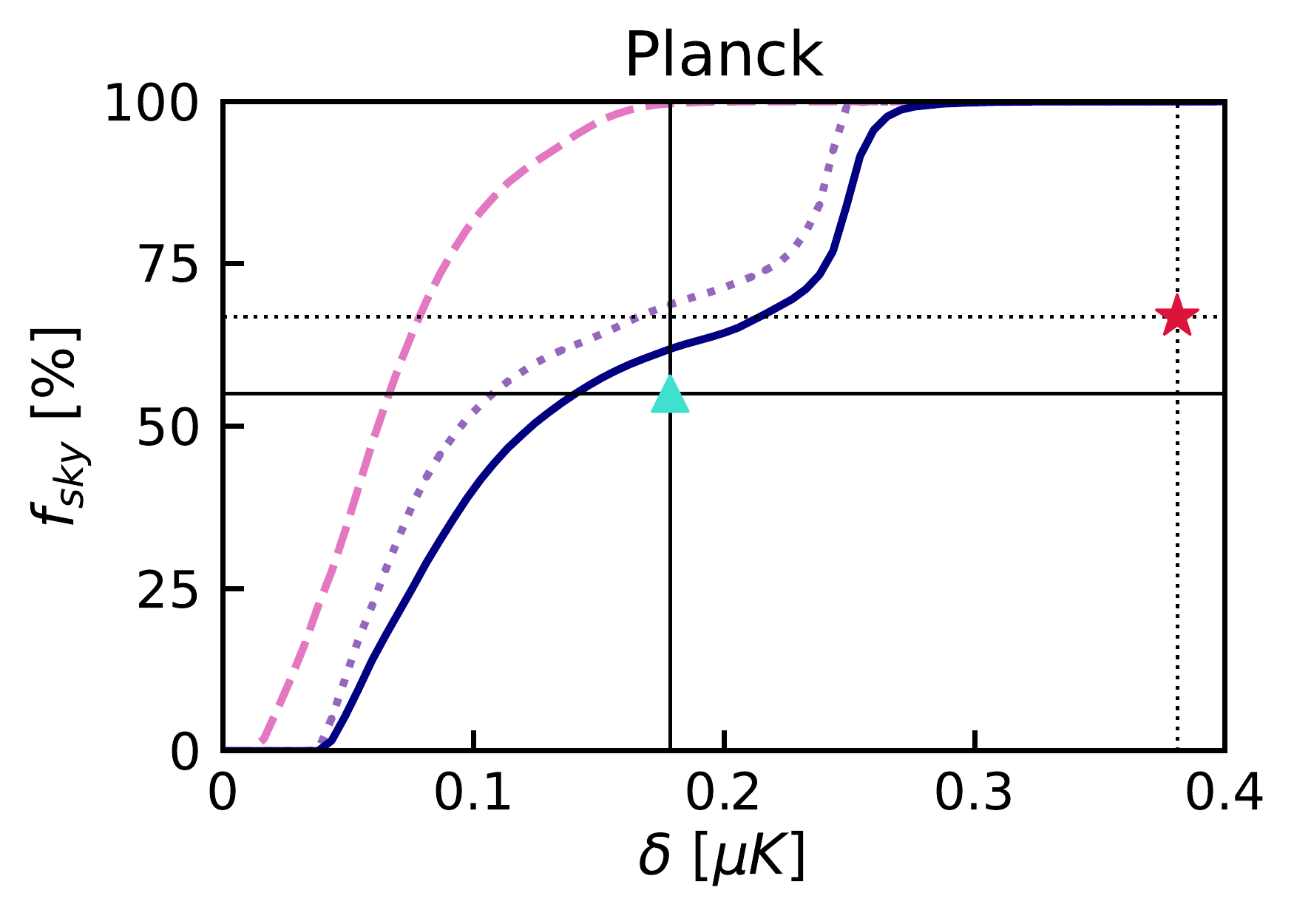}
    \includegraphics[width=0.32\linewidth]{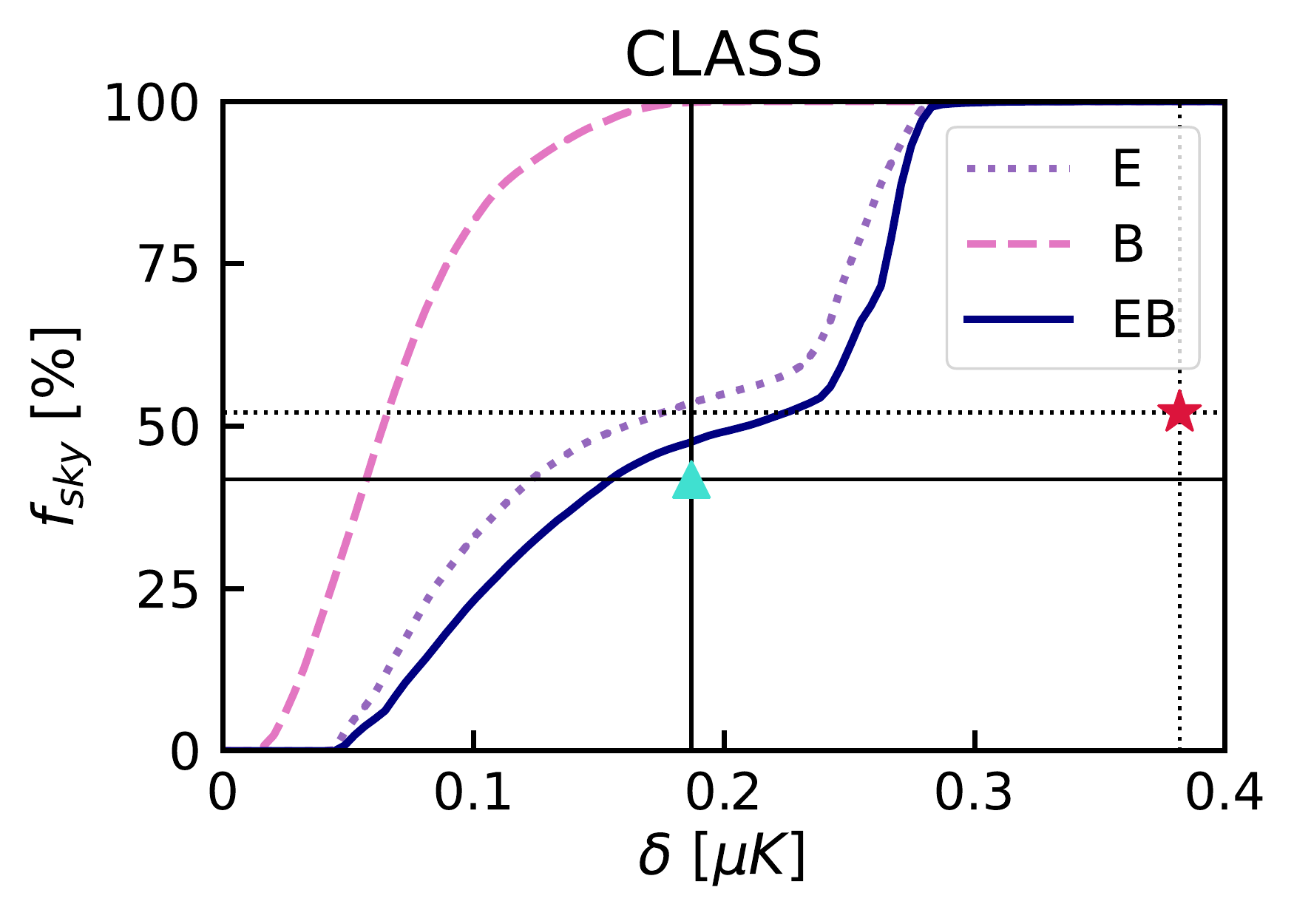}
    \includegraphics[width=0.32\linewidth]{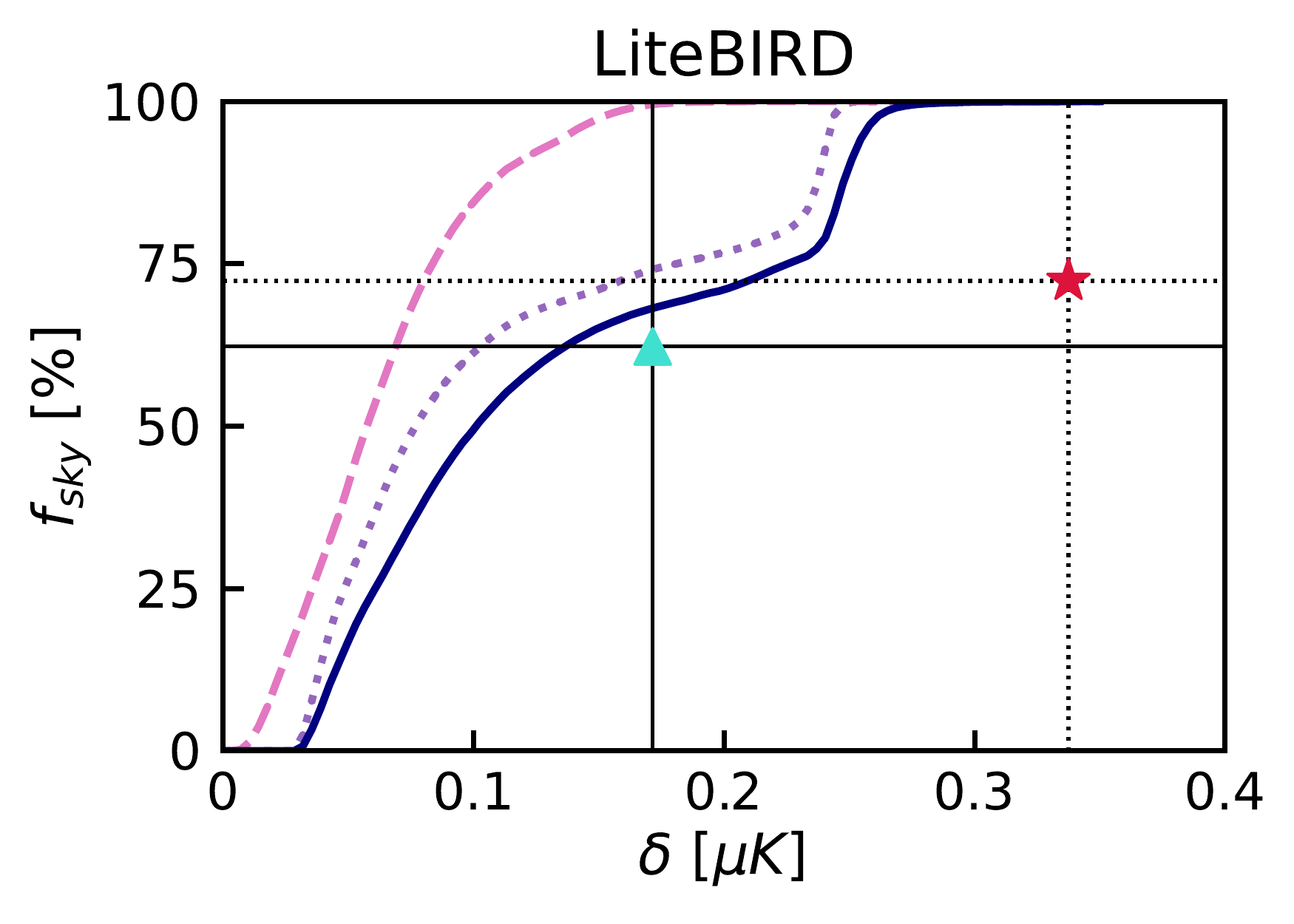}\\[5pt]
    \centering
    \includegraphics[width=0.32\linewidth]{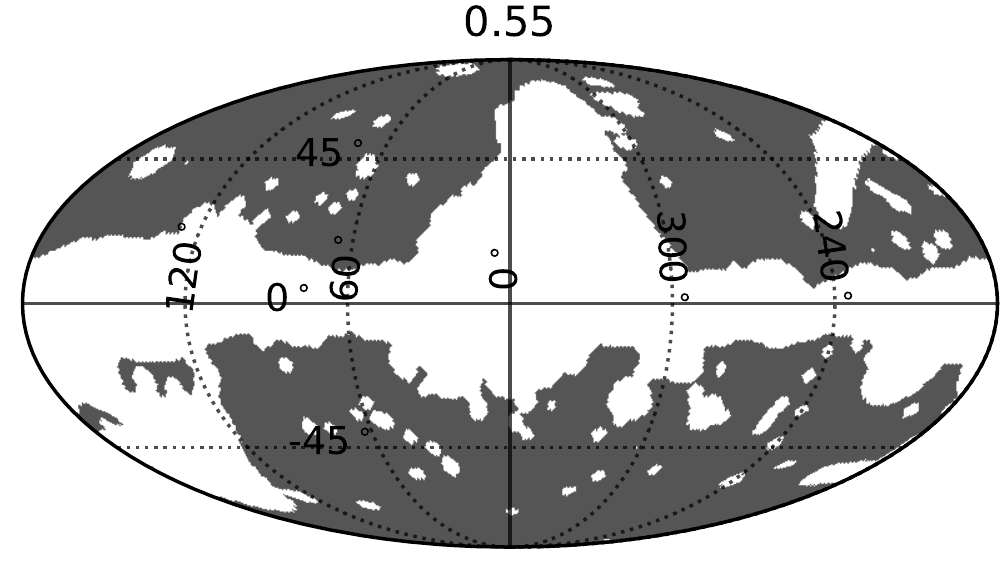}
    \includegraphics[width=0.32\linewidth]{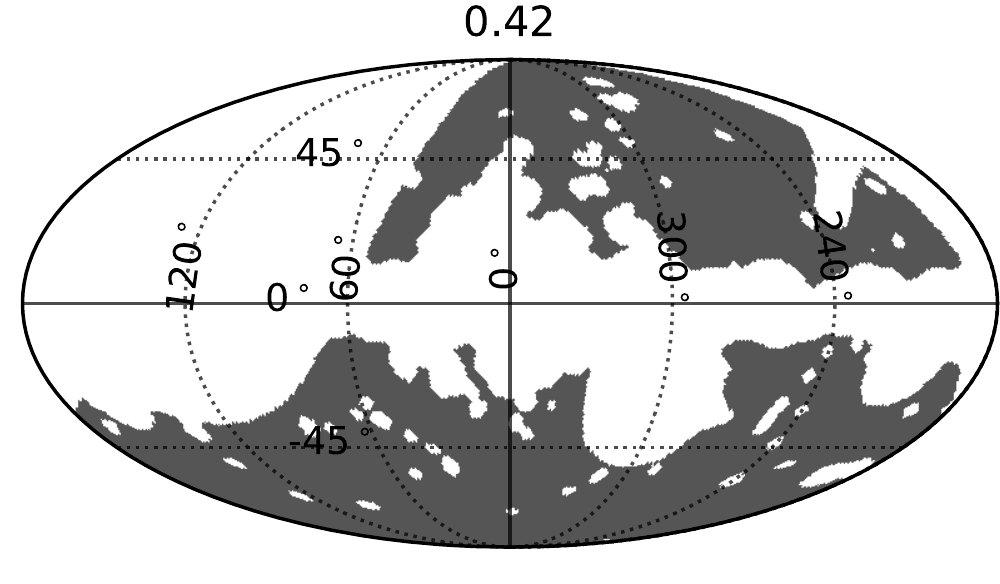}
    \includegraphics[width=0.32\linewidth]{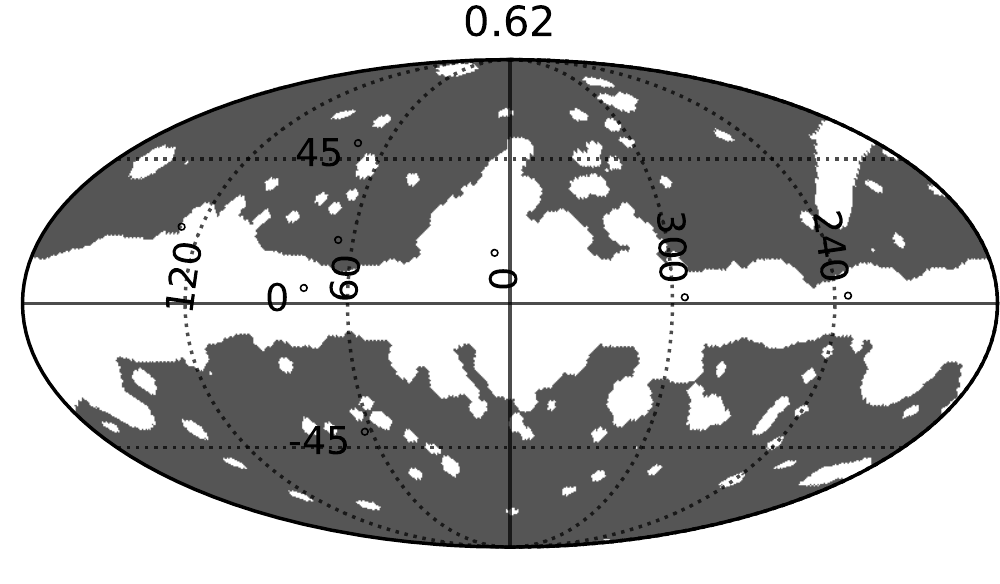}
    \caption{\textit{Upper panels}: Maximal RMS residuals for \planck{}, CLASS, and \litebird{} from left-to-right. 
    Plotted lines show the cumulative sky area with maximum deviation (Equation \ref{eq:Emask}) for E-modes ($\delta_E$, purple dotted), B-modes ($\delta_B$, pink dashed), and the combination of E- and B-modes ($\delta_{EB}$, navy solid).
    The red stars and thin dotted lines show the maximum deviation and sky fraction within the default polarization masks. 
    The turquoise triangles with solid lines show the maximum $\delta_{EB}$ and sky fraction within the E-mode masks shown in the bottom panels. 
    (The triangles fall below the navy solid line due to the mask apodization and the removal of small isolated unmasked regions).
    \textit{Bottom panels}: E-mode masks and sky fraction for each experiment.}
    \label{fig:E-masks}
\end{figure*}

We constructed the E-mode mask following the method in Appendix A.5 of \citet{akrami2020planck}.

The reconstruction of E and B-mode maps using masks described in section \ref{sec:simulations} introduces E/B-mixing, so we need to mask out extra regions of the E-mode map to reduce the impacts of compromised modes.
To do this, we generated the root mean square (RMS) residual maps from CMB $+$ noise simulations for the three different experiments.
The RMS residual map we used was defined to be $\delta_{EB}=\sqrt{(\delta_E)^2+(\delta_B)^2}$ with
\begin{equation}
    \begin{aligned}
    \delta_E &= \text{STD}(\tilde E - E^*),\\
    \delta_B &= \text{STD}(\tilde B - B^*),
    \end{aligned}\label{eq:Emask}
\end{equation}
where $\mathrm{STD}$ means taking the standard deviation per pixel, $\tilde E$ and $\tilde B$ are the E- and B-mode maps reconstructed beginning with the masked sky, and $E^*$ and $B^*$ are those based on the full sky.
We constructed residual maps $\delta_E$ and $\delta_B$ from all the simulations we generated ($10^4$ for each experiment).
The E-mode masks were created by first requiring the pixel values of the averaged RMS residual map to not exceed $0.2~\mathrm{\mu K}$. 
The thresholds were chosen based on the upper panels of Figure \ref{fig:E-masks}, which shows the maximal RMS residuals for E, B, and the combination of E and B modes. 
The masks were then smoothed using a Gaussian smoothing function with FWHM=160 arcmin. We set pixels in the smoothed maps with values smaller than 0.9 to be 0 (masked), otherwise 1 (unmasked).
After smoothing and setting thresholds, we removed a few islands of isolated non-zero pixels (with radius $<4^\circ$) that remained in our masks.
Due to the mask apodization and removal of small isolated unmasked regions, the triangles which represent the maximum $\delta_{EB}$ and sky fraction of the E-mode masks are not exactly on the $\delta_{EB}$ curves.

\begin{figure*}[t]
    \centering
    \includegraphics[width=\linewidth]{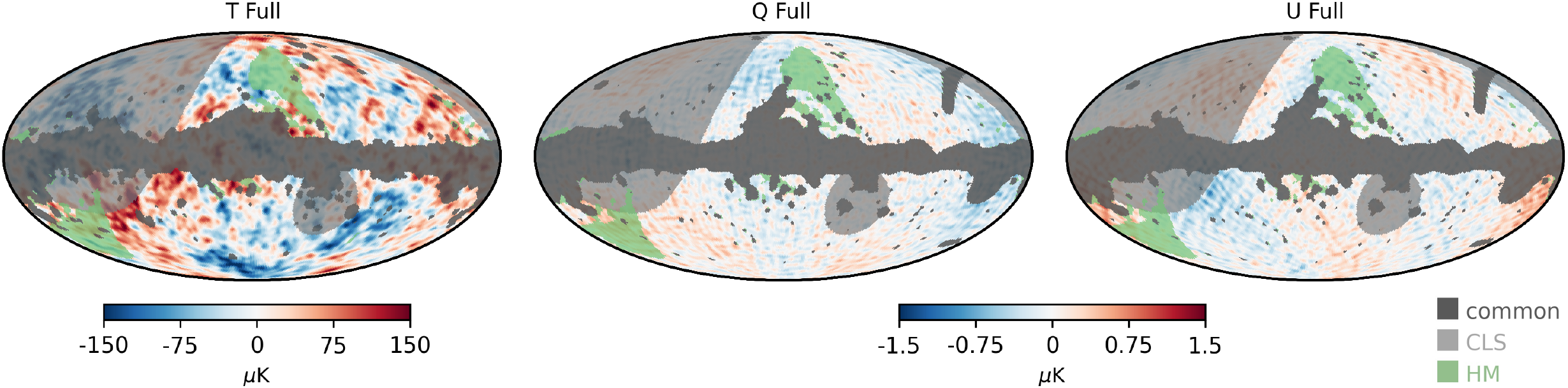}
    \caption{The Special CMB maps and corresponding masks at \texttt{HEALPix} resolution NSIDE=64, with T, Q and U maps from left-to-right. 
    Dark gray regions reflect the \planck{} common masks, light gray regions reflect the CLASS declination limits, and green regions are the parts of the \planck{} HM missing pixel masks that are outside of the \planck{} common masks.}
    \label{fig:spec_maps}
\end{figure*}

We checked the stability of our choice by decreasing the thresholds by $0.02~\mathrm{\mu K}$, and found the PTE for the amplitude of the E-mode hemispherical-power dipole ($d^{EE}$, Section \ref{sec:hpa}) measured in \planck{} \smica{} data was changed by $\lesssim2\%$, which does not change the significance level. 
(The E-mode mask was only used in analyzing the hemispherical power asymmetry.)
For the CLASS and \litebird{} forecasts, decreasing the thresholds by $0.02~\mathrm{\mu K}$ will shift the $d^{EE}$ medians of simulations by $\lesssim3\%$.
The lower panels of Figure \ref{fig:E-masks} shows the resulting E-mode masks.

\section{Visualization of the Special CMB E-mode anomalies}\label{sec:appendixb}
\begin{figure}[t]
    \centering
    \includegraphics[width=\linewidth]{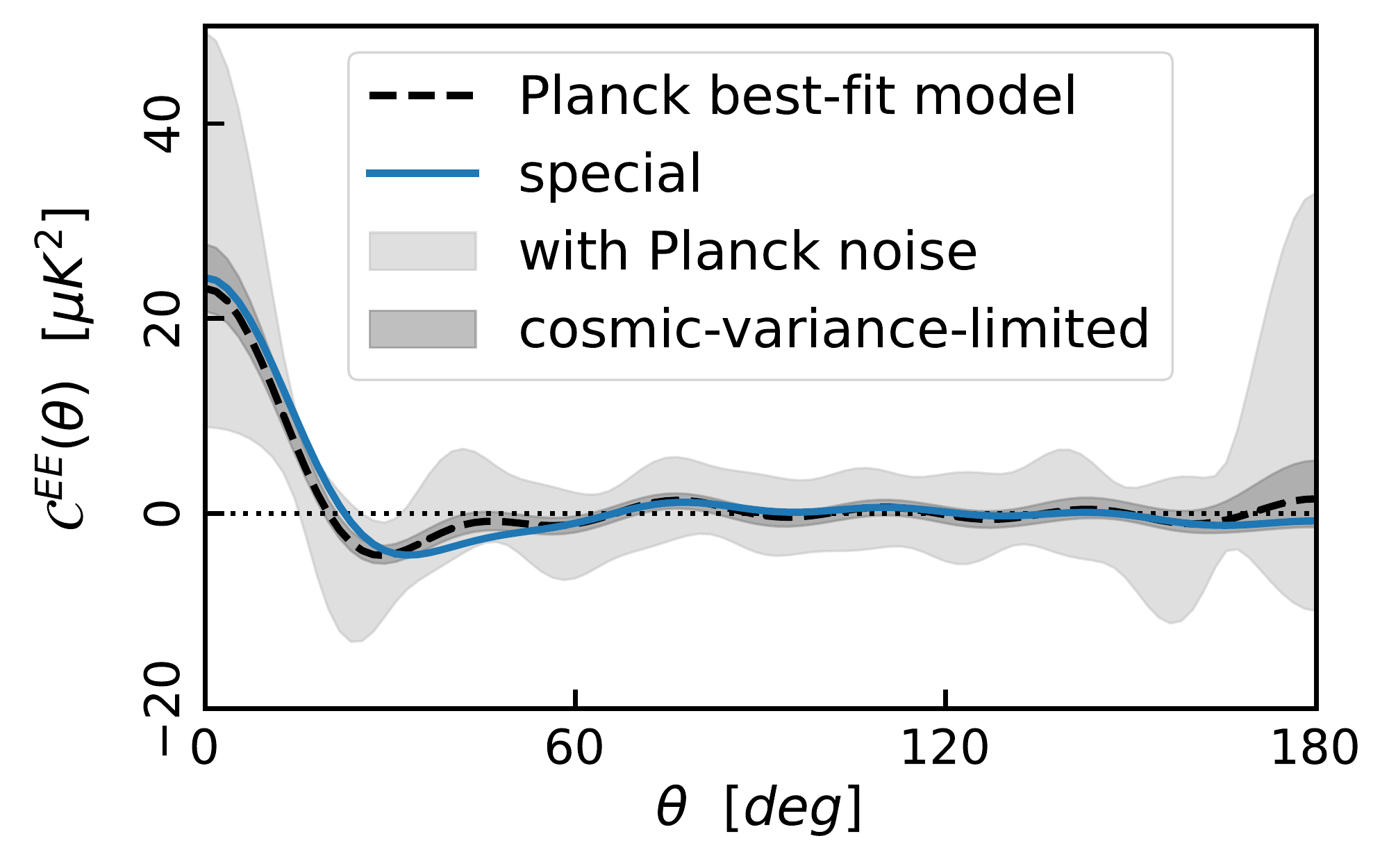}
    \caption{The E-mode two-point angular correlation functions computed using equation \ref{eq:Eaps2cf}, with $\ell_\mathrm{max}=10$. Blue solid line is the correlation functions from the Special CMB. 
    Black dashed line is that from \planck{} best-fit model, light gray region is the 1$\sigma$ range of CMB + \planck{} \smica{} noise simulations, and dark gray regions the $1\sigma$ range of cosmic-variance-limited simulations. 
    The PTE of $\log_{10}S_{1/2}^{EE}$ estimator is 96.3\%.}
    \label{fig:spec_corrEE}
\end{figure}

\begin{figure}[h]
    \centering
    \includegraphics[width=0.73\linewidth]{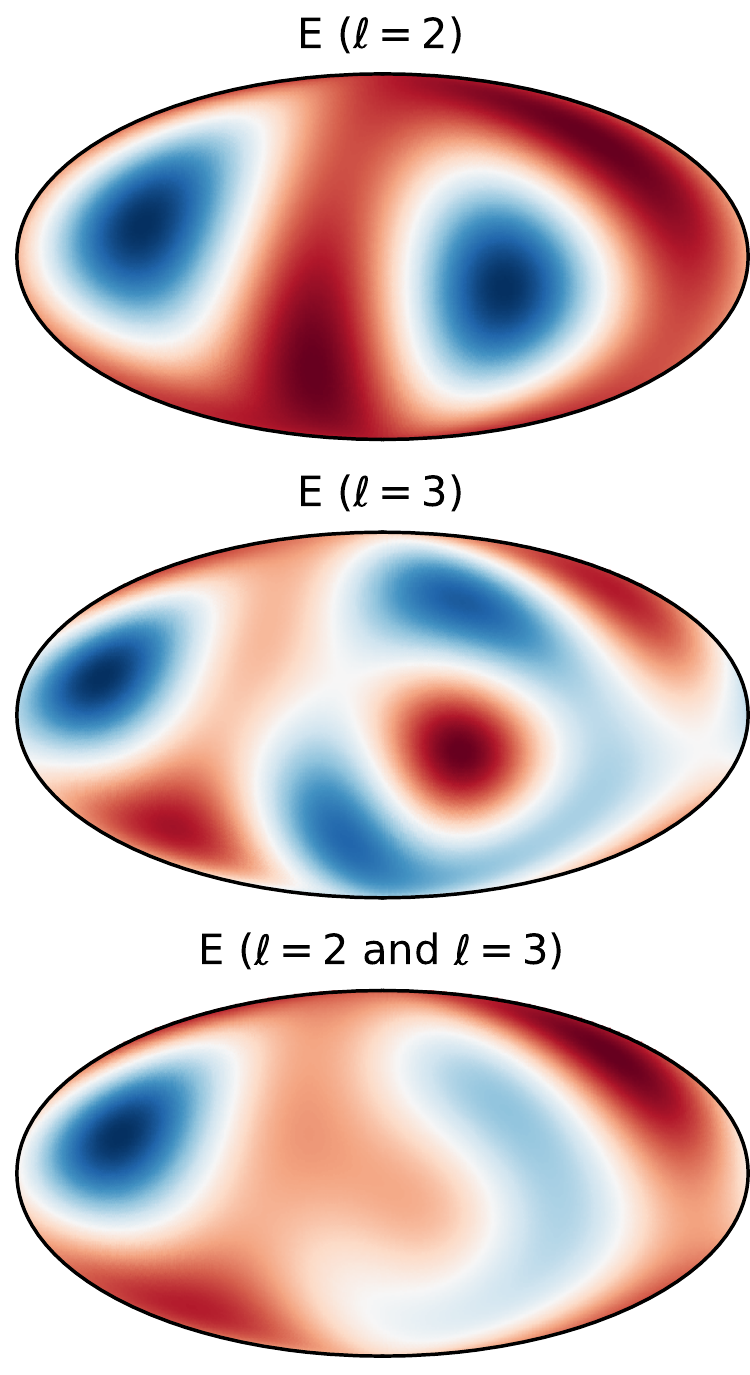}
    \caption{The quadrupole (left), octupole (middle), and their superposition (right) of the Special CMB E-mode map. The PTE of $L_{23}^{2(EE)}$ estimator is 1.4\%.}
    \label{fig:spec_QO}
\end{figure}

\begin{figure}[t]
    \centering
    \includegraphics[width=\linewidth]{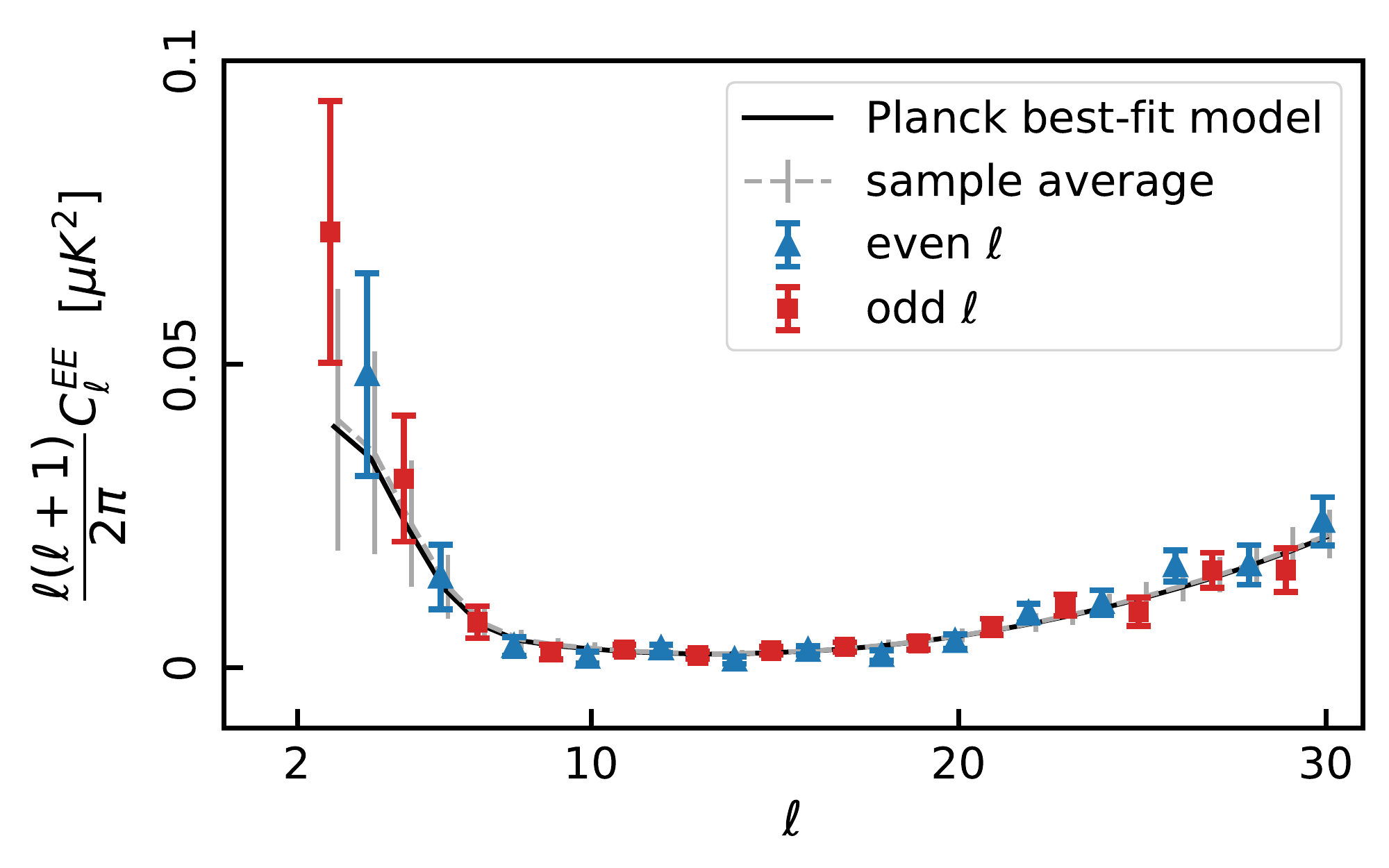}
    \caption{The power spectra for the NSIDE=16 Special CMB E-mode map, showing the first 30 multipoles.
    The even multipoles plotted in blue triangles, and odd in red squares. 
    The gray dashed line are the averages of the cosmic-variance-limited power spectra.
    The error bars for both data and simulations are the standard deviations of the simulations.
    The \planck{} best-fit model is shown with the black curve.
    Both data points and the sample average were horizontally shifted slightly relative to one another to provide a clearer view. The PTE of $D_{24}^{EE}$ estimator is 84.6\%.
    \label{fig:spec_pp}
    }
\end{figure}

\begin{figure*}[t]
    \centering
    \includegraphics[width=\linewidth]{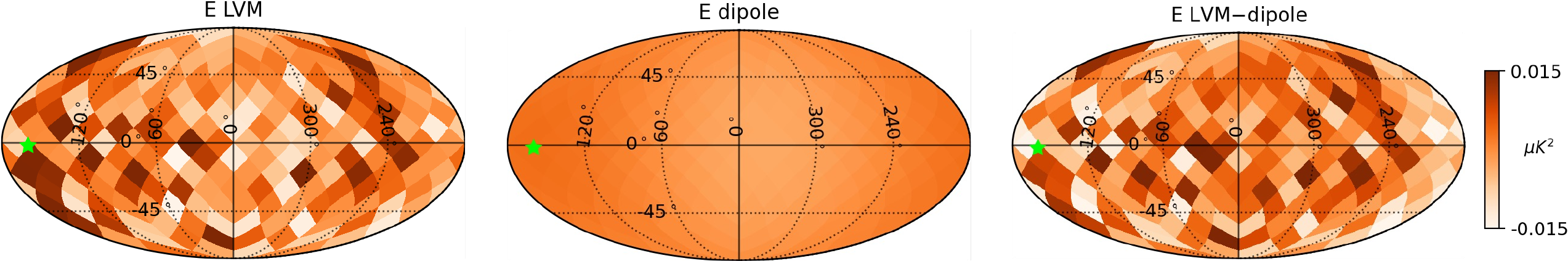}
    \caption{E-mode hemispherical power dipole estimation. \textit{Left panel}: the LVM (downgraded from NSIDE=16 to NSIDE=4) of the Special CMB map, with the average map of the LVM from the corresponding simulations (signal + noise) removed. 
    \textit{Middle panel}: the fitted dipole at NSIDE=4. 
    \textit{Right panel}: the dipole-subtracted LVM of Special CMB map. 
    The green stars mark the directions of the fitted dipoles. 
    The PTE of the dipole amplitude is 5.6\%.}
    \label{fig:spec_LVM}
\end{figure*}

In this appendix we show the Special CMB maps in Figure \ref{fig:spec_maps} and the visualization of the Special CMB E-mode anomalies in Figures \ref{fig:spec_corrEE}--\ref{fig:spec_LVM}, in analogy to Figures \ref{fig:corTT}--\ref{fig:LVM}.
The spectrum of the Special CMB in Figure \ref{fig:spec_pp} does not show a preference for all odd multipoles but only the first few. 
This is because in the definition of the $D_{24}^{EE}$ estimator (Equation \ref{eq:pp_DEE}), the first few multipoles are more important as they provide the largest level of power within the first 24 multipoles.

~
~
~

\section{Constrained universe}\label{sec:appendixc}
We followed the approach in \citet{copi2013large} for generating the constrained CMB realizations and hereby walk through some of the details.
According to the $\Lambda$CDM theory, the spherical harmonic coefficients $a_{\ell m}^T$, $a_{\ell m}^E$, and $a_{\ell m}^B$ of the CMB maps follow a complex normal distribution with zero mean and covariance matrix ($\mathrm{\Sigma}_\ell$)
\begin{equation}
    \mathrm{\Sigma_\ell}=
    \begin{bmatrix}
    C^{TT}_\ell & C^{TE}_\ell & 0\\[2.5pt]
    C^{TE}_\ell & C^{EE}_\ell & 0\\[2.5pt]
    0 & 0 & C^{BB}_\ell
    \end{bmatrix},
\end{equation}
where $C_\ell^{XY}$ (${XY}$ in $\{TT, TE, EE, BB\}$) are from \planck{}'s best-fit $\Lambda$CDM model. 
Its Cholesky decomposition $\mathrm\Sigma_\ell\equiv L_\ell\cdot L_\ell^T$ gives
\begin{equation}
    L_\ell=
    \begin{bmatrix}
    \sqrt{C_\ell^{TT}} & 0 & 0\\
    \frac{C_\ell^{TE}}{\sqrt{C_\ell^{TT}}} & \sqrt{C_\ell^{EE}-\frac{(C_\ell^{TE})^2}{C_\ell^{TT}}} & 0\\
    0 & 0 & C_\ell^{BB}
    \end{bmatrix}.
\end{equation}
To generate a unconstrained realization of $\vec{\hat{a}}_{\ell m}=(\hat a_{\ell m}^T,~\hat a_{\ell m}^E,~\hat a_{\ell m}^B)^T$ that have the desired covariance matrix, one needs to multiply a complex standard normal random vector $\vec \zeta=(\zeta_1,~\zeta_2,~\zeta_3)^T$ with the $L_\ell$ matrix. 
The constrained CMB realizations may be generated by setting $\hat a_{\ell m}^T$ to be what has been measured in the \planck{} \smica{} map (denoted with $a_{\ell m}^{T_\mathrm{data}}$). The constrained CMB realizations were sampled as
\begin{equation}
    \begin{aligned}
    \hat a_{\ell m}^T&=a_{\ell m}^{T_\mathrm{data}},\\
    \hat a_{\ell m}^E&=\frac{C_\ell^{TE}}{C_\ell^{TT}}a_{\ell m}^{T_\mathrm{data}}+\sqrt{C_\ell^{EE}-\frac{(C_\ell^{TE})^2}{C_\ell^{TT}}}\zeta_2,\\
    \hat a_{\ell m}^B&=\sqrt{C_\ell^{BB}}\zeta_3.
    \end{aligned}\label{eq:almconstr}
\end{equation}
We followed the convention in \texttt{HEALPix} \citep{gorski2005cosmology} by requiring that $a_{\ell m}^X=(-1)^m\left(a_{\ell -m}^X\right)^*$ where $m\geq0$ and $X\in\{ E, B\}$.

\begin{figure}
    \centering
    \includegraphics[width=1.03\linewidth]{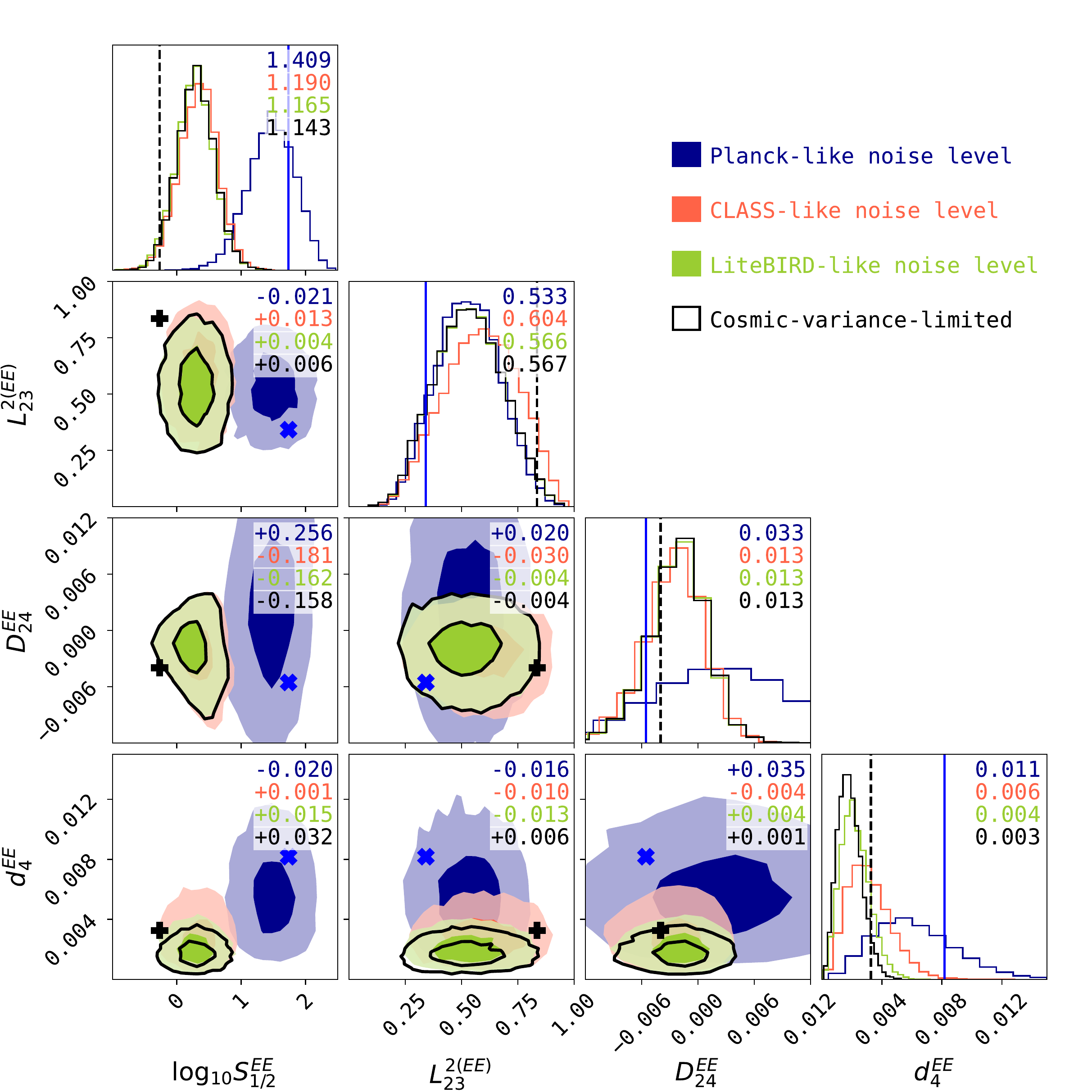}
    \caption{Confidence-curve matrix of polarization anomaly estimators based on the constrained simulation study, with contours and histograms from \planck{} (blue), CLASS (pink), \litebird{} (green), and the cosmic-variance-limited Ideal case (black); see Table \ref{tab:95conf} for a summary of anomaly names and estimators.
    The numbers on the top-right of off-diagonal panels are the Pearson correlation coefficients, and contours show $1\sigma$ and $2\sigma$ significance levels.
    The numbers on the top-right of diagonal panels are the 95\% interval widths.
    The blue crosses and vertical lines represent \planck{} \smica{} measurements, and black crosses and dashed vertical lines represent E-mode measurements from the Special CMB (see Section \ref{ssec:specialCMB} for more about the Special CMB).}
    \label{fig:constrainedtest}
\end{figure}

We generated $10^5$ constrained CMB realizations and looked into the distributions of E-mode anomaly estimators after combining these CMB realizations with different noise simulations as has been done in the main text.
The results are displayed in Table \ref{tab:constraindconf} and Figure \ref{fig:constrainedtest}.
We found that these new results are largely consistent with those for the unconstrained universe (Table \ref{tab:95conf} and Figure \ref{fig:corner}). 
The biggest difference we noticed is the median of $\log_{10}S_{1/2}^{EE}$ in the Ideal case, which shifted from $0.19$ (unconstrained) to $0.30$ (constrained).
This is expected because the variance of $\hat a_{\ell m}^E$ sampled according to Equation \ref{eq:almconstr} no longer equals to $C_\ell^{EE}$:
\begin{equation}
    \begin{aligned}
    \left\langle \hat a_{\ell m}^E\cdot \left(\hat a_{\ell m}^E\right)^*\right\rangle
    =&\left(\frac{C_\ell^{TE}}{C_\ell^{TT}}\right)^2\left\langle a_{\ell m}^{T_\mathrm{data}}\cdot \left(a_{\ell m}^{T_\mathrm{data}}\right)^*\right\rangle\\
    &+\left(\sqrt{C_\ell^{EE}-\frac{\left(C_\ell^{TE}\right)^2}{C_\ell^{TT}}}\right)^2\left\langle \zeta_2\cdot\zeta_2^* \right\rangle\\
    =& C_\ell^{EE}+\frac{\left(C_\ell^{TE}\right)^2}{C_\ell^{TT}}\left(\frac{C_\ell^{TT_\mathrm{data}}}{C_\ell^{TT}}-1\right),
    \end{aligned}
\end{equation}
where $C_\ell^{TT_\mathrm{data}}$ is the power spectra of \planck{} \smica{}, and $\left\langle a_{\ell m}^{T_\mathrm{data}}\cdot \zeta_2^*\right\rangle=\left\langle \zeta_2\cdot \left(a_{\ell m}^{T_\mathrm{data}}\right)^*\right\rangle=0$ was assumed. 
The fact that $C_\ell^{TT_\mathrm{data}}$ fluctuate around $C_\ell^{TT}$ caused the integral of two-point correlation function to be larger than that from the unconstrained simulations.

\begin{deluxetable*}{@{\extracolsep{4pt}}ccccccccc}[!]
\tablenum{7}
\tablecaption{Bias and error based on the constrained universe simulation set.}\label{tab:constraindconf}
\tablewidth{0pt}
\tablehead{
\colhead{Estimator} & \colhead{\smica{} (PTE)$^a$} & \multicolumn{2}{c}{\planck{}} & \multicolumn{2}{c}{CLASS} & \multicolumn{2}{c}{\litebird{}} & \colhead{Ideal}\\
\cline{3-4} \cline{5-6} \cline{7-8} \colhead{} & \colhead{} & \colhead{bias$^b$} & \colhead{error$^b$} & \colhead{bias$^b$} & \colhead{error$^b$} & \colhead{bias$^b$} & \colhead{error$^b$} & \colhead{}}
\startdata
$\log_{10}S_{1/2}^{EE}$ & $1.69$ $({19.6\%})$   & $0.98$  & $1.2$ & $0.04$  & $1.0$ & $-0.01$  & $1.0$ & $0.30_{-0.57}^{+0.57}$ \\
    $L_{23}^{2(EE)}$        & $0.34$ $(90.4\%)$   & $-0.01$ & $0.9$ & $0.10$ & $1.1$ & $0.00$ & $1.0$ & $0.53_{-0.28}^{+0.29}$ \\
    $D^{EE}_{24}$           & $-0.005$ $({85.8\%})$ & $0.33$   & ${2.6}$ & $-0.04$  & $1.1$ & $-0.00$  & $1.0$ & $-0.002_{-0.007}^{+0.006}$ \\
    $d^{EE}_4~[\mu K^2]$    & $0.008$ $(22.6\%)$  & $1.34$  & $3.4$ & $0.39$  & $1.8$ & $0.11$  & $1.2$ & $0.002_{-0.001}^{+0.002}$ \\
\enddata
\tablenotetext{}{$^a$We quote the PTEs of the \smica{} measurements (column 2) in the brackets. PTEs were obtained by comparing to the \planck{} simulations.\\
$^b$The bias relative to the Ideal value and the error are unitless (normalized by the Ideal 95\% error).}
\end{deluxetable*}

\bibliography{main.bib, planck_bib.bib}{}

\begin{thebibliography}{}
\expandafter\ifx\csname natexlab\endcsname\relax\def\natexlab#1{#1}\fi
\providecommand{\url}[1]{\href{#1}{#1}}
\providecommand{\dodoi}[1]{doi:~\href{http://doi.org/#1}{\nolinkurl{#1}}}
\providecommand{\doeprint}[1]{\href{http://ascl.net/#1}{\nolinkurl{http://ascl.net/#1}}}
\providecommand{\doarXiv}[1]{\href{https://arxiv.org/abs/#1}{\nolinkurl{https://arxiv.org/abs/#1}}}

\bibitem[{{Abramo} {et~al.}(2006){Abramo}, {Sodr{\'e}}, \&
  {Wuensche}}]{abramo2006anomalies}
{Abramo}, L.~R., {Sodr{\'e}}, Laerte, J., \& {Wuensche}, C.~A. 2006, \prd, 74,
  083515, \dodoi{10.1103/PhysRevD.74.083515}

\bibitem[{{Addamo} {et~al.}(2021){Addamo}, {Ade}, {Baccigalupi}, {Baldini},
  {Battaglia}, {Battistelli}, {Ba{\`u}}, {de Bernardis}, {Bersanelli},
  {Biasotti}, {Boscaleri}, {Caccianiga}, {Caprioli}, {Cavaliere}, {Cei},
  {Cleary}, {Columbro}, {Coppi}, {Coppolecchia}, {Cuttaia}, {D'Alessandro}, {De
  Gasperis}, {De Petris}, {Fafone}, {Farsian}, {Ferrari Barusso}, {Fontanelli},
  {Franceschet}, {Gaier}, {Galli}, {Gatti}, {Genova-Santos}, {Gerbino},
  {Gervasi}, {Ghigna}, {Grosso}, {Gruppuso}, {Gualtieri}, {Incardona}, {Jones},
  {Kangaslahti}, {Krachmalnicoff}, {Lamagna}, {Lattanzi},
  {L{\'o}pez-Caraballo}, {Lumia}, {Mainini}, {Maino}, {Mandelli}, {Maris},
  {Masi}, {Matarrese}, {May}, {Mele}, {Mena}, {Mennella}, {Molina}, {Molinari},
  {Morgante}, {Natale}, {Nati}, {Natoli}, {Pagano}, {Paiella}, {Panico},
  {Paonessa}, {Paradiso}, {Passerini}, {Perez-de-Taoro}, {Peverini},
  {Pezzotta}, {Piacentini}, {Piccirillo}, {Pisano}, {Polenta}, {Poletti},
  {Presta}, {Realini}, {Reyes}, {Rocchi}, {Rubino-Martin}, {Sandri}, {Sartor},
  {Schillaci}, {Signorelli}, {Siri}, {Soria}, {Spinella}, {Tapia}, {Tartari},
  {Taylor}, {Terenzi}, {Tomasi}, {Tommasi}, {Tucker}, {Vaccaro}, {Vigano},
  {Villa}, {Virone}, {Vittorio}, {Volpe}, {Watkins}, {Zacchei}, {Zannoni}, \&
  {LSPE Collaboration}}]{addamo2021large}
{Addamo}, G., {Ade}, P.~A.~R., {Baccigalupi}, C., {et~al.} 2021, \jcap, 2021,
  008, \dodoi{10.1088/1475-7516/2021/08/008}

\bibitem[{{Akrami} {et~al.}(2014){Akrami}, {Fantaye}, {Shafieloo}, {Eriksen},
  {Hansen}, {Banday}, \& {G{\'o}rski}}]{akrami2014power}
{Akrami}, Y., {Fantaye}, Y., {Shafieloo}, A., {et~al.} 2014, \apjl, 784, L42,
  \dodoi{10.1088/2041-8205/784/2/L42}

\bibitem[{{Aluri} {et~al.}(2017){Aluri}, {Ralston}, \&
  {Weltman}}]{aluri2017alignments}
{Aluri}, P.~K., {Ralston}, J.~P., \& {Weltman}, A. 2017, \mnras, 472, 2410,
  \dodoi{10.1093/mnras/stx2112}

\bibitem[{{Bennett} {et~al.}(2003){Bennett}, {Halpern}, {Hinshaw}, {Jarosik},
  {Kogut}, {Limon}, {Meyer}, {Page}, {Spergel}, {Tucker}, {Wollack}, {Wright},
  {Barnes}, {Greason}, {Hill}, {Komatsu}, {Nolta}, {Odegard}, {Peiris},
  {Verde}, \& {Weiland}}]{Bennett_2003}
{Bennett}, C.~L., {Halpern}, M., {Hinshaw}, G., {et~al.} 2003, \apjs, 148, 1,
  \dodoi{10.1086/377253}

\bibitem[{{Bennett} {et~al.}(2011){Bennett}, {Hill}, {Hinshaw}, {Larson},
  {Smith}, {Dunkley}, {Gold}, {Halpern}, {Jarosik}, {Kogut}, {Komatsu},
  {Limon}, {Meyer}, {Nolta}, {Odegard}, {Page}, {Spergel}, {Tucker}, {Weiland},
  {Wollack}, \& {Wright}}]{bennett2011seven}
{Bennett}, C.~L., {Hill}, R.~S., {Hinshaw}, G., {et~al.} 2011, \apjs, 192, 17,
  \dodoi{10.1088/0067-0049/192/2/17}

\bibitem[{{Billi} {et~al.}(2019){Billi}, {Gruppuso}, {Mandolesi}, {Moscardini},
  \& {Natoli}}]{billi2019polarisation}
{Billi}, M., {Gruppuso}, A., {Mandolesi}, N., {Moscardini}, L., \& {Natoli}, P.
  2019, Physics of the Dark Universe, 26, 100327,
  \dodoi{10.1016/j.dark.2019.100327}

\bibitem[{{Bond} {et~al.}(1998){Bond}, {Jaffe}, \& {Knox}}]{bond1998estimating}
{Bond}, J.~R., {Jaffe}, A.~H., \& {Knox}, L. 1998, \prd, 57, 2117,
  \dodoi{10.1103/PhysRevD.57.2117}

\bibitem[{{Caswell} {et~al.}(2019){Caswell}, {Droettboom}, {Lee}, {Hunter},
  {Firing}, {Stansby}, {Klymak}, {Sales De Andrade}, {Hoffmann}, {Hedegaard
  Nielsen}, {Varoquaux}, {Root}, {Elson}, {May}, {Dale}, {Lee}, {Sepp{\"a}nen},
  {McDougall}, {Straw}, {Hobson}, {Gohlke}, {Yu}, {Ma}, {Vincent}, {Silvester},
  {Moad}, {Ernest}, {Katins}, {Kniazev}, \& {Ivanov}}]{caswell2019matplotlib}
{Caswell}, T.~A., {Droettboom}, M., {Lee}, A., {et~al.} 2019,
  {matplotlib/matplotlib v3.1.2}, v3.1.2,  Zenodo,
  \dodoi{10.5281/zenodo.3563226}

\bibitem[{{Cayuso} \& {Johnson}(2020)}]{cayuso2020towards}
{Cayuso}, J.~I., \& {Johnson}, M.~C. 2020, \prd, 101, 123508,
  \dodoi{10.1103/PhysRevD.101.123508}

\bibitem[{{Cheng} {et~al.}(2016){Cheng}, {Zhao}, {Huang}, \&
  {Santos}}]{cheng2016preferred}
{Cheng}, C., {Zhao}, W., {Huang}, Q.-G., \& {Santos}, L. 2016, Physics Letters
  B, 757, 445, \dodoi{10.1016/j.physletb.2016.04.030}

\bibitem[{{Chiocchetta} {et~al.}(2021){Chiocchetta}, {Gruppuso}, {Lattanzi},
  {Natoli}, \& {Pagano}}]{chiocchetta2020lack}
{Chiocchetta}, C., {Gruppuso}, A., {Lattanzi}, M., {Natoli}, P., \& {Pagano},
  L. 2021, \jcap, 2021, 015, \dodoi{10.1088/1475-7516/2021/08/015}

\bibitem[{{Chon} {et~al.}(2004){Chon}, {Challinor}, {Prunet}, {Hivon}, \&
  {Szapudi}}]{chon2004fast}
{Chon}, G., {Challinor}, A., {Prunet}, S., {Hivon}, E., \& {Szapudi}, I. 2004,
  \mnras, 350, 914, \dodoi{10.1111/j.1365-2966.2004.07737.x}

\bibitem[{{Cicoli} {et~al.}(2014){Cicoli}, {Downes}, {Dutta}, {Pedro}, \&
  {Westphal}}]{cicoli2014just}
{Cicoli}, M., {Downes}, S., {Dutta}, B., {Pedro}, F.~G., \& {Westphal}, A.
  2014, \jcap, 2014, 030, \dodoi{10.1088/1475-7516/2014/12/030}

\bibitem[{{Copi} {et~al.}(2006){Copi}, {Huterer}, {Schwarz}, \&
  {Starkman}}]{copi2006large}
{Copi}, C.~J., {Huterer}, D., {Schwarz}, D.~J., \& {Starkman}, G.~D. 2006,
  \mnras, 367, 79, \dodoi{10.1111/j.1365-2966.2005.09980.x}

\bibitem[{{Copi} {et~al.}(2007){Copi}, {Huterer}, {Schwarz}, \&
  {Starkman}}]{copi2007uncorrelated}
---. 2007, \prd, 75, 023507, \dodoi{10.1103/PhysRevD.75.023507}

\bibitem[{{Copi} {et~al.}(2009){Copi}, {Huterer}, {Schwarz}, \&
  {Starkman}}]{copi2009no}
---. 2009, \mnras, 399, 295, \dodoi{10.1111/j.1365-2966.2009.15270.x}

\bibitem[{{Copi} {et~al.}(2013){Copi}, {Huterer}, {Schwarz}, \&
  {Starkman}}]{copi2013large}
---. 2013, \mnras, 434, 3590, \dodoi{10.1093/mnras/stt1287}

\bibitem[{{Copi} {et~al.}(2015{\natexlab{a}}){Copi}, {Huterer}, {Schwarz}, \&
  {Starkman}}]{copi2015large}
---. 2015{\natexlab{a}}, \mnras, 449, 3458, \dodoi{10.1093/mnras/stv501}

\bibitem[{{Copi} {et~al.}(2015{\natexlab{b}}){Copi}, {Huterer}, {Schwarz}, \&
  {Starkman}}]{copi2015lack}
---. 2015{\natexlab{b}}, \mnras, 451, 2978, \dodoi{10.1093/mnras/stv1143}

\bibitem[{{Copi} {et~al.}(2004){Copi}, {Huterer}, \&
  {Starkman}}]{copi2004multipole}
{Copi}, C.~J., {Huterer}, D., \& {Starkman}, G.~D. 2004, \prd, 70, 043515,
  \dodoi{10.1103/PhysRevD.70.043515}

\bibitem[{{Copi} {et~al.}(2016){Copi}, {O'Dwyer}, \& {Starkman}}]{copi2016isw}
{Copi}, C.~J., {O'Dwyer}, M., \& {Starkman}, G.~D. 2016, \mnras, 463, 3305,
  \dodoi{10.1093/mnras/stw2163}

\bibitem[{{Cruz} {et~al.}(2005){Cruz}, {Mart{\'\i}nez-Gonz{\'a}lez}, {Vielva},
  \& {Cay{\'o}n}}]{cruz2005detection}
{Cruz}, M., {Mart{\'\i}nez-Gonz{\'a}lez}, E., {Vielva}, P., \& {Cay{\'o}n}, L.
  2005, \mnras, 356, 29, \dodoi{10.1111/j.1365-2966.2004.08419.x}

\bibitem[{{Dahal} {et~al.}(2022){Dahal}, {Appel}, {Datta}, {Brewer}, {Ali},
  {Bennett}, {Bustos}, {Chan}, {Chuss}, {Cleary}, {Couto}, {Denis},
  {D{\"u}nner}, {Eimer}, {Espinoza}, {Essinger-Hileman}, {Golec}, {Harrington},
  {Helson}, {Iuliano}, {Karakla}, {Li}, {Marriage}, {McMahon}, {Miller},
  {Novack}, {N{\'u}{\~n}ez}, {Osumi}, {Padilla}, {Palma}, {Parker}, {Petroff},
  {Reeves}, {Rhoades}, {Rostem}, {Valle}, {Watts}, {Weiland}, {Wollack}, \&
  {Xu}}]{dahal2021four}
{Dahal}, S., {Appel}, J.~W., {Datta}, R., {et~al.} 2022, \apj, 926, 33,
  \dodoi{10.3847/1538-4357/ac397c}

\bibitem[{{de Oliveira-Costa} {et~al.}(2004){de Oliveira-Costa}, {Tegmark},
  {Zaldarriaga}, \& {Hamilton}}]{de2004significance}
{de Oliveira-Costa}, A., {Tegmark}, M., {Zaldarriaga}, M., \& {Hamilton}, A.
  2004, \prd, 69, 063516, \dodoi{10.1103/PhysRevD.69.063516}

\bibitem[{{Dvorkin} {et~al.}(2008){Dvorkin}, {Peiris}, \&
  {Hu}}]{dvorkin2008testable}
{Dvorkin}, C., {Peiris}, H.~V., \& {Hu}, W. 2008, \prd, 77, 063008,
  \dodoi{10.1103/PhysRevD.77.063008}

\bibitem[{{Eriksen} {et~al.}(2004){Eriksen}, {Hansen}, {Banday}, {G{\'o}rski},
  \& {Lilje}}]{eriksen2004asymmetries}
{Eriksen}, H.~K., {Hansen}, F.~K., {Banday}, A.~J., {G{\'o}rski}, K.~M., \&
  {Lilje}, P.~B. 2004, \apj, 605, 14, \dodoi{10.1086/382267}

\bibitem[{{Essinger-Hileman} {et~al.}(2014){Essinger-Hileman}, {Ali}, {Amiri},
  {Appel}, {Araujo}, {Bennett}, {Boone}, {Chan}, {Cho}, {Chuss}, {Colazo},
  {Crowe}, {Denis}, {D{\"u}nner}, {Eimer}, {Gothe}, {Halpern}, {Harrington},
  {Hilton}, {Hinshaw}, {Huang}, {Irwin}, {Jones}, {Karakla}, {Kogut}, {Larson},
  {Limon}, {Lowry}, {Marriage}, {Mehrle}, {Miller}, {Miller}, {Moseley},
  {Novak}, {Reintsema}, {Rostem}, {Stevenson}, {Towner}, {U-Yen}, {Wagner},
  {Watts}, {Wollack}, {Xu}, \& {Zeng}}]{essinger2014class}
{Essinger-Hileman}, T., {Ali}, A., {Amiri}, M., {et~al.} 2014, in Society of
  Photo-Optical Instrumentation Engineers (SPIE) Conference Series, Vol. 9153,
  Millimeter, Submillimeter, and Far-Infrared Detectors and Instrumentation for
  Astronomy VII, ed. W.~S. {Holland} \& J.~{Zmuidzinas}, 91531I,
  \dodoi{10.1117/12.2056701}

\bibitem[{{Frejsel}(2015)}]{frejsel2015large}
{Frejsel}, A.~M. 2015, PhD thesis, University of Copenhagen, Denmark

\bibitem[{{Frommert} \& {En{\ss}lin}(2010)}]{frommert2010axis}
{Frommert}, M., \& {En{\ss}lin}, T.~A. 2010, \mnras, 403, 1739,
  \dodoi{10.1111/j.1365-2966.2010.16255.x}

\bibitem[{{Gandilo} {et~al.}(2016){Gandilo}, {Ade}, {Benford}, {Bennett},
  {Chuss}, {Dotson}, {Eimer}, {Fixsen}, {Halpern}, {Hilton}, {Hinshaw},
  {Irwin}, {Jhabvala}, {Kimball}, {Kogut}, {Lowe}, {McMahon}, {Miller},
  {Mirel}, {Moseley}, {Pawlyk}, {Rodriguez}, {Sharp}, {Shirron}, {Staguhn},
  {Sullivan}, {Switzer}, {Taraschi}, {Tucker}, \&
  {Wollack}}]{gandilo2016primordial}
{Gandilo}, N.~N., {Ade}, P. A.~R., {Benford}, D., {et~al.} 2016, in Society of
  Photo-Optical Instrumentation Engineers (SPIE) Conference Series, Vol. 9914,
  Millimeter, Submillimeter, and Far-Infrared Detectors and Instrumentation for
  Astronomy VIII, ed. W.~S. {Holland} \& J.~{Zmuidzinas}, 99141J,
  \dodoi{10.1117/12.2231109}

\bibitem[{{Gordon} {et~al.}(2005){Gordon}, {Hu}, {Huterer}, \&
  {Crawford}}]{gordon2005spontaneous}
{Gordon}, C., {Hu}, W., {Huterer}, D., \& {Crawford}, T. 2005, \prd, 72,
  103002, \dodoi{10.1103/PhysRevD.72.103002}

\bibitem[{{G{\'o}rski} {et~al.}(2005){G{\'o}rski}, {Hivon}, {Banday},
  {Wandelt}, {Hansen}, {Reinecke}, \& {Bartelmann}}]{gorski2005cosmology}
{G{\'o}rski}, K.~M., {Hivon}, E., {Banday}, A.~J., {et~al.} 2005, \apj, 622,
  759, \dodoi{10.1086/427976}

\bibitem[{{Groeneboom} \& {Eriksen}(2009)}]{groeneboom2008bayesian}
{Groeneboom}, N.~E., \& {Eriksen}, H.~K. 2009, \apj, 690, 1807,
  \dodoi{10.1088/0004-637X/690/2/1807}

\bibitem[{{Gruppuso}(2014)}]{gruppuso2014two}
{Gruppuso}, A. 2014, \mnras, 437, 2076, \dodoi{10.1093/mnras/stt1937}

\bibitem[{{Gruppuso} {et~al.}(2011){Gruppuso}, {Finelli}, {Natoli}, {Paci},
  {Cabella}, {de Rosa}, \& {Mandolesi}}]{gruppuso2011new}
{Gruppuso}, A., {Finelli}, F., {Natoli}, P., {et~al.} 2011, \mnras, 411, 1445,
  \dodoi{10.1111/j.1365-2966.2010.17773.x}

\bibitem[{{Hanany} {et~al.}(2019){Hanany}, {Alvarez}, {Artis}, {Ashton},
  {Aumont}, {Aurlien}, {Banerji}, {Barreiro}, {Bartlett}, {Basak}, {Battaglia},
  {Bock}, {Boddy}, {Bonato}, {Borrill}, {Bouchet}, {Boulanger}, {Burkhart},
  {Chluba}, {Chuss}, {Clark}, {Cooperrider}, {Crill}, {De Zotti},
  {Delabrouille}, {Di Valentino}, {Didier}, {Dor{\'e}}, {Eriksen}, {Errard},
  {Essinger-Hileman}, {Feeney}, {Filippini}, {Fissel}, {Flauger}, {Fuskeland},
  {Gluscevic}, {Gorski}, {Green}, {Hensley}, {Herranz}, {Hill}, {Hivon},
  {Hlo{\v{z}}ek}, {Hubmayr}, {Johnson}, {Jones}, {Jones}, {Knox}, {Kogut},
  {L{\'o}pez-Caniego}, {Lawrence}, {Lazarian}, {Li}, {Madhavacheril}, {Melin},
  {Meyers}, {Murray}, {Negrello}, {Novak}, {O'Brient}, {Paine}, {Pearson},
  {Pogosian}, {Pryke}, {Puglisi}, {Remazeilles}, {Rocha}, {Schmittfull},
  {Scott}, {Shirron}, {Stephens}, {Sutin}, {Tomasi}, {Trangsrud}, {van
  Engelen}, {Vansyngel}, {Wehus}, {Wen}, {Xu}, {Young}, \&
  {Zonca}}]{hanany2019pico}
{Hanany}, S., {Alvarez}, M., {Artis}, E., {et~al.} 2019, arXiv e-prints,
  arXiv:1902.10541.
\newblock \doarXiv{1902.10541}

\bibitem[{{Hansen} {et~al.}(2004){Hansen}, {Cabella}, {Marinucci}, \&
  {Vittorio}}]{hansen2004asymmetries}
{Hansen}, F.~K., {Cabella}, P., {Marinucci}, D., \& {Vittorio}, N. 2004, \apjl,
  607, L67, \dodoi{10.1086/421904}

\bibitem[{{Hansen} {et~al.}(2012){Hansen}, {Kim}, {Frejsel}, {Ramazanov},
  {Naselsky}, {Zhao}, \& {Burigana}}]{hansen2012can}
{Hansen}, M., {Kim}, J., {Frejsel}, A.~M., {et~al.} 2012, \jcap, 2012, 059,
  \dodoi{10.1088/1475-7516/2012/10/059}

\bibitem[{{Harrington} {et~al.}(2016){Harrington}, {Marriage}, {Ali}, {Appel},
  {Bennett}, {Boone}, {Brewer}, {Chan}, {Chuss}, {Colazo}, {Dahal}, {Denis},
  {D{\"u}nner}, {Eimer}, {Essinger-Hileman}, {Fluxa}, {Halpern}, {Hilton},
  {Hinshaw}, {Hubmayr}, {Iuliano}, {Karakla}, {McMahon}, {Miller}, {Moseley},
  {Palma}, {Parker}, {Petroff}, {Pradenas}, {Rostem}, {Sagliocca}, {Valle},
  {Watts}, {Wollack}, {Xu}, \& {Zeng}}]{harrington2016cosmology}
{Harrington}, K., {Marriage}, T., {Ali}, A., {et~al.} 2016, in Society of
  Photo-Optical Instrumentation Engineers (SPIE) Conference Series, Vol. 9914,
  Millimeter, Submillimeter, and Far-Infrared Detectors and Instrumentation for
  Astronomy VIII, ed. W.~S. {Holland} \& J.~{Zmuidzinas}, 99141K,
  \dodoi{10.1117/12.2233125}

\bibitem[{{Harrington} {et~al.}(2021){Harrington}, {Datta}, {Osumi}, {Ali},
  {Appel}, {Bennett}, {Brewer}, {Bustos}, {Chan}, {Chuss}, {Cleary}, {Denes
  Couto}, {Dahal}, {D{\"u}nner}, {Eimer}, {Essinger-Hileman}, {Hubmayr}, {Raul
  Espinoza Inostroza}, {Iuliano}, {Karakla}, {Li}, {Marriage}, {Miller},
  {N{\'u}{\~n}ez}, {Padilla}, {Parker}, {Petroff}, {Pradenas M{\'a}rquez},
  {Reeves}, {Flux{\'a} Rojas}, {Rostem}, {Augusto Nunes Valle}, {Watts},
  {Weiland}, {Wollack}, {Xu}, \& {Class Collaboration}}]{harrington2020two}
{Harrington}, K., {Datta}, R., {Osumi}, K., {et~al.} 2021, \apj, 922, 212,
  \dodoi{10.3847/1538-4357/ac2235}

\bibitem[{{Hazumi} {et~al.}(2020){Hazumi}, {Ade}, {Adler}, {Allys}, {Arnold},
  {Auguste}, {Aumont}, {Aurlien}, {Austermann}, {Baccigalupi}, {Banday},
  {Banjeri}, {Barreiro}, {Basak}, {Beall}, {Beck}, {Beckman}, {Bermejo}, {de
  Bernardis}, {Bersanelli}, {Bonis}, {Borrill}, {Boulanger}, {Bounissou},
  {Brilenkov}, {Brown}, {Bucher}, {Calabrese}, {Campeti}, {Carones}, {Casas},
  {Challinor}, {Chan}, {Cheung}, {Chinone}, {Cliche}, {Colombo}, {Columbro},
  {Cubas}, {Cukierman}, {Curtis}, {D'Alessandro}, {Dachlythra}, {De Petris},
  {Dickinson}, {Diego-Palazuelos}, {Dobbs}, {Dotani}, {Duband}, {Duff},
  {Duval}, {Ebisawa}, {Elleflot}, {Eriksen}, {Errard}, {Essinger-Hileman},
  {Finelli}, {Flauger}, {Franceschet}, {Fuskeland}, {Galloway}, {Ganga}, {Gao},
  {Genova-Santos}, {Gerbino}, {Gervasi}, {Ghigna}, {Gjerl{\o}w}, {Gradziel},
  {Grain}, {Grupp}, {Gruppuso}, {Gudmundsson}, {de Haan}, {Halverson},
  {Hargrave}, {Hasebe}, {Hasegawa}, {Hattori}, {Henrot-Versill{\'e}}, {Herman},
  {Herranz}, {Hill}, {Hilton}, {Hirota}, {Hivon}, {Hlozek}, {Hoshino}, {de la
  Hoz}, {Hubmayr}, {Ichiki}, {Iida}, {Imada}, {Ishimura}, {Ishino}, {Jaehnig},
  {Kaga}, {Kashima}, {Katayama}, {Kato}, {Kawasaki}, {Keskitalo}, {Kisner},
  {Kobayashi}, {Kogiso}, {Kogut}, {Kohri}, {Komatsu}, {Komatsu}, {Konishi},
  {Krachmalnicoff}, {Kreykenbohm}, {Kuo}, {Kushino}, {Lamagna}, {Lanen},
  {Lattanzi}, {Lee}, {Leloup}, {Levrier}, {Linder}, {Louis}, {Luzzi},
  {Maciaszek}, {Maffei}, {Maino}, {Maki}, {Mandelli}, {Martinez-Gonzalez},
  {Masi}, {Matsumura}, {Mennella}, {Migliaccio}, {Minami}, {Mitsuda},
  {Montgomery}, {Montier}, {Morgante}, {Mot}, {Murata}, {Murphy}, {Nagai},
  {Nagano}, {Nagasaki}, {Nagata}, {Nakamura}, {Namikawa}, {Natoli}, {Nerval},
  {Nishibori}, {Nishino}, {Noviello}, {O'Sullivan}, {Ogawa}, {Ogawa}, {Oguri},
  {Ohsaki}, {Ohta}, {Okada}, {Okada}, {Pagano}, {Paiella}, {Paoletti},
  {Patanchon}, {Peloton}, {Piacentini}, {Pisano}, {Polenta}, {Poletti},
  {Prouv{\'e}}, {Puglisi}, {Rambaud}, {Raum}, {Realini}, {Reinecke},
  {Remazeilles}, {Ritacco}, {Roudil}, {Rubino-Martin}, {Russell}, {Sakurai},
  {Sakurai}, {Sandri}, {Sasaki}, {Savini}, {Scott}, {Seibert}, {Sekimoto},
  {Sherwin}, {Shinozaki}, {Shiraishi}, {Shirron}, {Signorelli}, {Smecher},
  {Stever}, {Stompor}, {Sugai}, {Sugiyama}, {Suzuki}, {Suzuki}, {Svalheim},
  {Switzer}, {Takaku}, {Takakura}, {Takakura}, {Takase}, {Takeda}, {Tartari},
  {Taylor}, {Terao}, {Thommesen}, {Thompson}, {Thorne}, {Toda}, {Tomasi},
  {Tominaga}, {Trappe}, {Tristram}, {Tsuji}, {Tsujimoto}, {Tucker}, {Ullom},
  {Vermeulen}, {Vielva}, {Villa}, {Vissers}, {Vittorio}, {Wehus}, {Weller},
  {Westbrook}, {Wilms}, {Winter}, {Wollack}, {Yamasaki}, {Yoshida}, {Yumoto},
  {Zannoni}, \& {Zonca}}]{hazumi2020litebird}
{Hazumi}, M., {Ade}, P.~A.~R., {Adler}, A., {et~al.} 2020, in Society of
  Photo-Optical Instrumentation Engineers (SPIE) Conference Series, Vol. 11443,
  Society of Photo-Optical Instrumentation Engineers (SPIE) Conference Series,
  114432F, \dodoi{10.1117/12.2563050}

\bibitem[{{Hinshaw} {et~al.}(1996){Hinshaw}, {Branday}, {Bennett}, {Gorski},
  {Kogut}, {Lineweaver}, {Smoot}, \& {Wright}}]{hinshaw1996two}
{Hinshaw}, G., {Branday}, A.~J., {Bennett}, C.~L., {et~al.} 1996, \apjl, 464,
  L25, \dodoi{10.1086/310076}

\bibitem[{{Hunter}(2007)}]{hunter2007matplotlib}
{Hunter}, J.~D. 2007, Computing in Science and Engineering, 9, 90,
  \dodoi{10.1109/MCSE.2007.55}

\bibitem[{{Kamionkowski} {et~al.}(1997){Kamionkowski}, {Kosowsky}, \&
  {Stebbins}}]{kamionkowski1997statistics}
{Kamionkowski}, M., {Kosowsky}, A., \& {Stebbins}, A. 1997, \prd, 55, 7368,
  \dodoi{10.1103/PhysRevD.55.7368}

\bibitem[{{Kim} \& {Naselsky}(2010)}]{kim2010anomalous}
{Kim}, J., \& {Naselsky}, P. 2010, \apjl, 714, L265,
  \dodoi{10.1088/2041-8205/714/2/L265}

\bibitem[{{Kim} {et~al.}(2012){Kim}, {Naselsky}, \& {Hansen}}]{kim2012symmetry}
{Kim}, J., {Naselsky}, P., \& {Hansen}, M. 2012, Advances in Astronomy, 2012,
  960509, \dodoi{10.1155/2012/960509}

\bibitem[{{Kogut} {et~al.}(2011){Kogut}, {Fixsen}, {Chuss}, {Dotson}, {Dwek},
  {Halpern}, {Hinshaw}, {Meyer}, {Moseley}, {Seiffert}, {Spergel}, \&
  {Wollack}}]{kogut2011primordial}
{Kogut}, A., {Fixsen}, D.~J., {Chuss}, D.~T., {et~al.} 2011, \jcap, 2011, 025,
  \dodoi{10.1088/1475-7516/2011/07/025}

\bibitem[{{Land} \& {Magueijo}(2005)}]{land2005universe}
{Land}, K., \& {Magueijo}, J. 2005, \prd, 72, 101302,
  \dodoi{10.1103/PhysRevD.72.101302}

\bibitem[{{Lee} {et~al.}(2020){Lee}, {Choi}, {G{\'e}nova-Santos}, {Hattori},
  {Hazumi}, {Honda}, {Ikemitsu}, {Ishida}, {Ishitsuka}, {Jo}, {Karatsu},
  {Kiuchi}, {Komine}, {Koyano}, {Kutsuma}, {Mima}, {Minowa}, {Moon}, {Nagai},
  {Nagasaki}, {Naruse}, {Oguri}, {Otani}, {Peel}, {Rebolo},
  {Rubi{\~n}o-Mart{\'\i}n}, {Sekimoto}, {Suzuki}, {Taino}, {Tajima}, {Tomita},
  {Uchida}, {Won}, \& {Yoshida}}]{lee2020groundbird}
{Lee}, K., {Choi}, J., {G{\'e}nova-Santos}, R.~T., {et~al.} 2020, Journal of
  Low Temperature Physics, 200, 384, \dodoi{10.1007/s10909-020-02511-5}

\bibitem[{{Li} {et~al.}(2019){Li}, {Chen}, {Cai}, \& {Mao}}]{li2019testing}
{Li}, B., {Chen}, Z., {Cai}, Y.-F., \& {Mao}, Y. 2019, \mnras, 487, 5564,
  \dodoi{10.1093/mnras/stz1619}

\bibitem[{{Monteser{\'\i}n} {et~al.}(2008){Monteser{\'\i}n}, {Barreiro},
  {Vielva}, {Mart{\'\i}nez-Gonz{\'a}lez}, {Hobson}, \&
  {Lasenby}}]{monteserin2008low}
{Monteser{\'\i}n}, C., {Barreiro}, R.~B., {Vielva}, P., {et~al.} 2008, \mnras,
  387, 209, \dodoi{10.1111/j.1365-2966.2008.13149.x}

\bibitem[{{Muir} {et~al.}(2018){Muir}, {Adhikari}, \&
  {Huterer}}]{muir2018covariance}
{Muir}, J., {Adhikari}, S., \& {Huterer}, D. 2018, \prd, 98, 023521,
  \dodoi{10.1103/PhysRevD.98.023521}

\bibitem[{{Muir} \& {Huterer}(2016)}]{muir2016reconstructing}
{Muir}, J., \& {Huterer}, D. 2016, \prd, 94, 043503,
  \dodoi{10.1103/PhysRevD.94.043503}

\bibitem[{{Mukherjee} \& {Souradeep}(2016)}]{mukherjee2016litmus}
{Mukherjee}, S., \& {Souradeep}, T. 2016, \prl, 116, 221301,
  \dodoi{10.1103/PhysRevLett.116.221301}

\bibitem[{{O'Dwyer} {et~al.}(2020){O'Dwyer}, {Copi}, {Nagy}, {Netterfield},
  {Ruhl}, \& {Starkman}}]{o2020hemispherical}
{O'Dwyer}, M., {Copi}, C.~J., {Nagy}, J.~M., {et~al.} 2020, \mnras, 499, 3563,
  \dodoi{10.1093/mnras/staa3049}

\bibitem[{{Panda} {et~al.}(2021){Panda}, {Aluri}, {Samal}, \&
  {Rath}}]{panda2021parity}
{Panda}, S., {Aluri}, P.~K., {Samal}, P.~K., \& {Rath}, P.~K. 2021,
  Astroparticle Physics, 125, 102493,
  \dodoi{10.1016/j.astropartphys.2020.102493}

\bibitem[{{P{\'e}rez-de-Taoro} {et~al.}(2016){P{\'e}rez-de-Taoro},
  {Aguiar-Gonz{\'a}lez}, {C{\'o}zar-Castellano}, {G{\'e}nova-Santos},
  {G{\'o}mez-Re{\~n}asco}, {Hoyland}, {Pel{\'a}ez-Santos}, {Poidevin},
  {Tramonte}, {Rebolo-L{\'o}pez}, {Rubi{\~n}o-Mart{\'\i}n},
  {S{\'a}nchez-de-la-Rosa}, {Vega-Moreno}, {Viera-Curbelo}, {Vignaga}, {Casas},
  {Martinez-Gonzalez}, {Ortiz}, {Aja}, {Artal}, {Cano-de-Diego},
  {de-la-Fuente}, {Mediavilla}, {Ter{\'a}n}, {Villa}, {Harper}, {McCulloch},
  {Melhuish}, {Piccirillo}, \& {Lasenby}}]{perez2016quijote}
{P{\'e}rez-de-Taoro}, M.~R., {Aguiar-Gonz{\'a}lez}, M., {C{\'o}zar-Castellano},
  J., {et~al.} 2016, in Society of Photo-Optical Instrumentation Engineers
  (SPIE) Conference Series, Vol. 9906, Ground-based and Airborne Telescopes VI,
  ed. H.~J. {Hall}, R.~{Gilmozzi}, \& H.~K. {Marshall}, 99061K,
  \dodoi{10.1117/12.2233225}

\bibitem[{{\sorthelp{Planck Collaboration 2018D}}{Planck Collaboration
  IV}(2020)}]{planck2016-l04}
{\sorthelp{Planck Collaboration 2018D}}{Planck Collaboration IV}. 2020, \aap,
  641, A4, \dodoi{10.1051/0004-6361/201833881}

\bibitem[{{\sorthelp{Planck Collaboration 2018E}}{Planck Collaboration
  V}(2020)}]{aghanim2020planck}
{\sorthelp{Planck Collaboration 2018E}}{Planck Collaboration V}. 2020, \aap,
  641, A5, \dodoi{10.1051/0004-6361/201936386}

\bibitem[{{\sorthelp{Planck Collaboration 2018G}}{Planck Collaboration
  VII}(2020)}]{akrami2020planck}
{\sorthelp{Planck Collaboration 2018G}}{Planck Collaboration VII}. 2020, \aap,
  641, A7, \dodoi{10.1051/0004-6361/201935201}

\bibitem[{{\sorthelp{Planck Collaboration IntZU}}{Planck Collaboration Int.
  XLVI}(2016)}]{aghanim2016planck}
{\sorthelp{Planck Collaboration IntZU}}{Planck Collaboration Int. XLVI}. 2016,
  \aap, 596, A107, \dodoi{10.1051/0004-6361/201628890}

\bibitem[{{Ramirez} \& {Schwarz}(2012)}]{ramirez2012predictions}
{Ramirez}, E., \& {Schwarz}, D.~J. 2012, \prd, 85, 103516,
  \dodoi{10.1103/PhysRevD.85.103516}

\bibitem[{{Schwarz} {et~al.}(2016){Schwarz}, {Copi}, {Huterer}, \&
  {Starkman}}]{schwarz2016cmb}
{Schwarz}, D.~J., {Copi}, C.~J., {Huterer}, D., \& {Starkman}, G.~D. 2016,
  Classical and Quantum Gravity, 33, 184001,
  \dodoi{10.1088/0264-9381/33/18/184001}

\bibitem[{{Shiraishi} {et~al.}(2016){Shiraishi}, {Mu{\~n}oz}, {Kamionkowski},
  \& {Raccanelli}}]{shiraishi2016violation}
{Shiraishi}, M., {Mu{\~n}oz}, J.~B., {Kamionkowski}, M., \& {Raccanelli}, A.
  2016, \prd, 93, 103506, \dodoi{10.1103/PhysRevD.93.103506}

\bibitem[{{Spergel} {et~al.}(2003){Spergel}, {Verde}, {Peiris}, {Komatsu},
  {Nolta}, {Bennett}, {Halpern}, {Hinshaw}, {Jarosik}, {Kogut}, {Limon},
  {Meyer}, {Page}, {Tucker}, {Weiland}, {Wollack}, \&
  {Wright}}]{spergel2003first}
{Spergel}, D.~N., {Verde}, L., {Peiris}, H.~V., {et~al.} 2003, \apjs, 148, 175,
  \dodoi{10.1086/377226}

\bibitem[{{Tegmark}(1997)}]{Tegmark1997}
{Tegmark}, M. 1997, \prd, 55, 5895, \dodoi{10.1103/PhysRevD.55.5895}

\bibitem[{{Tegmark} \& {de Oliveira-Costa}(2001)}]{Tegmark2001}
{Tegmark}, M., \& {de Oliveira-Costa}, A. 2001, \prd, 64, 063001,
  \dodoi{10.1103/PhysRevD.64.063001}

\bibitem[{{Tegmark} {et~al.}(2003){Tegmark}, {de Oliveira-Costa}, \&
  {Hamilton}}]{tegmark2003high}
{Tegmark}, M., {de Oliveira-Costa}, A., \& {Hamilton}, A.~J. 2003, \prd, 68,
  123523, \dodoi{10.1103/PhysRevD.68.123523}

\bibitem[{{van der Walt} {et~al.}(2011){van der Walt}, {Colbert}, \&
  {Varoquaux}}]{van2011numpy}
{van der Walt}, S., {Colbert}, S.~C., \& {Varoquaux}, G. 2011, Computing in
  Science and Engineering, 13, 22, \dodoi{10.1109/MCSE.2011.37}

\bibitem[{{Vanneste} {et~al.}(2018){Vanneste}, {Henrot-Versill{\'e}}, {Louis},
  \& {Tristram}}]{vanneste2018quadratic}
{Vanneste}, S., {Henrot-Versill{\'e}}, S., {Louis}, T., \& {Tristram}, M. 2018,
  \prd, 98, 103526, \dodoi{10.1103/PhysRevD.98.103526}

\bibitem[{{Vielva} {et~al.}(2004){Vielva}, {Mart{\'\i}nez-Gonz{\'a}lez},
  {Barreiro}, {Sanz}, \& {Cay{\'o}n}}]{vielva2004detection}
{Vielva}, P., {Mart{\'\i}nez-Gonz{\'a}lez}, E., {Barreiro}, R.~B., {Sanz},
  J.~L., \& {Cay{\'o}n}, L. 2004, \apj, 609, 22, \dodoi{10.1086/421007}

\bibitem[{{Wandelt} {et~al.}(2001){Wandelt}, {Hivon}, \&
  {G{\'o}rski}}]{Wandelt2001}
{Wandelt}, B.~D., {Hivon}, E., \& {G{\'o}rski}, K.~M. 2001, \prd, 64, 083003,
  \dodoi{10.1103/PhysRevD.64.083003}

\bibitem[{{Yoho} {et~al.}(2015){Yoho}, {Aiola}, {Copi}, {Kosowsky}, \&
  {Starkman}}]{yoho2015microwave}
{Yoho}, A., {Aiola}, S., {Copi}, C.~J., {Kosowsky}, A., \& {Starkman}, G.~D.
  2015, \prd, 91, 123504, \dodoi{10.1103/PhysRevD.91.123504}

\bibitem[{{Yoho} {et~al.}(2014){Yoho}, {Copi}, {Starkman}, \&
  {Kosowsky}}]{yoho2014probing}
{Yoho}, A., {Copi}, C.~J., {Starkman}, G.~D., \& {Kosowsky}, A. 2014, \mnras,
  442, 2392, \dodoi{10.1093/mnras/stu942}

\bibitem[{{Zaldarriaga} \& {Seljak}(1997)}]{zaldarriaga1997all}
{Zaldarriaga}, M., \& {Seljak}, U. 1997, \prd, 55, 1830,
  \dodoi{10.1103/PhysRevD.55.1830}

\bibitem[{{Zibin} \& {Contreras}(2017)}]{zibin2017testing}
{Zibin}, J.~P., \& {Contreras}, D. 2017, \prd, 95, 063011,
  \dodoi{10.1103/PhysRevD.95.063011}

\end{thebibliography}

\end{document}